\documentclass[11pt,a4paper,fleqn,epsf]{article}

\usepackage{graphicx}
\usepackage{amsmath}
\usepackage{amssymb}
\usepackage{longtable}
\usepackage{cite}

\newcommand{\be}{\begin{equation}}
\newcommand{\ee}{\end{equation}}
\newcommand{\bea}{\begin{eqnarray}}
\newcommand{\eea}{\end{eqnarray}}
\newcommand{\bi}{\begin{itemize}}
\newcommand{\ei}{\end{itemize}}
\newcommand{\ben}{\begin{enumerate}}
\newcommand{\een}{\end{enumerate}}
\newcommand{\bt}{\begin{tabbing}}
\newcommand{\et}{\end{tabbing}}

\newcommand{\calO}{{\mathcal O}}

\newcommand{\sgn}{\mbox{sgn}}
\newcommand{\tr}{\mbox{tr}}

\newcommand{\dtau}{\Delta \tau}
\newcommand{\Hw}{H_{\rm W}}
\newcommand{\Dw}{D_{\rm W}}
\newcommand{\rmW}{{\rm W}}

\title{ 
\begin{flushright}
\small 
KEK-CP-208 \\
YITP-08-7\\[15mm]
\end{flushright}
Two-flavor QCD simulation with exact chiral symmetry
}

\author{
S.~Aoki$^{a,b}$, 
H.~Fukaya$^c$,
S.~Hashimoto$^{d,e}$, 
K-I.~Ishikawa$^f$, 
K.~Kanaya$^a$, 
\\[1mm]
T.~Kaneko$^{d,e}$, 
H.~Matsufuru$^d$, 
M.~Okamoto$^d$, 
M.~Okawa$^f$, 
T.~Onogi$^g$, 
\\[1mm]
A.~Ukawa$^{a,h}$, 
N.~Yamada$^{d,e}$,
T.~Yoshi\'e$^{a,h}$
\\[1mm]
(JLQCD Collaboration)
\\
\\
\llap{$^a$}
\it 
Graduate School of Pure and Applied Sciences, 
University of Tsukuba, 
\\
\it 
Ibaraki 305-8571, Japan
\\[1mm]
\llap{$^b$}
\it 
Riken BNL Research Center, 
Brookhaven National Laboratory,
\\
\it
Upton, New York 11973, USA
\\[1mm]
\llap{$^c$}
\it 
The Niels Bohr Institute,
The Niels Bohr International Academy,
\\ 
\it
Blegdamsvej 17 DK-2100 Copenhagen {\O},
Denmark
\\[1mm]
\llap{$^d$}
\it 
High Energy Accelerator Research Organization (KEK),
Ibaraki 305-0801, Japan 
\\[1mm]
\llap{$^e$}
\it 
School of High Energy Accelerator Science, 
The Graduate University 
\\
\it 
for Advanced Studies (Sokendai),
Ibaraki 305-0801, Japan
\\[1mm]
\llap{$^f$}
\it 
Department of Physics, Hiroshima University, Hiroshima 739-8526, Japan
\\[1mm]
\llap{$^g$}
\it 
Yukawa Institute for Theoretical Physics, Kyoto University, 
Kyoto 606-8502, Japan
\\[1mm]
\llap{$^h$}
\it 
Center for Computational Sciences,
University of Tsukuba, Ibaraki 305-8577, Japan
}

\date{\today}

\begin{document}
\maketitle
\setlength{\baselineskip}{15pt}

\begin{abstract}

We perform numerical simulations of lattice QCD with 
two flavors of dynamical overlap quarks,
which have exact chiral symmetry on the lattice.
While this fermion discretization is computationally demanding,
we demonstrate the feasibility
to simulate reasonably large and fine lattices 
by a careful choice of the lattice action and algorithmic improvements.
Our production runs are carried out 
on a $16^3 \times 32$ lattice at a single lattice spacing around 0.12~fm.
We explore the sea quark mass region down to $m_{s}/6$, 
where $m_{s}$ is the physical strange quark mass,
for a good control of the chiral extrapolation 
in future calculations of physical observables.
We describe in detail our setup and algorithmic properties 
of the production simulations 
and present results for the static quark potential 
to fix the lattice scale and the locality of the overlap operator.
\end{abstract}


\maketitle

\newpage

\section{Introduction}
\label{sec:intro}

Since lattice QCD emerged as a quantitative tool 
to study non-per\-tur\-ba\-tive aspects of the strong interaction,
enormous efforts have been made to calculate physical observables 
with controlled systematic uncertainties
by large-scale simulations on increasingly finer and larger lattices.
In particular, 
recent algorithmic improvements 
\cite{PHMC:odd_Nf,PHMC:JLQCD,mtsMD,mass_precond1,mass_precond2,DDHMC:1,DDHMC:2,RHMC}
as well as development of computer technology
enable us to approach the chiral regime of QCD
\cite{Spectrum:Nf3:impG+impKS:MILC,Prod_Run:Nf3:RG+Clv:PACS-CS,Prod_Run:Nf2:Plq+Wlsn:CERN,Spectrum:Nf2:Plq+Clv:QCDSF,Spectrum:Nf2:Sym+tmW:ETM,Spectrum:Nf3:RG+DW:RBC+UKQCD,Lat07:JLQCD:Hashimoto,Lat07:JLQCD:Matsufuru,Prod_Run:Nf3:Sym+smrClv:DFHKKKLLSV}
and to include all three light flavors of quarks dynamically 
\cite{Spectrum:Nf3:impG+impKS:MILC,Prod_Run:Nf3:RG+Clv:PACS-CS,Spectrum:Nf3:RG+DW:RBC+UKQCD,Lat07:JLQCD:Hashimoto,Lat07:JLQCD:Matsufuru,QuarkMass:Nf3:CP-PACS+JLQCD,Prod_Run:Nf3:Sym+smrClv:DFHKKKLLSV}.


The remaining crucial step towards {\it the} simulation of QCD 
is to preserve chiral symmetry on the lattice.
Chiral perturbation theory (ChPT) based on this symmetry 
provides a theoretical guidance
to the chiral extrapolation of physical observables
to the quark mass in the real world.
The explicit symmetry breaking in the conventional lattice actions for quarks
distorts chiral behavior of the observables
and 
introduces additional free parameters into ChPT
\cite{WChPT:1,WChPT:2,WChPT:3,SChPT:1,SChPT:2,SChPT:3}.
This makes the chiral extrapolation unstable 
unless 
one simulates a sufficiently wide region of the quark mass 
and the lattice spacing.
Lattice operator mixing 
is another serious obstacle to precise calculations
of hadronic matrix elements, such as the kaon $B$ parameter.
We also note that 
massive Wilson-type Dirac operators are not protected from 
their (near-)zero modes due to the symmetry breaking.
Studies on their spectrum \cite{Algo:a_effects:1,Algo:a_effects:2}
suggest that it is safe to simulate fine and large lattices 
with this type of fermion discretization.


The five dimensional domain-wall formulation \cite{DWF:K,DWF:S,DWF:FS}
restores chiral symmetry in the limit of infinitely large size $L_s$
in the fifth dimension,
while it is $L_s/a$ times more costly 
with respect to the Wilson-type fermions. 
The RBC and UKQCD collaborations 
\cite{Spectrum:Nf3:RG+DW:RBC+UKQCD,Spectrum:Nf2:DBW2+DW:RBC}
have been pursuing large-scale simulations
employing $L_s/a \! \approx \! 10$\,--\,20,
with which 
the symmetry breaking is reduced to 
a level of a few MeV in terms of the additive quark mass renormalization.


There has been no large-scale simulations with negligible symmetry breaking,
which is of course desirable 
and opens new possibility to study unexplored subjects
such as the calculation of the chiral condensate and the pion mass splitting 
through the difference between vector and axial vector current correlators
$\langle VV - AA \rangle$.
The JLQCD collaboration has started such realistic simulations of QCD
employing the overlap formulation \cite{Overlap:NN,Overlap:N}.
Its Dirac operator is
\bea
   D(m)
   & = &
   \left(m_0+\frac{m}{2}\right)
  +\left(m_0-\frac{m}{2}\right)\,
      \gamma_5\, \sgn\left[ H_W(-m_0) \right],
   \label{eqn:overlap}
\eea
where 
$m$ is the quark mass and 
$H_{\rm W}\!=\!\gamma_5\,D_{\rm W}$ 
is the Hermitian Wilson-Dirac operator with a large negative mass $-m_0$.
This formulation exactly\footnote{
Dirac operators satisfying the GW relation approximately 
have been proposed in Refs.\cite{FP,ChiImp,HCF}
and 
have been employed in simulations in 
Refs.\cite{Prod_Run:Nf3:HHNHS:FP,Prod_Run:Nf2:LMO:ChiImp}.
}
satisfies the Ginsparg-Wilson (GW) relation \cite{GWR}
\bea
   \{\gamma_5,D(0)\}
   & = &
   \frac{1}{m_0}\,D(0)\,\gamma_5 D(0),
   \label{eqn:GWR}
\eea
and, hence, has exact chiral symmetry on the lattice
\cite{GWF:HLN,GWF:H,lat_chial_sym}.
Note that, in practical simulations,
the lattice spacing should be sufficiently small to guarantee its locality.


The overlap fermions are however computationally more demanding 
than the domain-wall fermions.
Simulations with dynamical overlap quarks
have been limited to small and coarse lattices
\cite{overlap:HMC:BHEN,overlap:HMC:FKS,overlap:HMC:CKFLS,overlap:HMC:DS,overlap:HMC:CKAFLS}.
The main difficulty arises from 
the discontinuity of the overlap action Eq.~(\ref{eqn:overlap}) 
when an eigenvalue of $\Hw$ changes its sign.
This substantially impairs the efficiency of 
the commonly used Hybrid Monte Carlo (HMC) algorithm 
unless the time consuming reflection/refraction procedure 
\cite{overlap:HMC:FKS} is implemented.
In our dynamical overlap simulations,
we avoid this overhead by suppressing zero modes of $\Hw$ 
with a modification of the lattice action proposed 
in Ref.\cite{exW+extmW:JLQCD}.
While this prevents us from sampling different topological sectors 
in a single simulation, 
the expectation values of physical observables in the QCD vacuum
can be estimated by simulations in fixed topological sectors
\cite{fixed_Q:BCNW,fixed_Q:AFHO}.


In this article, 
we perform the first large-scale simulations with dynamical overlap 
quarks in two-flavor QCD.
The above mentioned setup to fix the topology
enables us to simulate a lattice spacing $a\!\approx\!0.125$~fm on a 2~fm box, 
which is comparable to those in recent studies with other discretizations.
By implementing recent algorithmic improvements
\cite{mtsMD,mass_precond1,mass_precond2},
the quark mass is reduced down to $m_s/6$
to control the chiral extrapolation of physical observables.
As an example,
we present our estimate of the lattice spacing 
through the Sommer scale $r_0$ \cite{r0} extrapolated to the chiral limit.
The locality of the Dirac operator is a non-trivial issue for the GW fermions
and is directly checked on the generated gauge ensembles.
Status report of these production runs can be found 
in Refs.\cite{Lat07:JLQCD:Matsufuru,Lat06:JLQCD:Matsufuru,Lat06:JLQCD:Hashimoto,Lat06:JLQCD:Kaneko,Lat06:JLQCD:Yamada}.
Simulations to study the $\epsilon$-regime at a slightly fine lattice spacing 
have been already presented in Refs.\cite{Lat06:JLQCD:Fukaya,Lat07:JLQCD:Fukaya,e-regime:sigma:Nf2:RG+Ovr:JLQCD:1,e-regime:sigma:Nf2:RG+Ovr:JLQCD:2,e-regime:meson:Nf2:RG+Ovr:JLQCD}.

This paper is organized as follows.
We introduce our lattice action in Sec.~\ref{sec:action}.
Section~\ref{sec:algorithm} is devoted to 
a description of our simulation algorithm.
In Sec.~\ref{sec:prod_run},
we present our choice of simulation parameters
and discuss algorithmic aspects of our production runs in detail.
We present results for the static quark potential 
and the locality of the overlap operator 
in Secs.~\ref{sec:pot} and \ref{sec:locality}, respectively.
Our conclusions are given in Sec.~\ref{sec:conclusion}.

\section{Lattice action}
\label{sec:action}

We employ the overlap quark action Eq.~(\ref{eqn:overlap})
which can be rewritten in terms of the massless Dirac operator as
\bea
   D(m)
   & = &
   \left( 1-\frac{m}{2\,m_0}\right)\, D(0) + m,
   \label{eqn:overlap:nzm}
   \\[2mm]
   D(0)
   & = &
   m_0\,\left( 1 + \gamma_5\,\sgn\left[ H_W(-m_0) \right] \right).
   \label{eqn:overlap:m0}
\eea
The parameter $m_0$ should be adjusted so that 
the overlap operator has good locality properties.
We set $m_0\!=\!1.6$, which was also employed in a previous 
simulation in quenched QCD around our target lattice spacing 
\cite{Spectrum:Nf0:Plq+Ovr:BBGHHLRS}.
The locality with our simulation setup is checked 
on generated gauge ensembles 
in Sec.\ref{sec:prod_run}.


A major problem with the unsmeared Wilson kernel $\Hw$
is the appearance of its (near-)zero modes on relatively coarse lattices.
This makes simulations costly and possibly spoils the locality of $D$.
One way to reduce the localized (near-)zero modes is 
the use of improved gauge actions 
leading to smooth gauge configurations \cite{smooth_conf}.
In our simulations, we employ the Iwasaki gauge action \cite{Iwasaki}
\bea
   S_g 
   & = &
   \beta
   \left\{c_0 \sum_{x,\,\mu <\nu} 
              \frac{1}{3}\, {\rm Re}\,\tr [ 1\!-\!P_{\mu\nu}(x) ]
         +c_1 \sum_{x,\,\mu ,\,\nu} w_{\mu\nu}^R(x_0) 
              \frac{1}{3}\, {\rm Re}\,\tr [ 1\!-\!R_{\mu\nu}(x) ] \right\},
   \label{eqn:Iwasaki}
\eea
where $\beta\!=\!6/g_0^2$, and $P_{\mu\nu}$ and $R_{\mu\nu}$ are 
$1 \times 1$ and $1 \times 2$ Wilson loops in the $(\mu,\nu)$ plane.
Their weights are $c_0\!=\!3.648$ and $c_1\!=\!-0.331$.
In our preparatory study in quenched QCD \cite{Lat06:JLQCD:Yamada},
we find that
(near-)zero modes are remarkably reduced 
and the overlap operator shows better locality properties
by switching the standard plaquette gauge action to this improved action.


We however need to {\it rule out} the appearance of exact zero modes
in order to avoid 
the time-consuming reflection/refraction step \cite{overlap:HMC:FKS}.
To this end, 
we introduce 
two copies of unphysical Wilson fermions with the large negative mass $-m_0$
\cite{exW+extmW:JLQCD,exW:Vranas,exW:RBC}
and two copies of twisted mass ghosts \cite{exW+extmW:JLQCD}
leading to the following auxiliary fermionic determinant
\bea
   \det[ \Delta_\rmW ]
   & = & 
   \frac{\det\left[ \Hw(-m_0)^2 \right]}
        {\det\left[ \Hw(-m_0)^2 + \mu^2 \right]}
   \label{eqn:exW+extmW}.
\eea
The numerator suppresses the zero modes 
during continuous evolutions of the gauge field such as HMC,
whereas 
effects of high modes of $\Hw$ are cancelled by the denominator.
The twisted mass parameter $\mu$ is tuned 
to compromise between the suppression of the zero modes and 
reduction of the $\beta$ shift due to the unphysical fermions.
It should be noted that 
these unphysical fields have a mass of $O(a^{-1})$
and hence they do not change the continuum limit of the theory.
Their effects can be simply considered as a 
modification of the gauge action by 
$\delta S_g \! = \! - \tr[\ln[ \Delta_\rmW ]]$.


The auxiliary determinant $\det[ \Delta_\rmW ]$ fixes 
the net topological charge $Q$ during the HMC update.
%
We note however that
local topological fluctuations are not suppressed in this setup:
actually the topological susceptibility is calculable 
in a topological sector as demonstrated 
in Ref.\cite{chi_t:Nf2:RG+Ovr:JLQCD+TWQCD}.
The correction due to the fixed global topology can be considered as a 
{\it finite size effect}, 
which is suppressed by the inverse space-time volume $1/V$
\cite{fixed_Q:BCNW,fixed_Q:AFHO}.
In addition,
the expectation values of physical observables in the QCD vacuum 
can be estimated by studying their $Q$ dependence 
from simulations in fixed topological sectors \cite{fixed_Q:BCNW,fixed_Q:AFHO}.
The conventional setup, on the other hand, can sample different
topological sectors through lattice dislocations,  {\it i.e.}
discontinuities of a gauge configuration.
They occur by short distance statistical fluctuations and also by the
appearance (or disappearance) of more physical topological objects,
such as instantons, of order of lattice spacing.
Both of these effects yielding the topological tunneling are
increasingly more suppressed as the continuum limit is approached.
Our setup to fix the topology 
(or modified algorithms 
such as in Refs.\cite{tunneling_HMC:GS,tunneling_HMC:CKLS})
is an interesting alternative
in future simulations near the continuum limit.
It is also noteworthy that 
our setup provides a framework useful to study the $\epsilon$-regime of QCD,
as demonstrated in Refs.\cite{e-regime:sigma:Nf2:RG+Ovr:JLQCD:1,e-regime:sigma:Nf2:RG+Ovr:JLQCD:2,e-regime:meson:Nf2:RG+Ovr:JLQCD}.

\section{Simulation algorithm}
\label{sec:algorithm}

\subsection{Multiplication of overlap operator}


A central building block in our HMC program 
is the multiplication of the overlap operator $D(m)$
to a given quark field vector $\phi$.
We evaluate the sign function $\sgn[H_W]$ in $D(m)$ 
with the low mode preconditioning.
Namely, we introduce a threshold $\lambda_{\rmW,\rm th}$ 
in the spectrum of $\Hw$
and normalized eigenmodes $u_k$ $(k\!=\!1,\ldots,N_{\rm ep})$ 
with their eigenvalues $|\lambda_{\rmW,k}| \! \leq \! \lambda_{\rmW, \rm th}$
are determined by the implicitly restarted Lanczos algorithm. 
We denote the number of the low modes thus extracted 
by $N_{\rm ep}$ in the following.
These modes are projected out in the multiplication of $\sgn[\Hw]$
\bea
   \sgn[H_{\rm W}] \phi
   & = & 
   \sum_{k=1}^{N_{\rm ep}}
   \sgn[ \lambda_{\rmW,k} ] \, u_k (u_k^\dagger \phi)
   +
   \sgn[H_W] (1-P_{\rm low})\, \phi,
   \label{eqn:algo:LMP}
\eea
where $P_{\rm low}\!=\!\sum_{k=1}^{N_{\rm ep}} u_k\,u_k^\dagger$ 
is the projection operator on to the eigenspace spanned by $\{u_k\}$.
We also determine the largest eigenvalue $|\lambda_{\rmW, \rm max}|$.
The contribution of higher modes $\sgn[H_W] (1-P_{\rm low}) \phi$
is then estimated by a minmax rational approximation 
\bea
   \sgn[\Hw] 
   & = &  
   \Hw 
   \left(
      p_0 + \sum_{l=1}^{N_{\rm pole}} \frac{p_l}{\Hw^2+q_l}
   \right)
   \label{eqn:algo:sgnHw:rational}
\eea
with the Zolotarev coefficients $p_l$, $q_l$ 
for the range $[\lambda_{\rmW, \rm th},\lambda_{\rmW, \rm max}]$ 
\cite{Zolotarev:1,Zolotarev:2}.
The multiple inversions for $(\Hw^2+q_l)^{-1}$ 
$(l\!=\!1,\ldots,N_{\rm pole})$ can be carried out simultaneously 
by the multi-shift conjugate gradient (CG) algorithm \cite{msCG:1,msCG:2}.
We keep $\lambda_{\rmW, \rm th}$ and $N_{\rm pole}$ constant 
while $N_{\rm ep}$ varies as a result of fixing $\lambda_{\rmW, \rm th}$.
This together with small statistical fluctuation of $\lambda_{\rmW, \rm max}$ 
makes the accuracy and the computational cost 
in the evaluation of $\sgn[\Hw]$ stable.

\subsection{Overlap solver}


We need to solve the linear equation 
\bea
   D(m)\,x = b
   \label{eqn:algo:leq}
\eea
for a given source vector $b$
in preparations of pseudo-fermions and 
calculations of the Molecular Dynamics (MD) forces in HMC.
In the early stage of our simulation,
we evaluate $D(m)^{-1}$ by the nested four dimensional (4D) CG algorithm,
which consists of multi-shift CG for $(\Hw^2+q_l)^{-1}$ 
as the inner solver
and CG for normal equations (CGNE) to evaluate $D(m)^{-1}$ as the outer solver.
As the outer solver proceeds,
the computational cost of the inner solver can be substantially
reduced by adjusting its stopping condition
\bea
   |(\Hw^2+q_l)x_i - b|^2 < \epsilon_i^{\rm ms},
   \label{eqn:algo:relCG:stpcnd:inner}
\eea
where $i$ is the iteration count for the outer solver
\cite{e-regime:GHLW,relCG}.
We employ the relaxed stopping condition outlined in Ref.\cite{relCG}.
This is based on the idea that, as the outer solver proceeds,
the correction to the solution vector $|x_i\!-\!x_{i-1}|$ becomes smaller 
and we do not have to evaluate $D(m)$ with too much accuracy.
Its implementation depends on the outer solver algorithm
and, for CG(NE), the condition is loosened as 
\bea
   \epsilon_i^{\rm ms}
   \propto 
   \sqrt{\zeta_i},
   \hspace{5mm}
   \zeta_i
   = 
   \zeta_{i-1} + \frac{1}{|r_{i-1}|^2},
   \label{eqn:algo:relCG:relaxing}
\eea   
where $r_{i-1}\!=\!D\,x_{i-1}\!-\!b$ 
is the residual for the outer solver at $(i\!-\!1)$-th iteration.
It was observed on small lattices \cite{relCG}
that this relaxation leads to 
roughly a factor of 2 reduction in the computational cost.


For a further improvement in the solver performance,
we later switch to the five dimensional (5D) solver 
proposed in Refs.~\cite{5D_solver:NN,5D_solver:Borici,5D_solver:Edwards}.
In the case of $N_{\rm pole}\!=\!2$ for instance, 
we consider the following 5D matrix to solve Eq.~(\ref{eqn:algo:leq})
\bea
   M_5(m)
   & = &
   \left(
      \begin{array}{llll|l}
         \Hw        &-\sqrt{q_2}&           &           &0                 \\
         -\sqrt{q_2}&-H_W       &           &           &\sqrt{p_2}        \\
                    &           &\Hw        &-\sqrt{q_1}&0                 \\
                    &           &-\sqrt{q_1}&-H_W       &\sqrt{p_1}        \\ 
         \hline
         0          &\sqrt{p_2} &0          &\sqrt{p_1} &R\gamma_5+p_0 \Hw \\
      \end{array}
   \right)
   = 
   \left(
   \begin{array}{l|l}
      A_{11}       & A_{12}
      \\ \hline
      A_{21}       & A_{22}(m)
   \end{array}
   \right)
   \label{eqn:algo:5Dsolv:M5}
\eea
where $p_l$ and $q_l$ are the coefficients 
in the Zolotarev approximation Eq.~(\ref{eqn:algo:sgnHw:rational})
and $R\!=\!(2m_0+m)/(2m_0-m)$.
The Schur decomposition 
\bea
   M_5(m)
   & = &
   \left(
   \begin{array}{ll}
      1                 & 0
      \\ 
      A_{21}A_{11}^{-1} & 1
   \end{array}
   \right)
   \left(
   \begin{array}{ll}
      A_{11}           & 0
      \\ 
      0                & S(m)
   \end{array}
   \right)
   \left(
   \begin{array}{ll}
      1                & A_{11}^{-1}A_{12}
      \\ 
      0                & 1
   \end{array}
   \right)
   \label{eqn:5Dsolv:schur_d}
\eea
contains the Hermitian overlap operator 
as the Schur compliment 
\bea
   S(m)
   & = & 
   A_{22}(m) - A_{21}\,A_{11}^{-1}\,A_{12}
   = 
   \left( m_0 - \frac{m}{2} \right)^{-1} \gamma_5 D(m).
   \label{eqn:5Dsolv:schur_c}
\eea
Its inverse $x\!=\!S(m)^{-1}b$ can be evaluated by solving the 5D equation 
\bea
   M_5(m) \, x_5 = b_5,
   \hspace{5mm}
   x_5
   =
   \left(
   \begin{array}{l}
      \phi
      \\ 
      x
   \end{array}
   \right),
   \hspace{5mm}
   b_5
   = 
   \left(
   \begin{array}{l}
      0
      \\ 
      b
   \end{array}
   \right).
   \label{eqn:algo:5Dsolv:5D_eq}
\eea

We observe that the convergence of this solver can be improved 
by a preconditioning based on the 5D structure $\tilde{M}_5^{-1}M_5$, 
where $\tilde{M}_5$ is obtained from $M_5$ by setting all gauge links to zero.
Note that $\tilde{M}_5$ is local, uniform in space-time,
and easy to invert through its LU decomposition and 
forward/backward substitutions.
This is naturally incorporated into the even-odd preconditioning 
\cite{5D_solver:Edwards}
\bea
   (1-M_{5,\rm ee}^{-1} \, M_{5,\rm eo} \, M_{5,\rm oo}^{-1} \,M_{5,\rm oe})
   x_{5, \rm e}
   =
   b_{5, \rm e}^\prime,
   \hspace{5mm}
   b_{5, \rm e}^\prime
   = 
   M_{5,\rm ee}^{-1}\, 
   (b_{5, \rm e} - M_{5,\rm eo}\, M_{5,\rm oo}^{-1} b_{5, \rm o}),
   \label{eqn:algo:5Dsolv:precond}
\eea
since $M_{5,{\rm ee}({\rm oo})}\!=\!\tilde{M}_{5,{\rm ee}({\rm oo})}$
where the subscripts ``e'' and ``o'' represent even and odd sites.
It turns out in Sec.~\ref{sec:prod_run} that 
this preconditioned solver is roughly a factor of 3 faster 
than the relaxed 4D solver.

The low mode preconditioning Eq.~(\ref{eqn:algo:LMP})
is however not straightforward with the even-odd preconditioning.
We switched it off in simulations with the 5D solver in this article,
but it is implemented in our latest simulations of three-flavor QCD
\cite{Lat07:JLQCD:Matsufuru,Lat07:JLQCD:Hashimoto}.

\subsection{HMC}


In our implementation of HMC, 
we employ a combination of the Hasenbusch preconditioning 
\cite{mass_precond1,mass_precond2}
and the multiple time scale integration for MD \cite{mtsMD},
which has been shown to be very effective in simulations with 
Wilson-type fermions 
\cite{mass_precond+mtsMD:QCDSF,mass_precond+mtsMD:tmQCD}.
In our HMC program with the 4D overlap solver, 
which is referred to as ``HMC-4D'' in the following,
the fermionic determinant is expressed as 
\bea
   \det\left[D(m)^2\right]
   & = & 
   \det\left[ D(m^\prime)^2 \right]\,
   \det\left[ \frac{D(m)^2}{D(m^\prime)^2} \right],
   \label{eqn:hmc:weight:4d}
\eea
where $m^\prime$ is the mass of the Hasenbusch preconditioner.
Two determinants 
as well as $\det[ \Delta_\rmW]$ from the extra-Wilson fermions
are evaluated by introducing three pseudo-fermions 
$\phi_1$, $\phi_2$, $\phi_{\rm W}$: namely
\bea
   \det\left[ D(m^\prime)^2 \right]
   & = &
   \int [d \phi_1^\dagger] [d \phi_1] 
   e^{-S_1}, 
   \hspace{5mm}
   S_1
   = 
   \phi_1^\dagger 
   \left\{ D(m^\prime)^\dagger D(m^\prime) \right\}^{-1} 
   \phi_1,
   \label{eqn:algo:hmc:pf_1}
   \\
   \det\left[ \frac{D(m)^2}{D(m^\prime)^2} \right]
   & = &
   \int [d \phi_2^\dagger] [d \phi_2] 
   e^{-S_2}, 
   \hspace{5mm}
   S_2
   =
   \phi_2^\dagger \,
   D(m^\prime) \left\{ D(m)^\dagger D(m) \right\}^{-1} D(m^\prime)^\dagger
   \phi_2,
   \label{eqn:algo:hmc:pf_2}
\eea
and
\bea
   &&
   \det[ \Delta_\rmW ] 
   = 
   \int [d \phi_\rmW^\dagger] [d \phi_\rmW] 
   e^{-S_\rmW},
   \\
   &&
   S_\rmW
   =
   \phi_\rmW^\dagger 
   D_{\rm tm}(-m_0,\mu) \left\{\Dw(-m_0)^\dagger \Dw(m_0)\right\}^{-1} 
   D_{\rm tm}(-m_0,\mu)^\dagger
   \phi_\rmW,
   \label{eqn:algo:hmc:pf_W}
\eea
where $D_{\rm tm}(-m_0,\mu)\!=\!\Dw(-m_0)+i\mu\gamma_5$ is 
the Dirac operator for the twisted mass Wilson fermions.

The expression of the force associated with the overlap pseudo-fermion 
is already available 
in Refs.~\cite{overlap:HMC:BHEN,overlap:HMC:FKS,overlap:HMC:DS,overlap:HMC:CKAFLS}.
Here, we explicitly write down only the simplest one from $S_1$
\bea
   \frac{d S_1}{d \tau}
   & = &
   - 
   \left( m_0^2 - m^{\prime 2}/4 \right)
   \psi_1^\dagger
   \left\{ 
      \frac{d\, \sgn[\Hw]}{d \tau}\,\gamma_5 
    + \gamma_5\frac{d\, \sgn[\Hw]}{d \tau}
   \right\}
   \psi_1
   \label{eqn:algo:hmc:f_1}
\eea     
where 
$\psi_1 \! = \! \left\{ D(m^\prime)^\dagger D(m^\prime) \right\}^{-1}\phi_1$ 
and the derivative of the sign function is given by 
\bea
   \frac{d\, \sgn[\Hw]}{d \tau}  
   & = &
   \frac{d \Hw}{d \tau}
   \left(
      p_0 + \sum_{l=1}^{N_{\rm pole}} \frac{p_l}{\Hw^2+q_l}
   \right)
   - 
   \sum_{l=1}^{N_{\rm pole}} 
   \frac{p_l\,\Hw}{\Hw^2+q_l} 
   \left\{ \frac{d \Hw}{d \tau} , \Hw \right\}
   \frac{1}{\Hw^2+q_l}.
   \label{eqn:algo:hmc:d_signHw}
\eea
Hence, 
we need to evaluate $\left\{ D(m^\prime)^\dagger D(m^\prime) \right\}^{-1}$ 
by the 4D relaxed CGNE and also have to invoke the multi-shift CG
to calculate $d\, \sgn[\Hw]/d \tau$. 
The forces $F_1$ and $F_2$ from the overlap pseudo-fermion actions 
$S_1$ and $S_2$
are much more expensive to evaluate 
than $F_\rmW$ from $S_\rmW$ and the gauge force $F_g$.

The use of the multiple time scale integration is therefore 
crucial to reduce the computational cost of the MD evolution.
We employ the following three nested loops
\bea
   T_2(\dtau) 
   & = & 
   T_{P,2}\left(\frac{\dtau}{2}\right)\,
   \left\{ 
      T_1\left(\frac{\dtau}{r_\phi} \right) 
   \right\}^{r_\phi}\,
   T_{P,2}\left(\frac{\dtau}{2}\right),
   \label{eqn:MD:T_2}
   \\ 
   T_1\left( \dtau \right) 
   & = & 
   T_{P,1}\left(\frac{\dtau}{2}\right)\,
   \left\{ 
      T_g\left(\frac{\dtau}{r_g} \right) 
   \right\}^{r_g}\,
   T_{P,1}\left(\frac{\dtau}{2}\right),
   \label{eqn:MD:T_1}
   \\
   T_g\left( \dtau \right)
   & = & 
   T_{P,g}\left( \frac{\dtau}{2} \right)\,
   T_{P,\rmW}\left( \frac{\dtau}{2} \right)\,
   T_{U}\left( \dtau \right)\,
   T_{P,\rmW}\left( \frac{\dtau}{2} \right),
   T_{P,g}\left( \frac{\dtau}{2} \right),
   \label{eqn:MD:U+W}
\eea
where 
$T_{U}(\dtau)$ evolves the gauge field by the MD step size $\dtau$, 
and $T_{P,X}(\dtau)$ updates the conjugate momentum 
with the MD force $F_X$ ($X\!=\!1,2,\rmW,g$).
We put $T_{P,\rmW}$ together with $T_{U}$ in the inner-most loop,
otherwise the suppression of the (near-)zero modes of $\Hw$ fails 
by a mismatch between the updated gauge configuration and $F_\rmW$.
The integration scheme can be largely accelerated by an appropriate choice of 
positive integers $r_\phi$ and $r_g$ 
when the magnitude of the forces are well separated from each other.
This point is one of the central issues in the next section.


The reflection/refraction step is designed to deal with the discontinuity 
in the Hamiltonian 
along the MD evolution of the gauge field,
and hence has to be included into $T_U$ in the inner-most loop.
This step requires a significant computational cost to accurately locate
at which point of the MD evolution the sign of an eigenvalue of $\Hw$ changes. 
In our simple implementation, moreover,
it involves two inversions of the overlap operator to evaluate 
the Hamiltonian just before and after the change of the sign\footnote{
Improved implementations of the reflection/refraction step 
are proposed in Refs.\cite{reflection:SD,reflection:EFKS}.
}.
This step therefore could lead to a considerable slow down of simulations.
The determinant $\det[ \Delta_\rmW ]$ enables us to 
avoid this serious overhead.


In HMC with the 5D solver,
which we call ``HMC-5D'' in the following,
we have to modify the implementation of HMC
due to the lack of the low mode preconditioning.
The coefficients $p_l$ and $q_l$ for the 5D solver are determined 
with an appropriate choice of $\lambda_{\rm W,th}$ and $N_{\rm pole}$,
which are kept fixed during our simulation.
This could lead to a sizable error in $\sgn[\Hw]$,
when $\Hw$ has eigenvalues smaller than $\lambda_{\rm W,th}$.
In order to keep the accuracy of $\sgn[\Hw]$ comparable to that in HMC-4D,
the fermionic determinant is modified as 
\bea
   \det\left[D(m)^2\right]
   & = & 
   \det\left[ D^\prime(m^\prime)^2 \right]
   \det\left[ \frac{D^\prime(m)^2}{D^\prime(m^\prime)^2} \right]
   \det\left[ \frac{D(m)^2}{D^\prime(m)^2} \right],
   \label{eqn:hmc:weight:5d}
\eea
where $D^\prime$ represent the {\it less accurate} overlap operator 
without the low mode preconditioning. 
The first two determinants are dealt with by the usual HMC steps,
whereas the last factor is taken into account by the noisy Metropolis test
\cite{noisy_Metropolis}. 
The probability is evaluated as
\bea
   P 
   = 
   \mbox{min}\left\{ 1, e^{-dS} \right\},
   \hspace{5mm}
   dS 
   = 
   | W^{-1}[U_{\rm new}]\,W[U_{\rm old}]\,\xi|^2 - |\xi|^2,
   \label{algo:hmc:nmt}
\eea   
where $\xi$ is a random Gaussian noise vector
and $W[U_{\rm new(old)}]$ is $D(m)/D^{\prime}(m)$ 
on the final (initial) gauge configuration.
Therefore, 
this step needs to invert both of $D(m)$ and $D^{\prime}(m)$
and spends a significant fraction of the total CPU time.

Another difference from HMC-4D is that 
$(D^\dagger D)^{-1}$ is evaluated 
by invoking CGNE {\it twice} for Eq.~(\ref{eqn:algo:5Dsolv:precond}),
since no 5D representation is available for $(D^\dagger D)^{-1}$.
It turns out that, however, 
CGNE is effective in inverting the preconditioned 5D matrix
in Eq.~(\ref{eqn:algo:5Dsolv:precond})
and switching CGNE to MINRES does not lead to 
a substantial reduction in the computational cost.


\subsection{Machine}

Our numerical simulations are carried out on the supercomputer system at KEK. 
This is a combination of 16 nodes of Hitachi SR11000 
and 10 racks of IBM Blue Gene/L,
whose peak speeds are about 2.15 and 57.3~TFLOPS, respectively.
Our measurement of the static quark potential is inexpensive 
and is carried out on the SR11000 computer.
The configuration generation with the above mentioned HMC algorithm 
is computationally intensive and is carried out on Blue Gene/L.
To increase the sustained speed as much as possible,
we employ an assembler code developed by the IBM Japan 
for the multiplication of $\Dw$.
This code makes the best use of the so-called double FPU instructions,
which process complex-arithmetic operations in double precision effectively 
using two arithmetic pipelines.
It also has a good scalability with respect to the number of computing nodes
by using a low-level interface for inter-node communications.
We find that this assembler code is 
roughly a factor of 3 faster than our naive Fortran code.

\section{Production Run}
\label{sec:prod_run}


\subsection{Simulation parameters}


We simulate QCD with two flavors of degenerate up and down quarks
employing the lattice action introduced in Sec.~\ref{sec:action}.
The twisted mass for the auxiliary determinant $\det[ \Delta_\rmW ]$
is set to $\mu\!=\!0.2$ from our studies in quenched QCD
\cite{exW+extmW:JLQCD,Lat06:JLQCD:Yamada}.
Numerical simulations are carried out on a 
$N_s^3 \times N_t\!=\!16^3 \times 32$ 
lattice at a single value of $\beta\!=\!2.30$, 
which is expected to correspond to our target lattice spacing 0.125~fm.
The box size $L$ should be around 2~fm.
In the trivial topological sector,
we simulate six sea quark masses 
listed in Tables~\ref{tbl:run:sim_param:4D} and ~\ref{tbl:run:sim_param:5D}.
From our analysis of the meson spectrum \cite{Spectrum:Nf2:RG+Ovr},
this choice covers a range from $m_s$ down to $m_s/6$ in physical units.
The statistics are 10,000 HMC trajectories at each quark mass 
with the unit trajectory length set to 0.5.
At $m\!=\!0.050$ which roughly corresponds to $\simeq m_s/2$,
we also accumulate 5,000 trajectories 
in the non-trivial topological sectors with $Q\!=\!-2$ and $-4$.
The initial gauge configuration for these runs is prepared 
as in Ref.~\cite{Nf0:Adm}.
The generated gauge configurations are stored on disks every 10 trajectories
for future measurements of physical observables.
These parameters are summarized in 
Tables~\ref{tbl:run:sim_param:4D} and ~\ref{tbl:run:sim_param:5D}.

\begin{table}[tb]
   \vspace{3mm}
   \begin{center}
   \begin{tabular}{ll|ll|llll|lll}
   \hline
   $m$            & $Q$           & 
   $N_{\rm pole}$ & $\lambda_{\rm W,th}$ &
   $m^\prime$     & $N_{\rm MD}$  & $r_\phi$  & $r_g$ & 
   HMC traj.      & $P_{\rm HMC}$
   & time[min]
   \\ \hline
   0.015  &  0 & 10 & 0.108 & 0.2 &  9 & 4 & 5  & 2800 & 0.875(7)  & 50 
   \\
   0.025  &  0 & 10 & 0.108 & 0.2 &  8 & 4 & 5  & 5200 & 0.900(3)  & 41
   \\
   0.035  &  0 & 10 & 0.108 & 0.4 &  6 & 5 & 6  & 4600 & 0.739(7)  & 28
   \\
   0.050  &  0 & 10 & 0.108 & 0.4 &  6 & 5 & 6  & 4800 & 0.781(5)  & 23 
   \\
   0.070  &  0 & 10 & 0.108 & 0.4 &  5 & 5 & 6  & 4500 & 0.818(7)  & 20 
   \\
   0.100  &  0 & 10 & 0.108 & 0.4 &  5 & 5 & 6  & 4600 & 0.852(5)  & 17 
   \\ \hline
   0.050  & $-2$ & 10 & 0.108 & 0.4 &  6 & 5 & 6  & 1100 & 0.762(13) & 24
   \\ \hline 
   \end{tabular}
   \caption{
      Parameters in the production simulations with HMC-4D.
      The right-most column shows the CPU time per 1 HMC trajectory 
      on one rack of Blue Gene/L. 
   }
   \vspace{0mm}
   \label{tbl:run:sim_param:4D}
   \end{center}
\end{table}

\begin{table}[htb]
   \vspace{3mm}
   \begin{center}
   \begin{tabular}{ll|ll|llll|lll}
   \hline
   $m$            & $Q$           & 
   $N_{\rm pole}$ & $\lambda_{\rm W,th}$ &
   $m^\prime$     & $N_{\rm MD}$  & $r_\phi$  & $r_g$ & 
   HMC traj.      & $P_{\rm HMC}$
   & time[min]
   \\ \hline
   0.015  &  0 & 10 & 0.108 & 0.2 & 13 & 6 & 8  & 7200 & 0.686(6) & 26
   \\
   0.025  &  0 & 10 & 0.108 & 0.2 & 10 & 6 & 8  & 4800 & 0.816(5) & 22 
   \\
   0.035  &  0 & 10 & 0.108 & 0.4 & 10 & 6 & 8  & 5400 & 0.875(5) & 19
   \\
   0.050  &  0 & 10 & 0.108 & 0.4 &  9 & 6 & 8  & 5200 & 0.879(5) & 15
   \\
   0.070  &  0 & 10 & 0.108 & 0.4 &  8 & 6 & 8  & 5500 & 0.917(4) & 13 
   \\
   0.100  &  0 & 10 & 0.108 & 0.4 &  7 & 6 & 8  & 5400 & 0.926(3) & 11 
   \\ \hline
   0.050  & $-2$ & 10 & 0.108 & 0.4 &  9 & 6 & 8  & 3900 & 0.882(5) & 15
   \\
   0.050  & $-4$ & 10 & 0.108 & 0.4 &  9 & 6 & 8  & 5000 & 0.872(5) & 15
   \\ \hline
   \end{tabular}
   \caption{
      Parameters in the production simulations with HMC-5D.
      The right-most column shows the CPU time per trajectory.
   }
   \vspace{0mm}
   \label{tbl:run:sim_param:5D}
   \end{center}
\end{table}

\begin{table}[htb]
   \vspace{3mm}
   \begin{center}
   \begin{tabular}{ll|ll|ll|llll|lll}
   \hline
   $\beta$ & $\mu$ & $m$ & $Q$ 
                               & $N_{\rm pole}$
                               & $\lambda_{\rm W,th}$ 
                               & $m^\prime$ 
                               & $N_{\rm MD}$
                               & $r_\phi$
                               & $r_g$
                               & HMC traj.
                               & $P_{\rm HMC}$
                               & time[min]
   \\ \hline
   2.45   &  0.0  & 0.090 &  --  & 12 & 0.096 & 0.4 & 6 & 5 & 6 &  300 & 0.78 & 46
   \\
   2.35   &  0.2  & 0.110 &    0 & 10 & 0.144 & 0.4 & 5 & 5 & 6 & 1200 & 0.87 & 12
   \\ \hline
   \end{tabular}
   \caption{
      Parameters in the test runs with and without 
      the determinant factor Eq.~(\ref{eqn:exW+extmW}).
      The right-most column shows the CPU time per trajectory.
   }
   \vspace{0mm}
   \label{tbl:run:sim_param:test}
   \end{center}
\end{table}


In the course of our calibration of the lattice spacing, 
we investigate the impact of the auxiliary determinant 
on the computational cost at a slightly finner lattice spacing 
at $(\beta,\mu)\!=\!(2.35,0.2)$.
A single quark mass around $m_s$ is simulated 
without the determinant (namely, $\mu\!=\!0.0$)
but with the reflection/refraction procedure.
This run can be compared to one of our simulations with $\mu\!=\!0.2$
at a similar lattice spacing,
which has been reported in Ref.\cite{e-regime:sigma:Nf2:RG+Ovr:JLQCD:2}.
We summarize parameters of these runs in Table~\ref{tbl:run:sim_param:test}.


\begin{figure}[btp]
\begin{center}
   \vspace{5mm}
   \includegraphics[width=0.48\linewidth,clip]{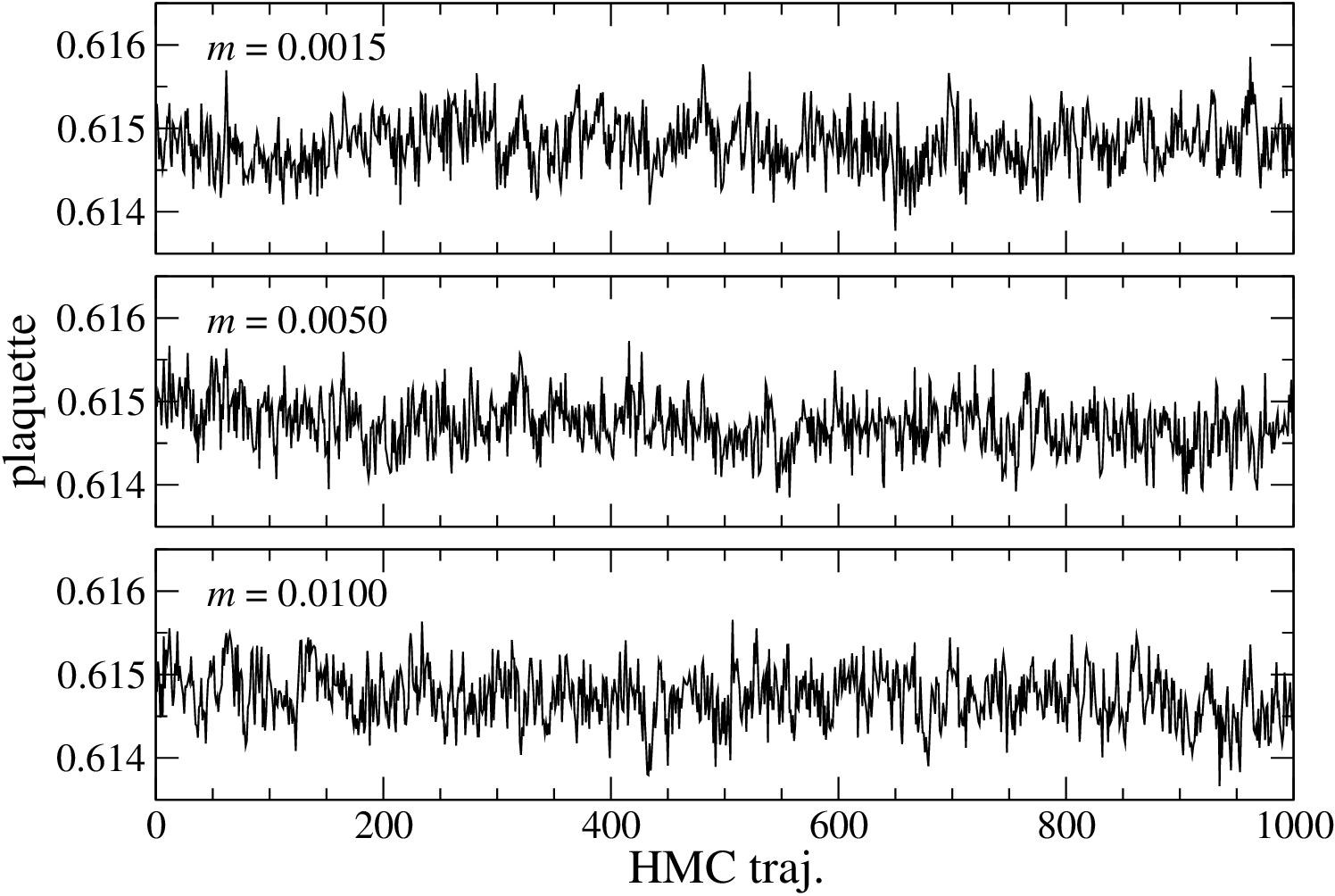}
   \hspace{2mm}
   \includegraphics[width=0.48\linewidth,clip]{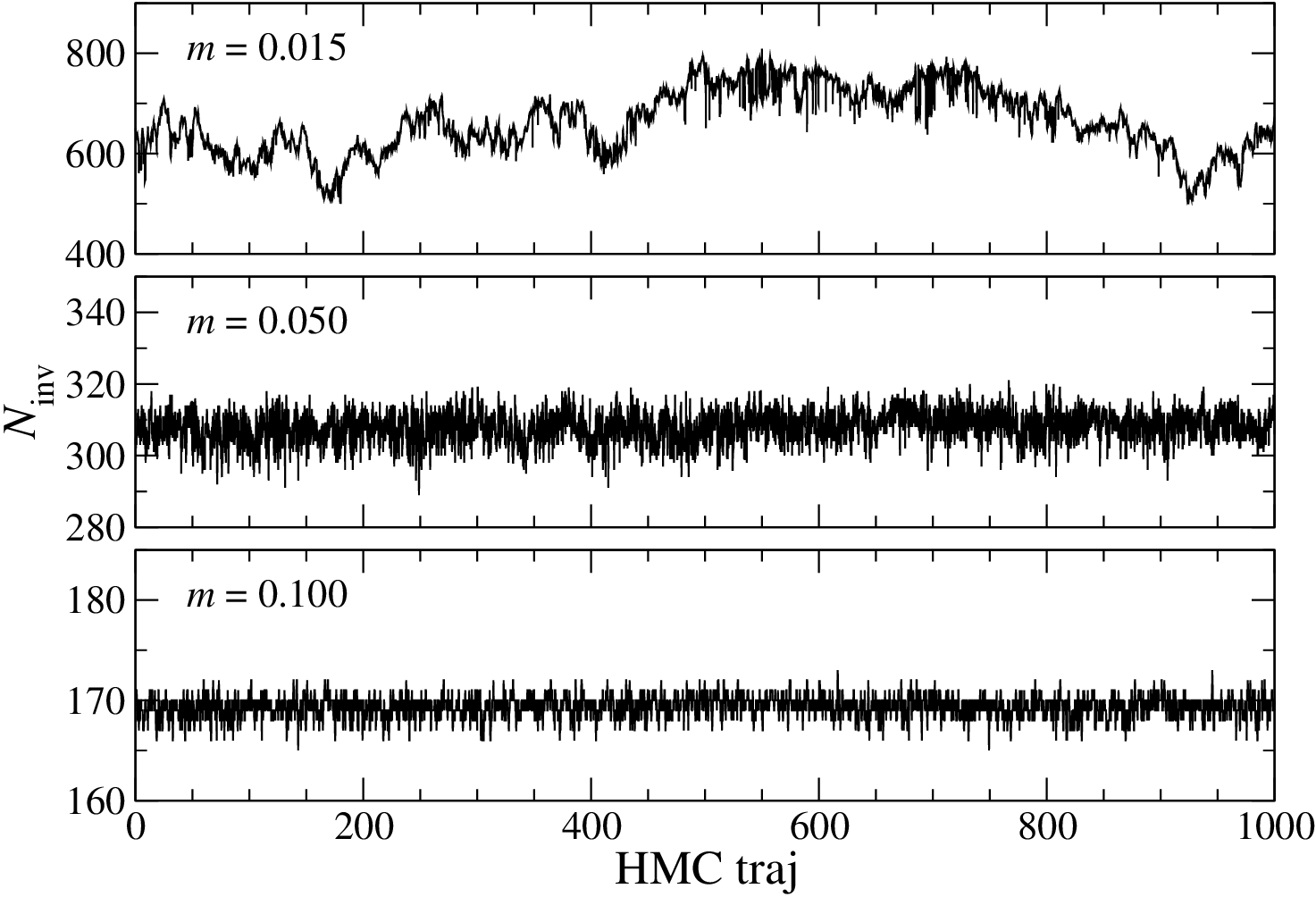}
   \caption{
      Monte Carlo history of plaquette (left panels) and 
      $N_{\rm inv}$ (right panels) 
      during first 1,000 trajectories with HMC-4D.
   }
   \label{fig:run:corr:history}
\end{center}
\end{figure}

\begin{figure}[btp]
\begin{center}
   \vspace{5mm}
   \includegraphics[width=0.48\linewidth,clip]{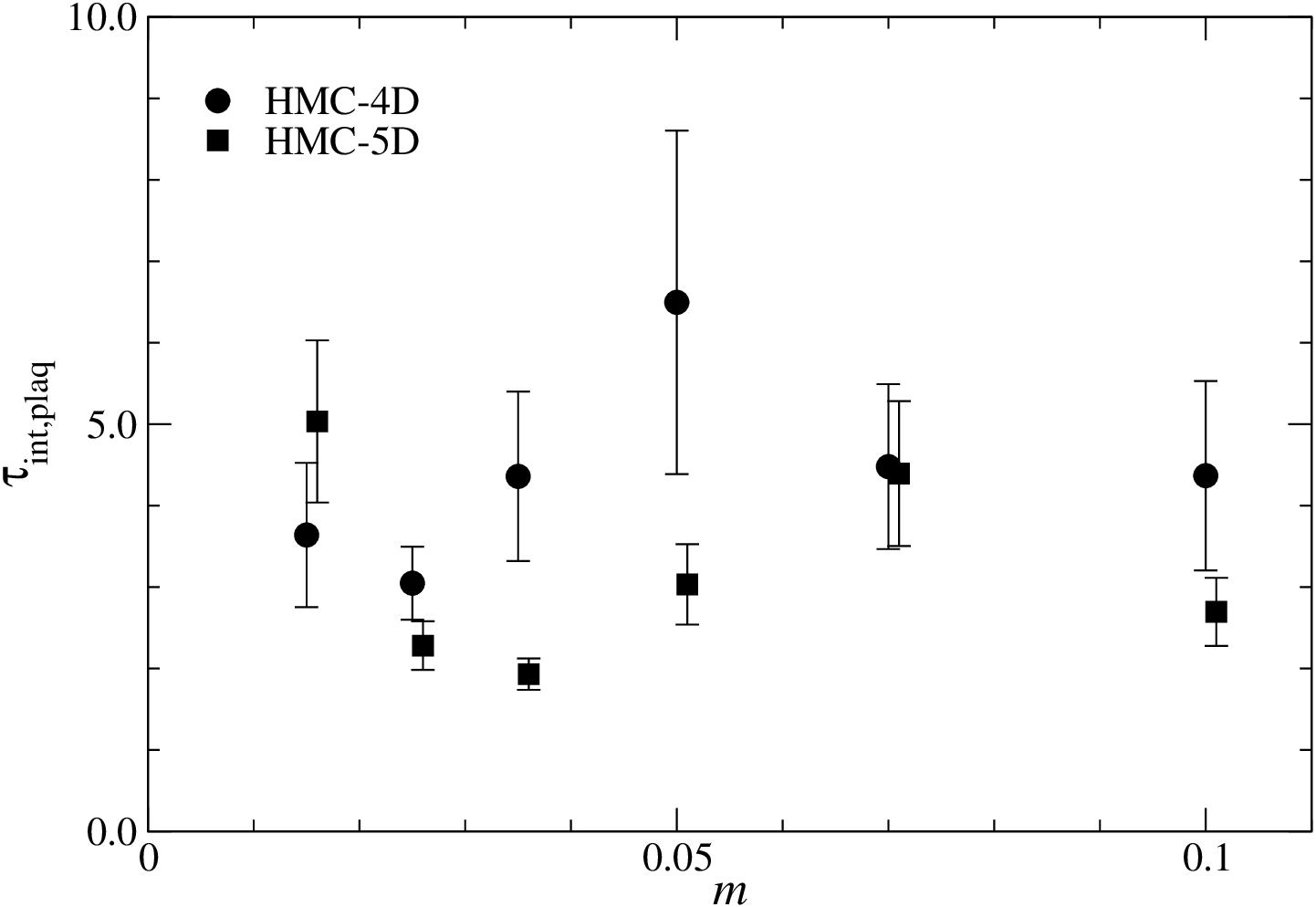}
   \hspace{2mm}
   \includegraphics[width=0.48\linewidth,clip]{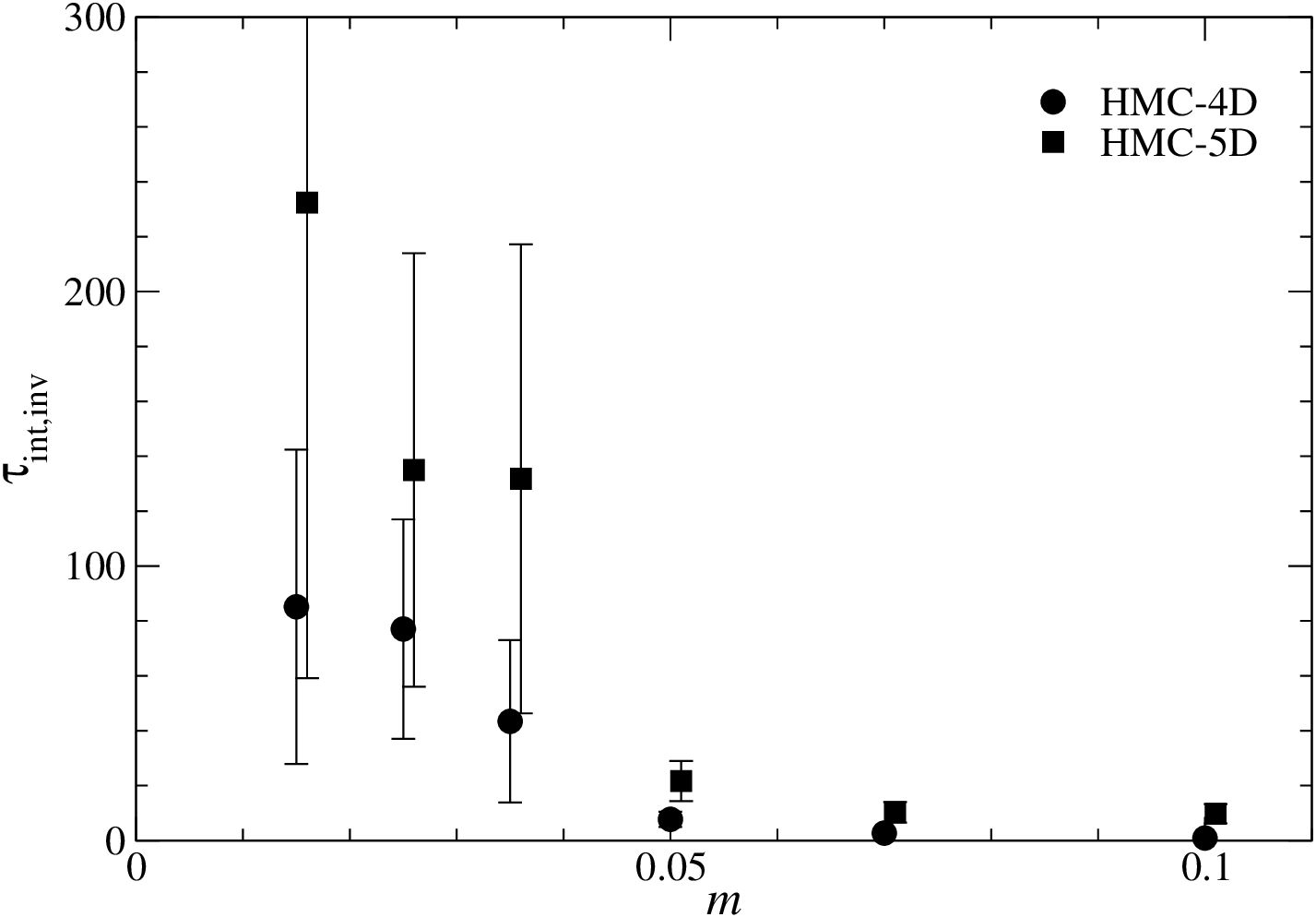}
   \caption{
      Autocorrelation times $\tau_{\rm int,plaq}$ (left panel)
      and $\tau_{\rm int,inv}$ (right panel) 
      as a function of quark mass $m$.
      Data for HMC-5D are slightly shifted 
      in the horizontal direction for clarity. 
   }
   \label{fig:run:corr:tau_int}
\end{center}
\end{figure}

\subsection{Autocorrelation}
\label{sec:corr}

We first discuss the autocorrelation in our simulations 
to fix the bin size used in the jackknife analysis 
in the subsequent sections.
In Fig.~\ref{fig:run:corr:history},
we plot the time history of the plaquette and 
the number of iterations $N_{\rm inv}$ to invert $D(m)$ 
in the simulations with HMC-4D.
It is observed that 
$N_{\rm inv}$ shows longer and larger fluctuations as $m$ decreases,
while such a tendency is not clear in the plaquette.

A conventional measure of the autocorrelation of an observable $\calO$ 
is the integrated autocorrelation time $\tau_{\rm int,\calO}$ 
defined by
\bea
   \tau_{\rm int,\calO}
   & = &
   \frac{1}{2} 
   + 
   \sum_{\tau=1}^{\infty} \rho_\calO(\tau)
   \label{eqn:run:corr:tau_int}
\eea
through the normalized autocorrelation function
\bea
   \rho_\calO(\tau)
   & = & 
   \frac{\Gamma_\calO(\tau)}{\Gamma_\calO(0)},
   \hspace{5mm}
   \Gamma_\calO(\tau) 
   =
   \langle 
      (\calO(\tau_0)- \langle \calO \rangle )
      (\calO(\tau_0+\tau)- \langle \calO \rangle )
   \rangle _{\tau_0},
   \label{eqn:run:corr:af}
\eea
where 
the Monte Carlo average (over $\tau_0$)
is explicitly indicated by the bracket 
$\langle \cdots \rangle_{(\tau_0)}$.
Practically, 
the sum in Eq.~(\ref{eqn:run:corr:tau_int}) has to be truncated
at a certain value of $\tau\!=\!\tau_{\rm max}$.
In this analysis,
we employ the condition adopted in Ref.\cite{DDHMC:2}:
namely, $\tau_{\rm max}$ is set to the minimum value of $\tau$ satisfying
\bea
   \rho_\calO(\tau) - \delta \rho_\calO(\tau) 
   \leq 
   0   
   \label{eqn:run:corr:MS-windos},
\eea
where 
$\delta \rho_\calO(\tau)$ is the standard deviation of $\rho_\calO(\tau)$ 
estimated by the Madras-Sokal formula
\cite{tau_int:MS:1,tau_int:MS:2}.

\begin{figure}[tbp]
\begin{center}
   \vspace{5mm}
   \includegraphics[width=0.48\linewidth,clip]{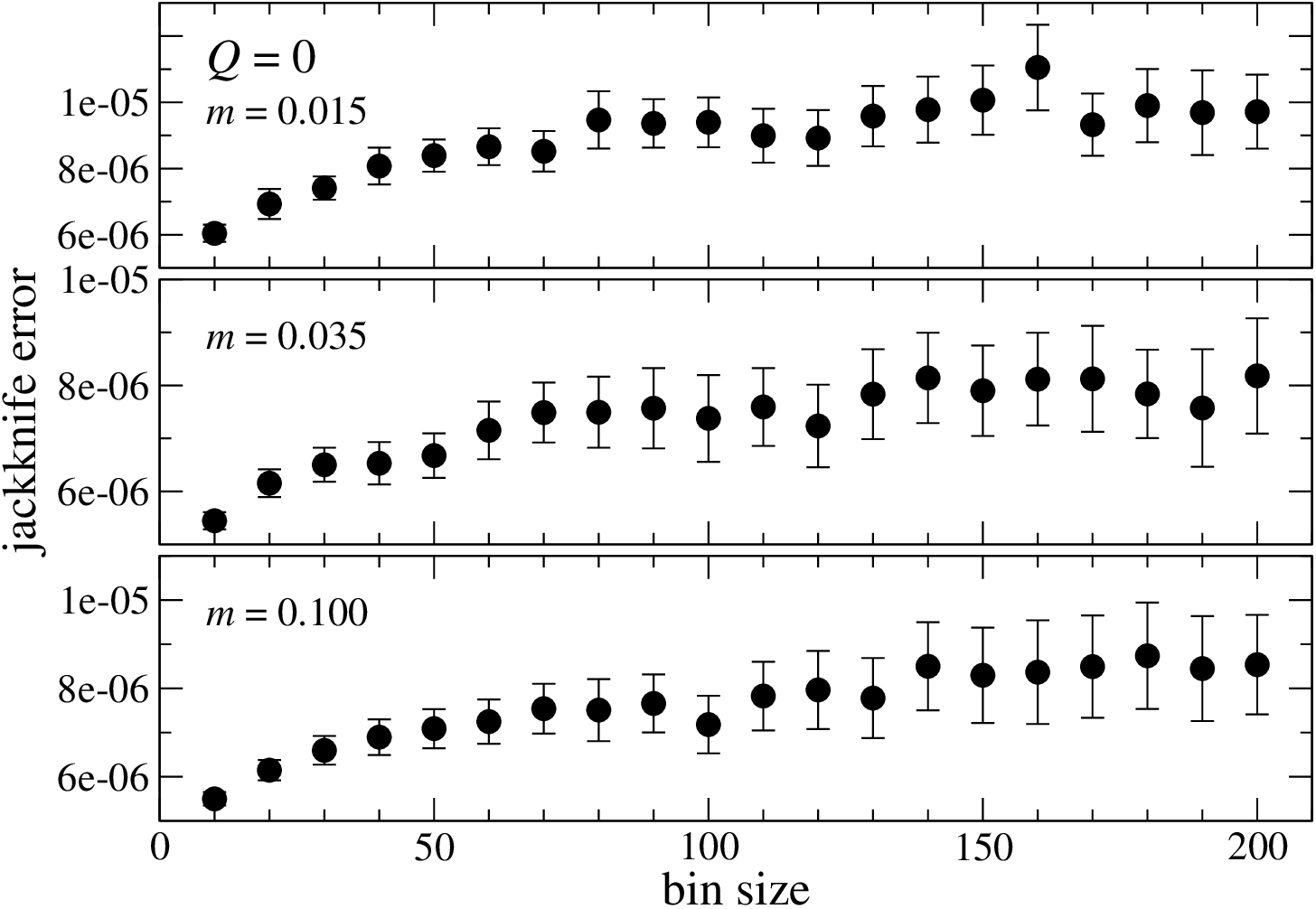}
   \hspace{2mm}
   \includegraphics[width=0.48\linewidth,clip]{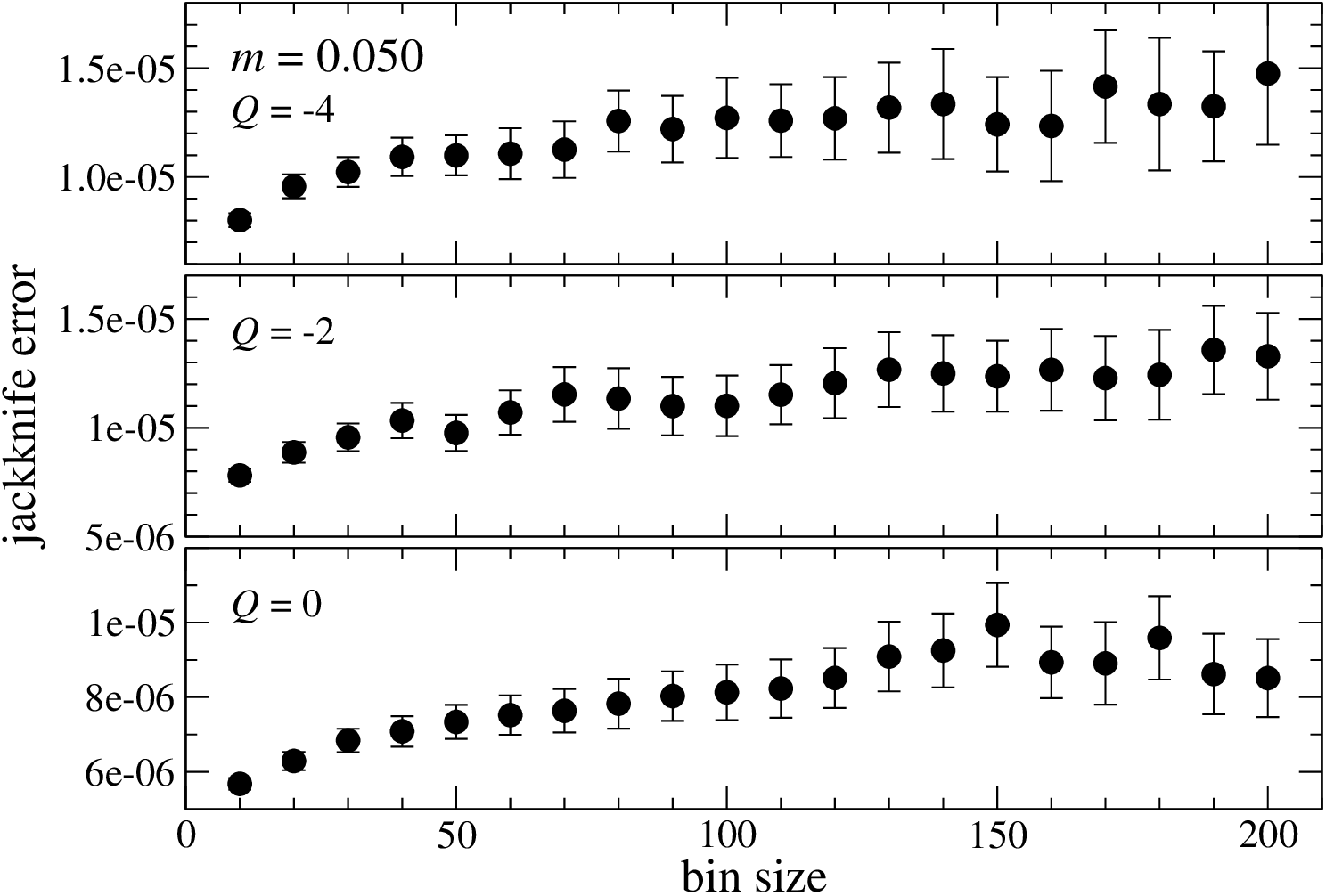}
   \caption{
      Jackknife error of combined data of plaquette as a function of bin size.
      Left and right panels show data at $Q\!=\!0$ and at $m\!=\!0.050$.
   }
   \label{fig:run:corr:jke:plaq}
\end{center}
\end{figure}

\begin{figure}[tbp]
\begin{center}
   \vspace{5mm}
   \includegraphics[width=0.48\linewidth,clip]{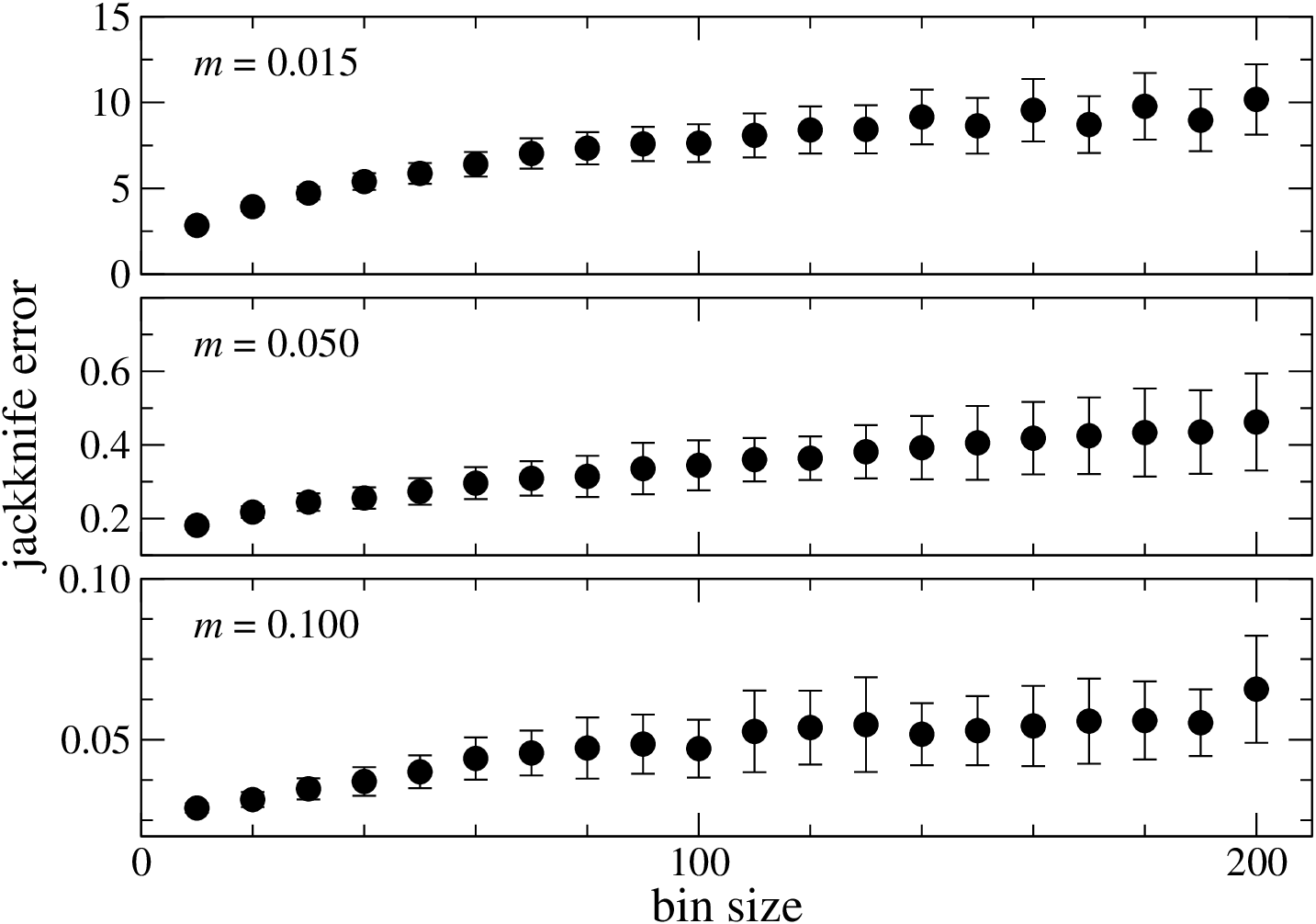}
   \caption{
      Jackknife error of $N_{\rm inv}$ in simulations with HMC-4D.
   }
   \label{fig:run:corr:jke:Ninv}
\end{center}
\end{figure}

In Fig.\ref{fig:run:corr:tau_int},
we plot $\tau_{\rm int,plaq}$ for the plaquette 
and $\tau_{\rm int,inv}$ for $N_{\rm inv}$ 
as a function of $m$.
It turns out that 
$\tau_{\rm int,plaq}$ is not large (about 5 trajectories) 
and 
has small $m$ dependence,
probably because it is a local quantity.
On the other hand,
$N_{\rm inv}$ is expected to be sensitive to the low modes of $D(m)$,
and in fact
$\tau_{\rm int,inv}$ increases to $O(100\mbox{\,--\,}200)$ trajectories
at small quark masses $m\! \lesssim \! 0.035$.
While these observations are consistent with Fig.~\ref{fig:run:corr:history},
it is clear that our statistics are not sufficiently large to estimate
$\tau_{\rm int,inv}$ accurately at small $m$.

Therefore, 
we also check the bin size dependence of the jackknife error
in Fig.~\ref{fig:run:corr:jke:plaq},
where two data of the plaquette obtained with HMC-4D and 5D are combined.
Roughly speaking, 
the jackknife error becomes stable
when the bin size is $\gtrsim 100$\,--\,200 trajectories
irrespective to the choice of $m$ and $Q$.
At similar bin sizes,
the jackknife error of $N_{\rm inv}$
also becomes stable as shown in Fig.~\ref{fig:run:corr:jke:Ninv}.
From these observations, 
we employ the bin size of 200 trajectories throughout this article,
unless otherwise stated.
Although 
only a limited number of bins are available to analyze 
algorithm-dependent quantities, such as $N_{\rm inv}$,
in simulations with HMC-4D at $(m,Q)\!=\!(0.015,0)$ and $(0.050,-2)$,
it turns out that 
decreasing the bin size leads to 
an even smaller statistical error for such quantities.
For a reference, 
the plaquette averaged over our full statistics 
and its jackknife errors are summarized
in Tables~\ref{tbl:run:plaq:Q0} and \ref{tbl:run:plaq:nzQ}.

\begin{table}[tb]
   \vspace{3mm}
   \begin{center}
   \begin{tabular}{l|llllll}
   \hline
   $m_{\rm sea}$ & 0.015 & 0.025 & 0.035 & 0.050 & 0.070 & 0.100 
   \\ \hline
   plaq.         & 0.614789(10) & 0.614777(9)  & 0.614764(8) 
                 & 0.614718(9)  & 0.614709(10) & 0.614667(9)
   \\ \hline
   \end{tabular}
   \caption{
      Average of plaquette from the production runs with $Q\!=\!0$.
   }
   \vspace{0mm}
   \label{tbl:run:plaq:Q0}
   \end{center}
\end{table}

\begin{table}[tb]
   \vspace{3mm}
   \begin{center}
   \begin{tabular}{l|l|l}
   \hline
   $(m_{\rm sea},Q)$  & (0.050,-2)    & (0.050,-4)
   \\ \hline
   plaq.              & 0.614762(13)  & 0.614704(15)
   \\ \hline
   \end{tabular}
   \caption{
      Average of plaquette at $Q\!\neq\!0$.
   }
   \vspace{0mm}
   \label{tbl:run:plaq:nzQ}
   \end{center}
\end{table}

\begin{figure}[tbp]
\begin{center}
   \vspace{5mm}
   \includegraphics[width=0.48\linewidth,clip]{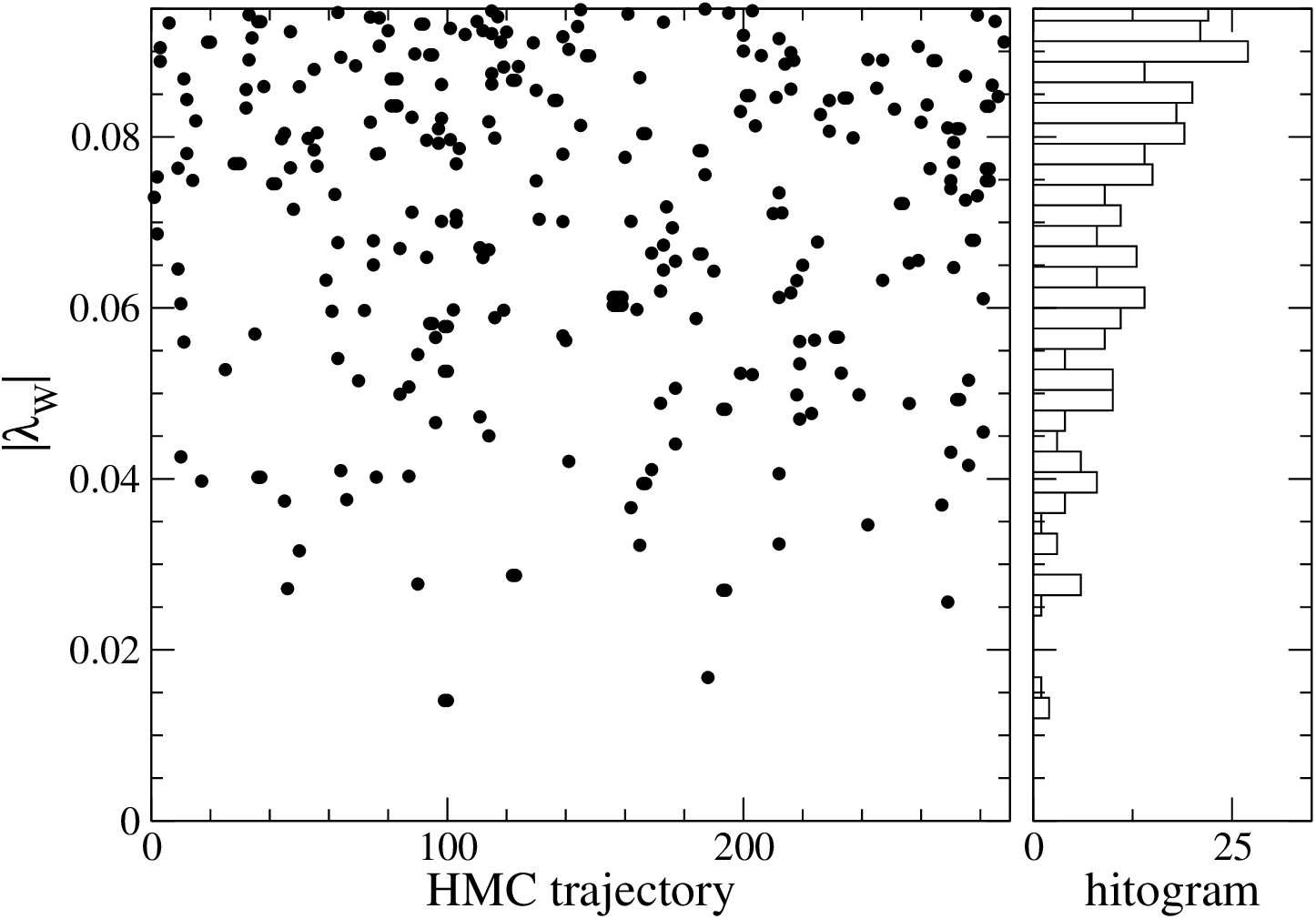}
   \hspace{2mm}
   \includegraphics[width=0.48\linewidth,clip]{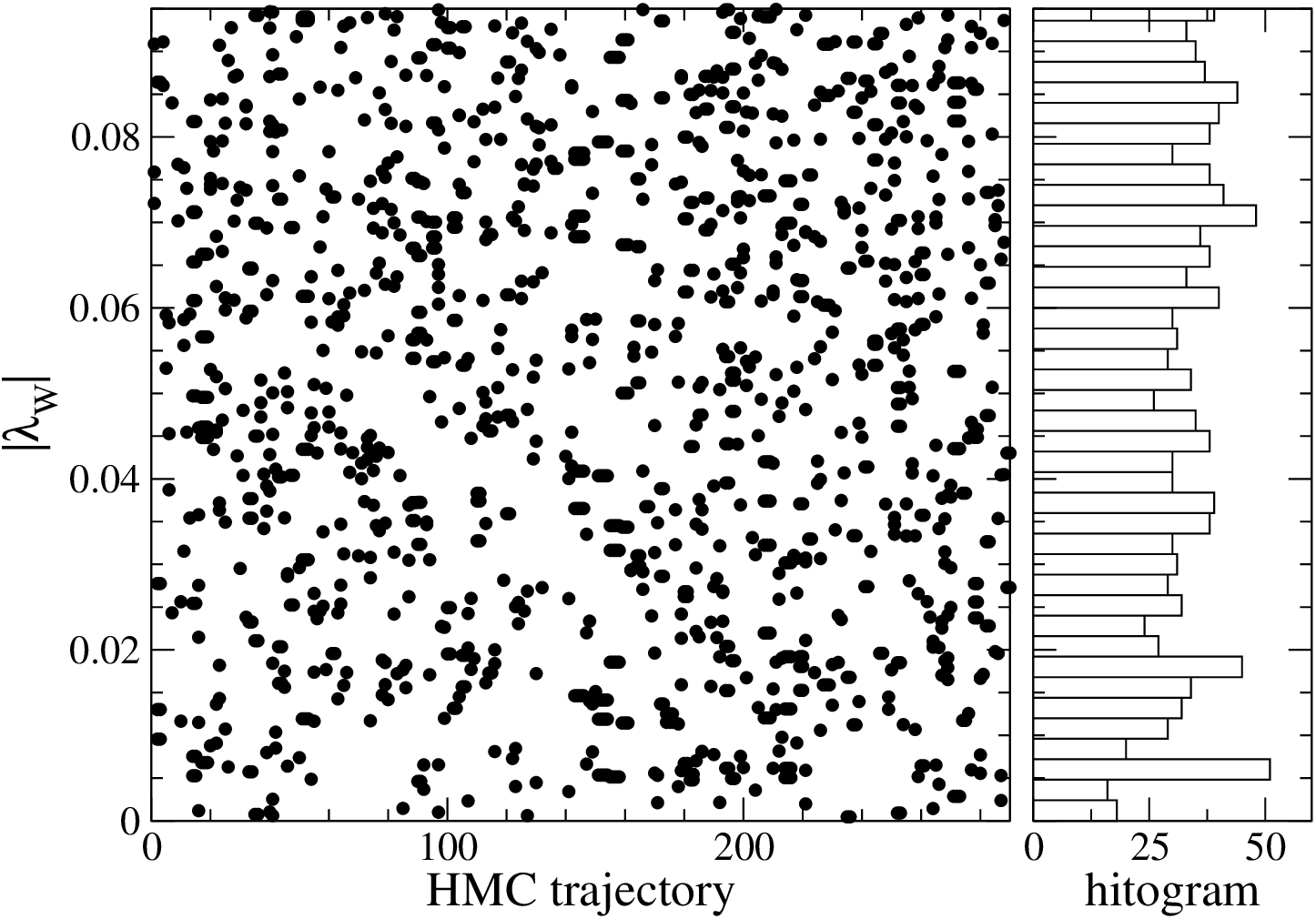}
   \vspace{-3mm}
   \caption{
      Eigenvalues of $\Hw$ smaller than 0.04 
      at each HMC trajectory and its histogram. 
      Left and right panels show data from test runs 
      at $(\beta,\mu)\!=\!(2.35,0.2$) and (2.45,0.0)
      (namely with and without the determinant $\det[ \Delta_\rmW ]$).
      We plot data during the first 300 trajectories in each simulation.
   }
   \label{fig:run:mult:eval:test}
\end{center}
\end{figure}


\subsection{Spectral density of $\Hw$ and $\sgn[\Hw]$}


In Fig.~\ref{fig:run:mult:eval:test},
we compare the low-lying spectrum $\{\lambda_\rmW\}$ of $\Hw$ 
in our test runs listed in Table~\ref{tbl:run:sim_param:test}.
Without the auxiliary determinant $\det[ \Delta_\rmW ]$, 
$|\lambda_\rmW|$ has an almost uniform density 
in the investigated region $|\lambda_\rmW| \! \lesssim \! 0.1$.
This causes the reflection (refraction)
occurring roughly 130 (14) times per 100 trajectories
at our lattice spacing which is only slightly coarser 
than those in recent simulations with Wilson-type actions.
In contrast,
near-zero modes are remarkably suppressed by the determinant.
We can safely turn off the reflection/refraction without
a serious loss in the acceptance rate
and observe about a factor of 4 reduction in CPU time 
as in Table~\ref{tbl:run:sim_param:test}.

\begin{figure}[tbp]
\begin{center}
   \vspace{5mm}
   \includegraphics[width=0.48\linewidth,clip]{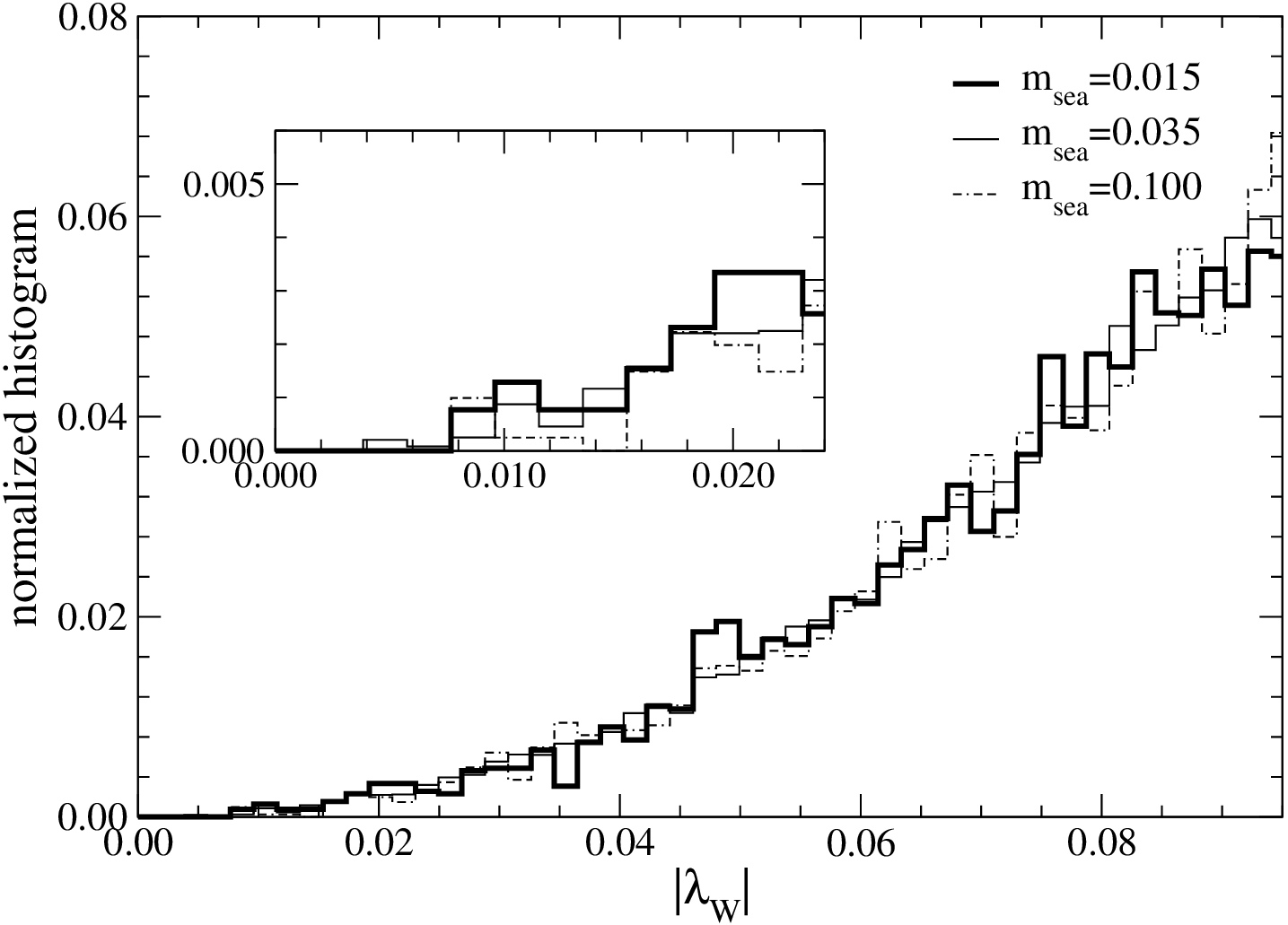}
   \vspace{-3mm}
   \caption{
      Histogram of low-lying engenvalue $|\lambda_\rmW|$ 
      in production runs at $m\!=\!0.015$ (thick solid line),
      0.050 (thin solid line), 
      and 0.100 (thin dot-dashed line).
      The small panel shows the same data near $\lambda_\rmW \!=\! 0$.
   }
   \label{fig:run:mult:eval:histo}
\end{center}
\end{figure}

\begin{figure}[tbp]
\begin{center}
   \vspace{5mm}
   \includegraphics[width=0.48\linewidth,clip]{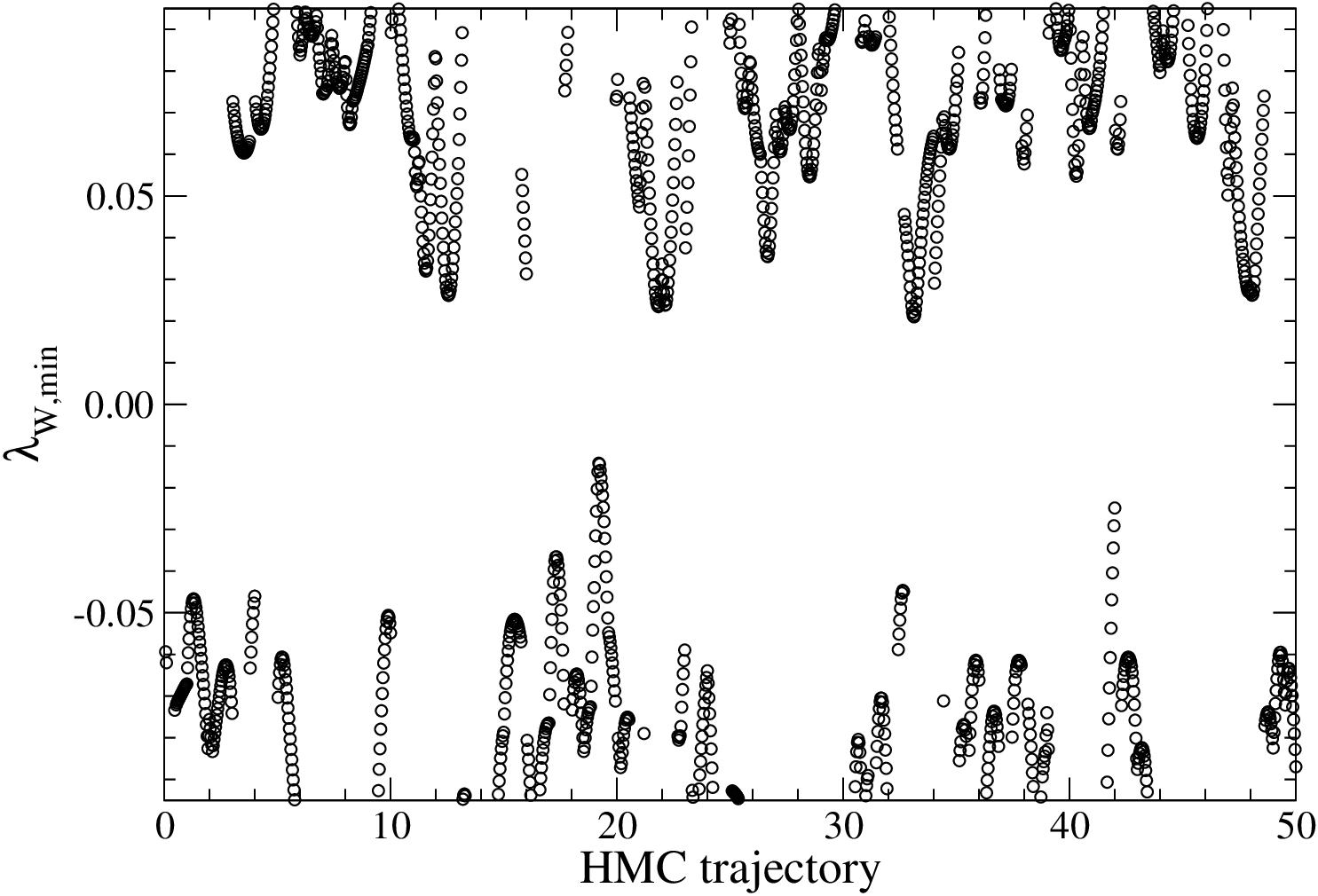}
   \vspace{-3mm}
   \caption{
      Lowest eigenvalue of $\Hw$ during MD evolution
      in first 50 trajectories at $\beta\!=\!2.30$
      and $m\!=\!0.050$.
   }
   \label{fig:run:mult:eval:MD}
\end{center}
\end{figure}

In Fig.~\ref{fig:run:mult:eval:histo}, we plot the eigenvalue distribution 
in our production simulations at several quark masses.
Near-zero modes are successfully suppressed
also in these high statistics runs,
and resulting distribution has small $m$ dependence, as it should.
Figure~\ref{fig:run:mult:eval:MD} shows 
the MD evolution of the lowest eigenvalue $\lambda_{\rm W,min}$.
We observe that 
$\lambda_{\rm W,min}$ approaching to $\lambda_\rmW\!=\!0$
is eventually bounced back 
by the repulsive force from the potential barrier generated by the determinant.

The suppression of near-zero modes enables us to 
take a relatively large value for the threshold $\lambda_{\rm W,th}$ 
and thus small $N_{\rm pole}$ in the multiplication of $D$.
We set $\lambda_{\rm W,th}\!=\!0.108$ at all values of $m$
based on the small $m$ dependence of the low mode distribution 
in Fig.~\ref{fig:run:mult:eval:histo}.
There are only a few eigenvalues below this threshold,
and hence it does not take too much time
to determine such low-lying modes
$(\lambda_{\rmW,k},u_{\rmW, k})$ $(k\!=\!1,\ldots,N_{\rm ep})$
with a strict condition 
\bea
   |(\Hw - \lambda_{\rmW,k})\,u_{\rmW,k}| < 10^{-13}
\eea
for the low-mode preconditioning Eq.~(\ref{eqn:algo:LMP}).
We set $N_{\rm pole}\!=\!10$ with which 
the accuracy of the Zolotarev approximation Eq.~(\ref{eqn:algo:sgnHw:rational})
is typically $|\sgn[\Hw]^2-1| \approx 10^{-7}$.


In simulations with HMC-5D, 
we consider saving the CPU time by loosening the approximation of 
$\sgn[\Hw]$ for $D^\prime(m^\prime)$ and $D^\prime(m)$
in Eq.~(\ref{eqn:hmc:weight:5d}),
since the noisy Metropolis test guarantees that 
gauge configurations are generated with the fermionic determinant
of the accurate overlap operator $\det[D(m)^2]$.
We set $(\lambda_{\rm W,th},N_{\rm pole})\! = \!(0.024,10)$
for the Hasenbusch preconditioner $D^\prime(m^\prime)$.
Since the error of the rational approximation scales as 
$\approx \exp[-\lambda_{\rm W,th}\,N_{\rm pole}]$,
this is less accurate compared to $D(m^\prime)$ with $(0.108,10)$ for HMC-4D.
While Fig.~\ref{fig:run:mult:eval:histo} shows that 
there appear a non-negligible number of eigenmodes 
below $\lambda_{\rm W,th}\!=\!0.024$,
HMC-5D achieves the reasonable acceptance rate 
listed in Table~\ref{tbl:run:sim_param:5D}.
This suggests that 
the error due to the rough approximation in $D^\prime$'s 
as well as that due to the lack of the low mode preconditioning 
are stochastically cancelled (in part) between 
$T_{P,1}$ and $T_{P,2}$ in the MD integration.
We set $(\lambda_{\rm W,th},N_{\rm pole})\! = \!(0.0024,16)$ for $D^\prime(m)$,
since relatively large $\lambda_{\rm W,th}$ 
leads to a substantially poor acceptance rate for the noisy Metropolis test.

\subsection{$\Hw^2$ and overlap solvers}
\label{sec:run:solver}



Our HMC programs involve various stopping conditions.
We need to specify conditions for the 4D overlap solver
\bea
   |Dx-b|\,/\,|b| < \epsilon_{\mbox{\scriptsize 4D,X}}
\eea
and those for the 5D solver
\bea
   | (1-M_{5,\rm ee}^{-1} \, M_{5,\rm eo} \, M_{5,\rm oo}^{-1} \,M_{5,\rm oe})
     x_{5,\rm e} - b_{5,\rm e}^\prime|\,/\,|b| 
   < \epsilon_{\mbox{\scriptsize 5D,X}},
\eea
where
the subscript $X$ is ``$\rm f$'' or ``$H$'' representing 
the condition for the calculation of the MD force or the Hamiltonian.
Note that 
$\epsilon_{\mbox{\scriptsize 5D,X}}$ is the condition 
fulfilled by the 5D preconditioned solution vector $x_{5,\rm e}$. 
Our numerical test suggests that 
$\epsilon_{\mbox{\scriptsize 5D,X}}$ 
should be stricter than $\epsilon_{\mbox{\scriptsize 4D,X}}$
by one or two orders of magnitude
so that the accuracy of the 4D piece $x$ in $x_5$ is comparable to 
that by the 4D solver with $\epsilon_{\mbox{\scriptsize 4D,X}}$.

The stopping condition for the multi-shift CG
inside the 4D overlap solver is automatically determined 
by Eq.~(\ref{eqn:algo:relCG:relaxing})
except for the initial residual $r_0\!=\!D\,x_0-b$.
An appropriately strict value $\leq 10^{-6}$ is employed
to calculate $r_0$ with the given initial guess $x_0$.
Therefore, we only need to specify conditions 
for multiplications of $D$ and calculations of $d\, \sgn[\Hw]/d \tau$
by Eq.~(\ref{eqn:algo:hmc:d_signHw}) 
\bea
   |(\Hw^2+q_l)x-b|\,/\,|b| \leq \epsilon_{\mbox{\scriptsize Y,X}}^{\rm ms}
   \hspace{5mm}
   (l=,1,\ldots,N_{\rm pole})
\eea
with $X\!=\!\mbox{``$\rm f$''}$ or ``$H$'.
Here the condition in HMC-4D (5D) is represented by 
the subscript $Y\!=\!\mbox{``4D''}$ (``5D'').
In our production run, 
we employ a rather strict condition 
$\epsilon_{\scriptsize Y,H}^{\rm ms} \! = \! \epsilon_{\scriptsize Y,H} \! = \! 10^{-10}$
for calculations of the Hamiltonian 
in order to carry out the Metropolis tests accurately.

\begin{figure}[tb]
\begin{center}
   \vspace{5mm}
   \includegraphics[width=0.48\linewidth,clip]{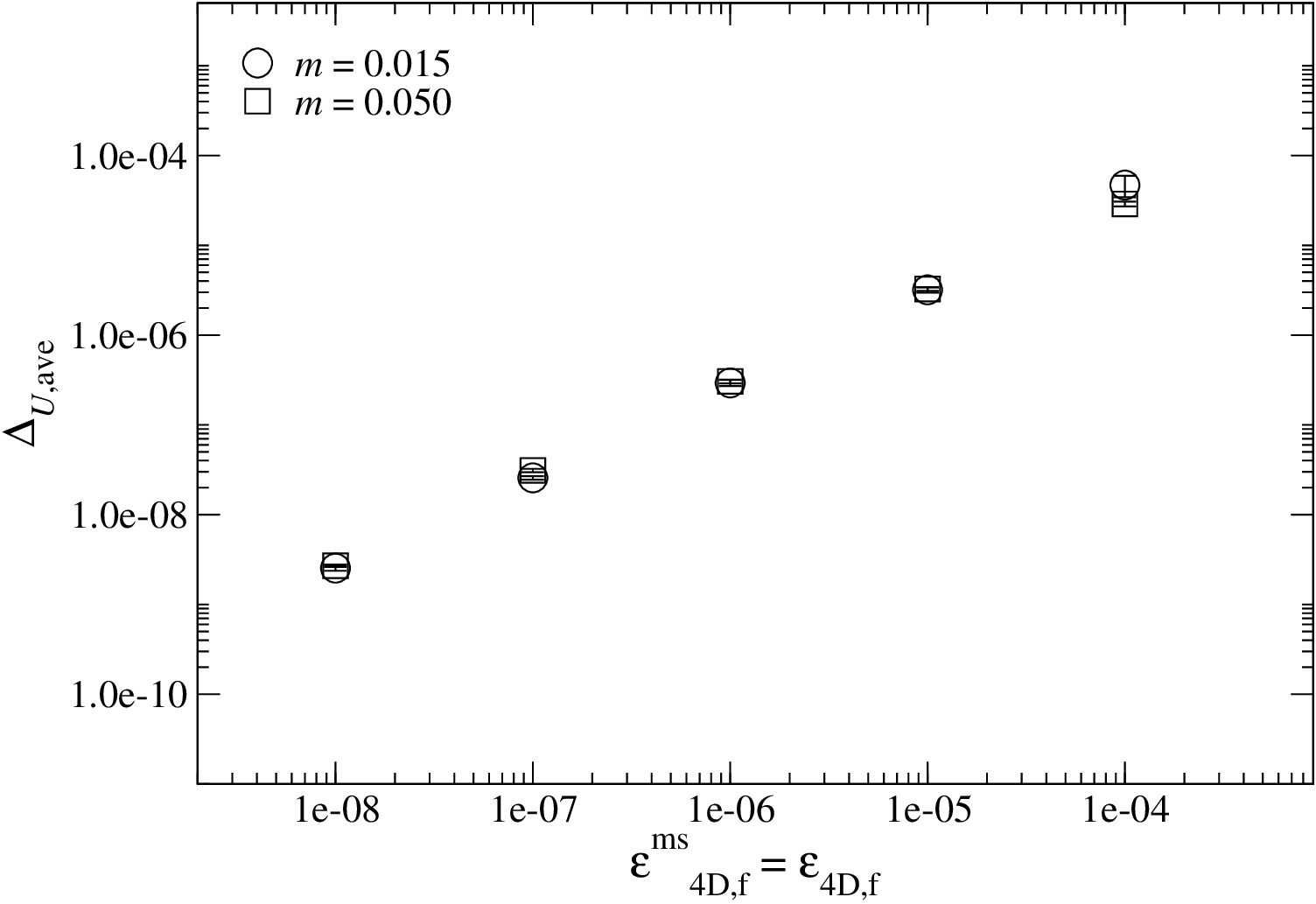}
   \hspace{2mm}
   \includegraphics[width=0.48\linewidth,clip]{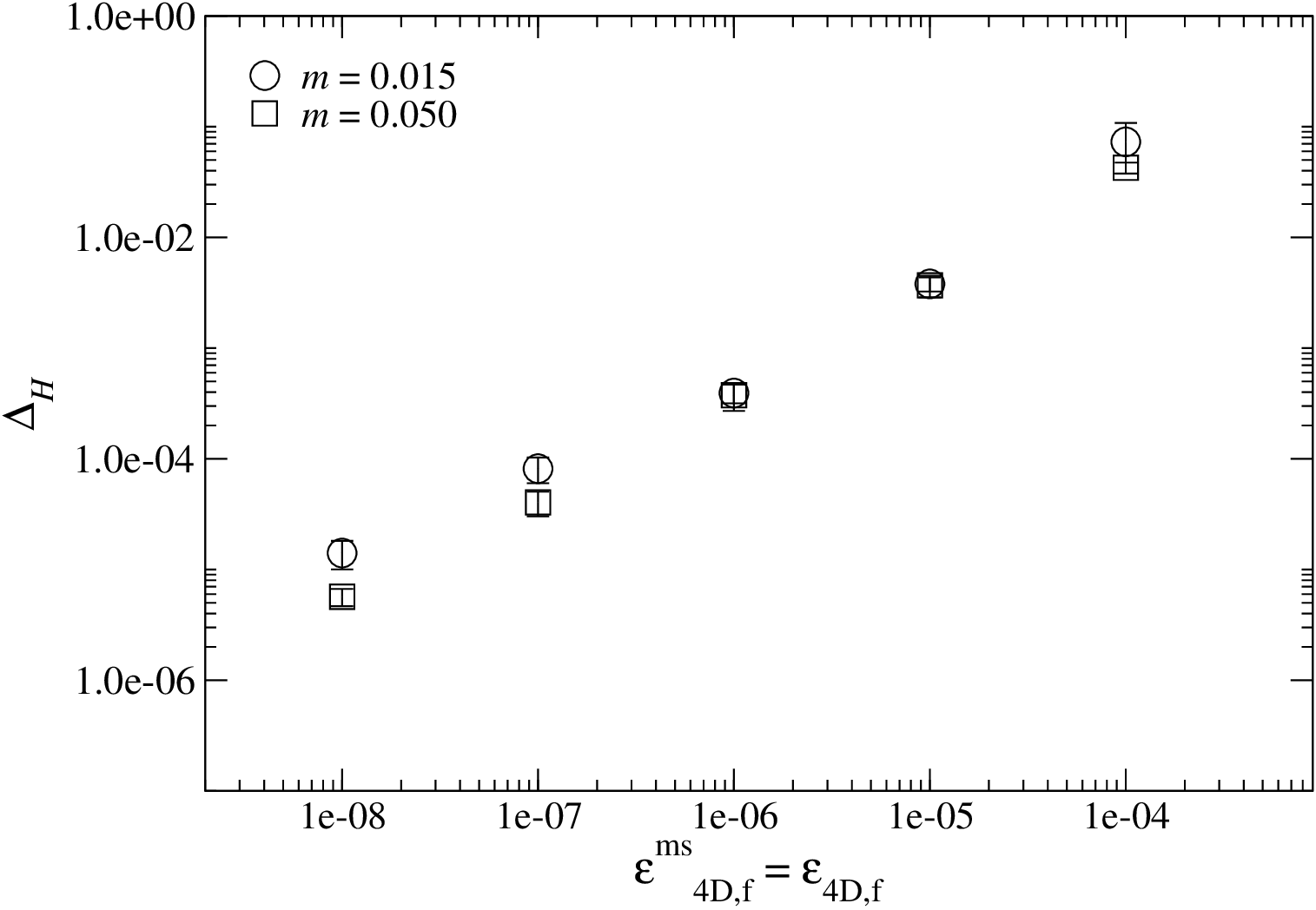}
   \vspace{-3mm}
   \caption{
      Measures of reversibility violation   
      $\Delta_{U,\rm ave}$ (left panel) and 
      $\Delta_H$ (right panel) for HMC-4D
      as a function of stopping conditions
      $\epsilon_{\rm 4D,f}^{\rm ms}\!=\!\epsilon_{\rm 4D,f}$.
   }
   \label{fig:run:solver:rev:4D}
\end{center}
\end{figure}

The choice of $\epsilon_{\scriptsize Y, \rm f}^{\rm ms}$ and 
$\epsilon_{Y, \rm f}$ is crucial to save the computational cost
but should be strict enough to make our HMC reversible.
The conventional measures of the reversibility violation 
in the gauge links $U_{x,\mu}$ and the Hamiltonian $H$ are
\bea
   \Delta_{U,\rm ave}
   & = &
   \sqrt{ \sum_{x,\mu,a,b}
          |U_{x,\mu}^{ab}(\tau+0.5-0.5) - U_{x,\mu}^{ab}(\tau)|^2
          / (36 N_s^3 N_t)
        },
   \label{eqn:run:solver:rev:dU}
   \\
   \Delta_H
   & = &
   |H(\tau+0.5-0.5) - H(\tau)|,
   \label{eqn:run:solver:rev:dH}
\eea
where $a$ and $b$ are color indices and note that 
our unit trajectory length is 0.5.
We pick up ten gauge configurations separated by 200 trajectories, 
and calculate $\Delta_{U,\rm ave}$ and $\Delta_H$
by updating them by one trajectory 
and then evolving back with the reversed momenta.
Figure~\ref{fig:run:solver:rev:4D}
shows these measures in HMC-4D,
for which we set $\epsilon_{\rm 4D,f}^{\rm ms}\!=\!\epsilon_{\rm 4D,f}$ 
for simplicity.
We observe a monotonous decrease in both measures
and small $m$ dependence of their size.
By employing 
\bea
   \epsilon_{\rm 4D,f}^{\rm ms} = \epsilon_{\rm 4D,f} =10^{-7},
\eea
at all quark masses,
the reversibility in our simulations is preserved 
at a level of $\Delta_{U,\rm ave} \! \lesssim \! 10^{-8}$ and 
$\Delta_H \! \lesssim \! 10^{-4}$,
which are comparable to those in previous large-scale simulations 
with the Wilson-type actions
\cite{Simulation:Nf2:SESAM,Spectrum:Nf2:RG+Clv:CP-PACS,Spectrum:Nf2:Plq+Clv:JLQCD}.
A similar study for HMC-5D leads us to set
\bea
   \epsilon_{\rm 5D,f}^{\rm ms} = 10^{-6}, \hspace{5mm}
   \epsilon_{\rm 5D,f}          = 10^{-7}
\eea
in order to maintain the reversibility at the same level to HMC-4D.


\begin{figure}[tbp]
\begin{center}
   \vspace{5mm}
   \includegraphics[width=0.48\linewidth,clip]{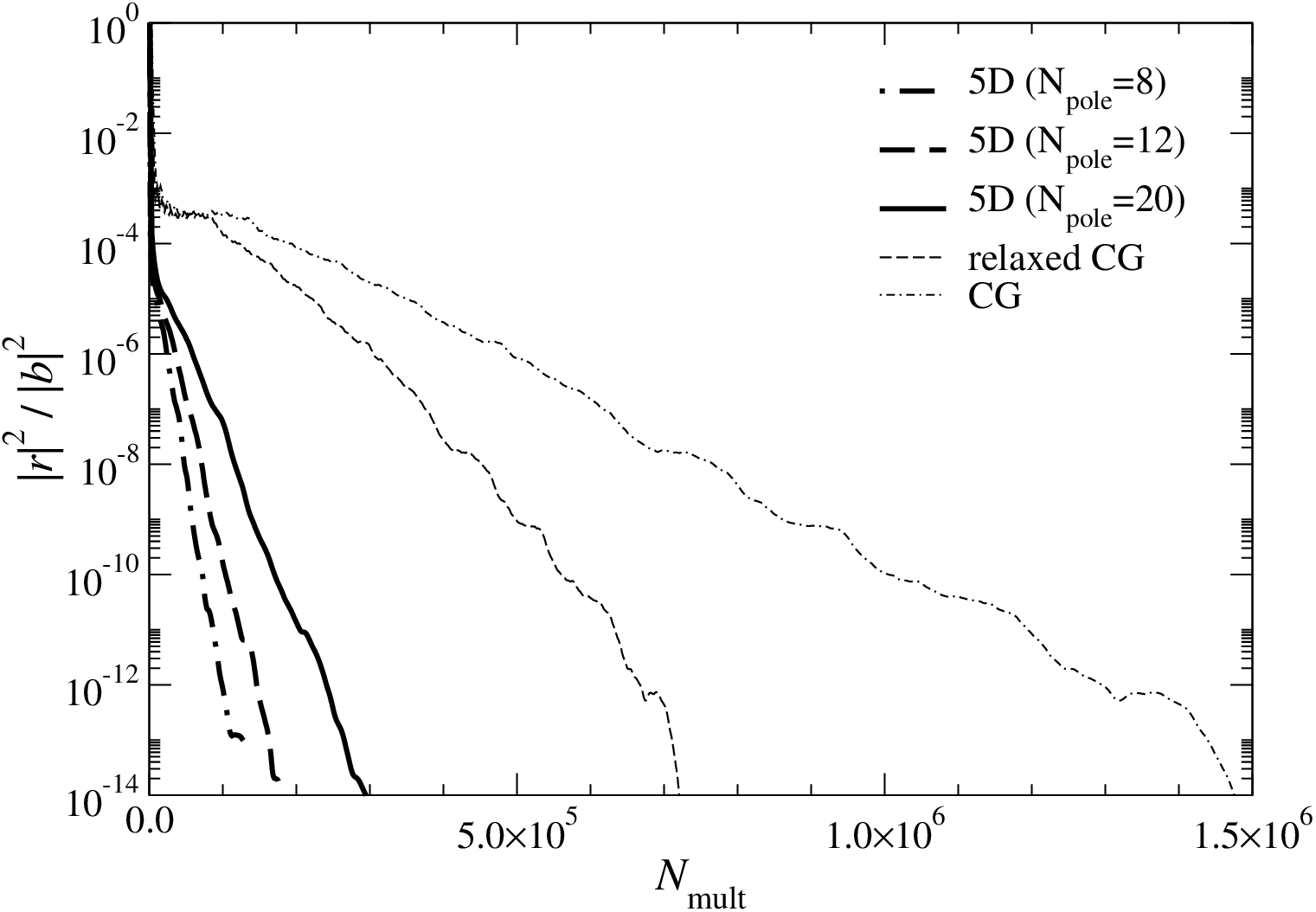}
   \vspace{-3mm}
   \caption{
      Residual of overlap solvers 
      as a function of the number of the $D_\rmW$ multiplication 
      at $m\!=\!0.025$. 
      Thick and thin lines show results for the 5D and 4D solvers.
      For the 5D solver,
      we plot the residual for the 4D piece $x$ of the 5D solution vector 
      $x_5$, namely $r\!=\!D(m)x\!-\!b$.
   }
   \label{fig:run:solver:time:4D_vs_5D}
\end{center}
\end{figure}


In Fig.~\ref{fig:run:solver:time:4D_vs_5D},
we compare the convergence of the 4D and 5D solvers
by plotting the normalized residual $|r|^2/|b|^2$
as a function of the number of the $D_\rmW$ multiplication $N_{\rm mult}$.
We take $m\!=\!0.025$ and turn off the low mode preconditioning 
by setting $\lambda_{\rmW, \rm th}\!=\!0.0$
for a fair comparison between the 4D and 5D algorithms.
The relaxed condition Eq.~(\ref{eqn:algo:relCG:relaxing})
works well on our 2~fm box and 
achieves about a factor of 2 speed up compared to the standard CG.
The 5D solver 
is even faster by about a factor of 3
mainly due to the preconditioning of Eq.~(\ref{eqn:algo:5Dsolv:precond}).
We observe an acceleration of similar magnitude also at other quark masses.


It is an interesting issue how the computational cost scales 
as a function of $m$.
The iteration count of our overlap solver $N_{\rm inv}$ 
depends on $m$ 
through eigenvalues $\lambda_k(m)$ of $D(m)$.
For simplicity,
we use the following approximation for low-lying eigenvalues
\bea
   \lambda_k(m) 
   & = & 
   m + i \tilde{\lambda}_k(0)
\eea
where 
$\tilde{\lambda}_k(0)$ is 
the $k$-th eigenvalue of the massless Dirac operator $D(0)$
projected to the imaginary axis
\bea
   \tilde{\lambda}_k(0)
   & = &
   \frac{\mbox{Im}[\lambda_k(0)]}{1-\mbox{Re}[\lambda_k(0)]/2},
\eea
and we ignore the small correction factor $1\!-\!m/(2m_0) \! \simeq \! 1$
in Eq.~(\ref{eqn:overlap:nzm}).
The $m$ dependence of $N_{\rm inv}$ in HMC-4D is then expected to be
\bea
   N_{\rm inv}
   & \propto &
   \frac{1}{(m^2 + \tilde{\lambda}_1(0)^2)^{\alpha/2}}
   \label{eqn:run:solver:Ninv}
\eea
with the power $\alpha \leq 1$ for CGNE.

\begin{figure}[tbp]
\begin{center}
   \vspace{5mm}
   \includegraphics[width=0.48\linewidth,clip]{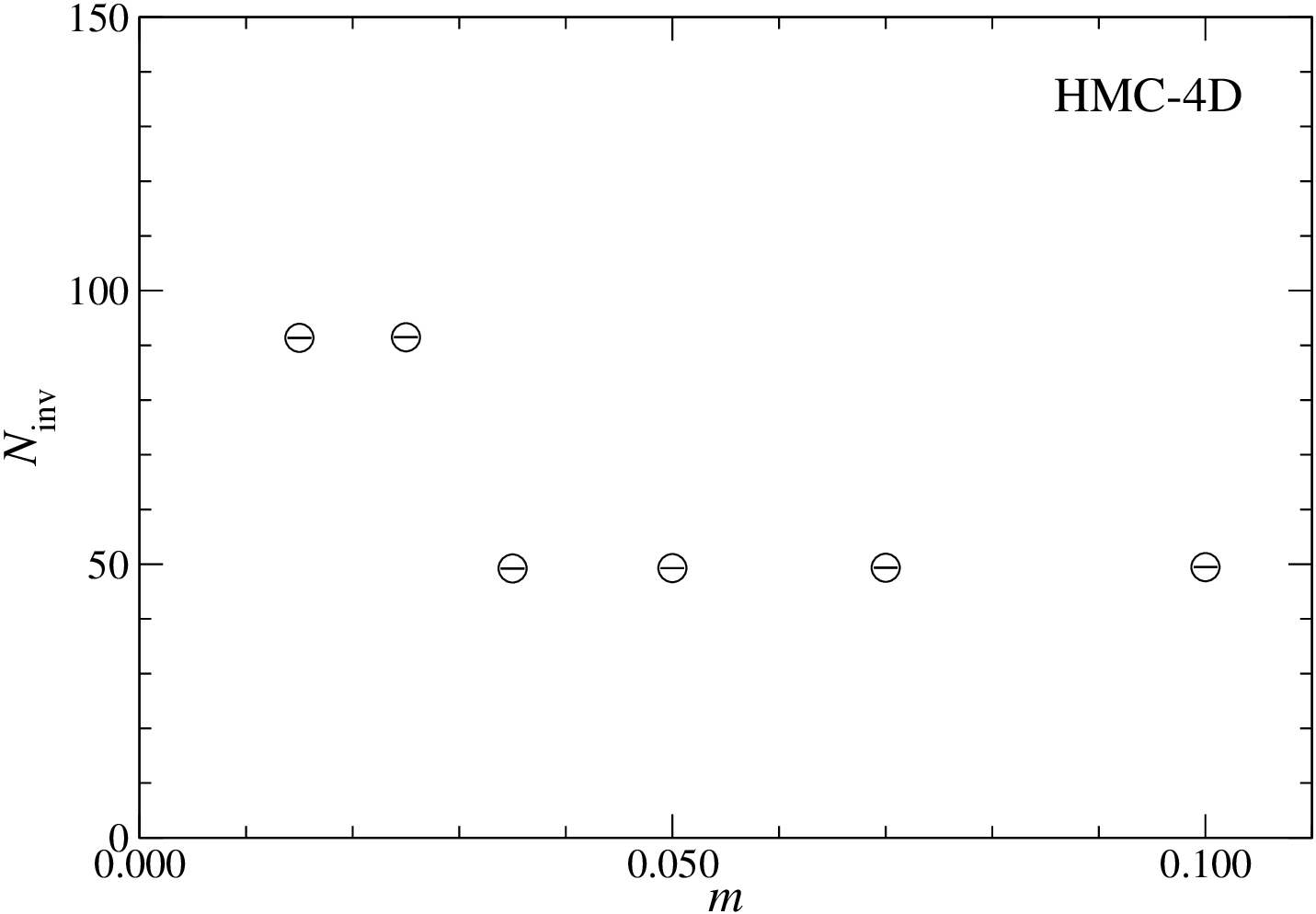}
   \hspace{2mm}
   \includegraphics[width=0.48\linewidth,clip]{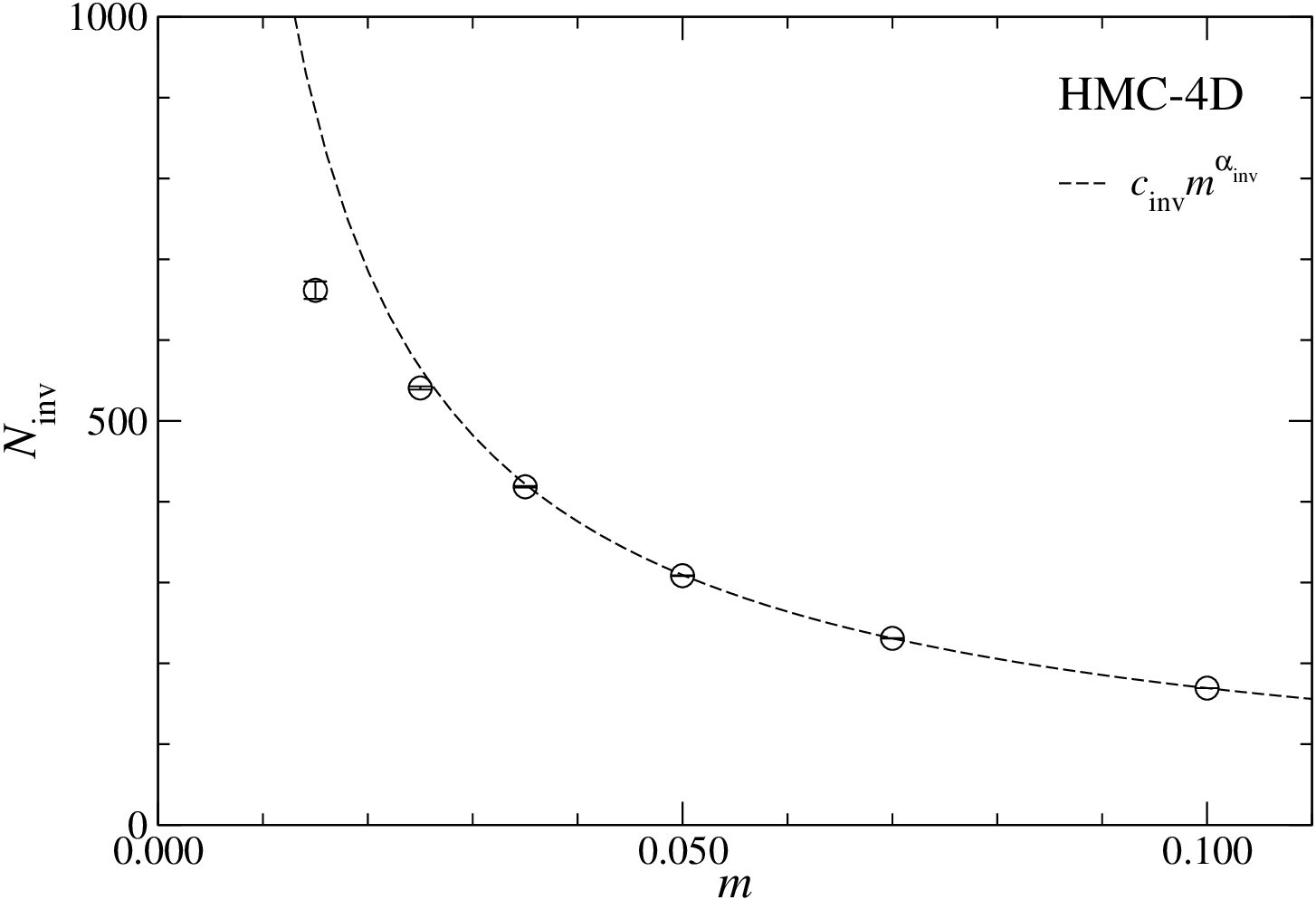}
   \vspace{-3mm}
   \caption{
      Quark mass dependence of CG iteration count $N_{\rm inv}$ 
      for 4D overlap solver.
      Left and right panels show data
      for the Hasenbusch preconditioner $D(m^\prime)$ and 
      the operator with the physical mass $D(m)$.
      The dashed line shows the fit to Eq.~(\ref{eqn:run:hmc:Ninv_vs_m}).
   }
   \label{fig:run:hmc:Ninv_vs_m:md:4d}
\end{center}
\end{figure}

\begin{table}[tbp]
   \vspace{3mm}
   \begin{center}
   \begin{tabular}{lll}
   \hline
   algorithm   & $c_{\rm inv}$  & $\alpha_{\rm inv}$ 
   \\ \hline
   HMC-4D      & 22.86(9)       & 0.869(1)
   \\
   HMC-5D      & 159(8)         & 0.64(2)
   \\ \hline
   \end{tabular}
   \caption{
      Fit parameters in Eq.~(\ref{eqn:run:hmc:Ninv_vs_m})
      for two algorithms HMC-4D and 5D.
   }
   \vspace{0mm}
   \label{tbl:run:hmc:Ninv:fit}
   \end{center}
\end{table}

The mass parameter in Eq.~(\ref{eqn:run:solver:Ninv}) 
should be $m^\prime$ for the Hasenbusch preconditioner $D(m^\prime)$.
Since we take large values for $m^\prime$, 
$N_{\rm inv}$ is governed by $m^\prime$ and has small $m$ dependence
as shown in Fig.~\ref{fig:run:hmc:Ninv_vs_m:md:4d}.
On the other hand, $N_{\rm inv}$ to invert $D(m)$ increases 
monotonously as $m$ decreases. 
Data at four heaviest quark masses are reasonably described by 
the scaling law Eq.~(\ref{eqn:run:solver:Ninv}) 
with $\tilde{\lambda}_1$ neglected 
\bea
   N_{\rm inv} = c_{\rm inv}\, m^{-\alpha_{\rm inv}}.
   \label{eqn:run:hmc:Ninv_vs_m}
\eea
Fit parameters are listed in Table~\ref{tbl:run:hmc:Ninv:fit}.
We note that the power $\alpha_{\rm inv}$ is close to its maximum value 
$\alpha_{\rm inv}\!=\!1$.

\begin{figure}[tbp]
\begin{center}
   \vspace{5mm}
   \includegraphics[width=0.48\linewidth,clip]{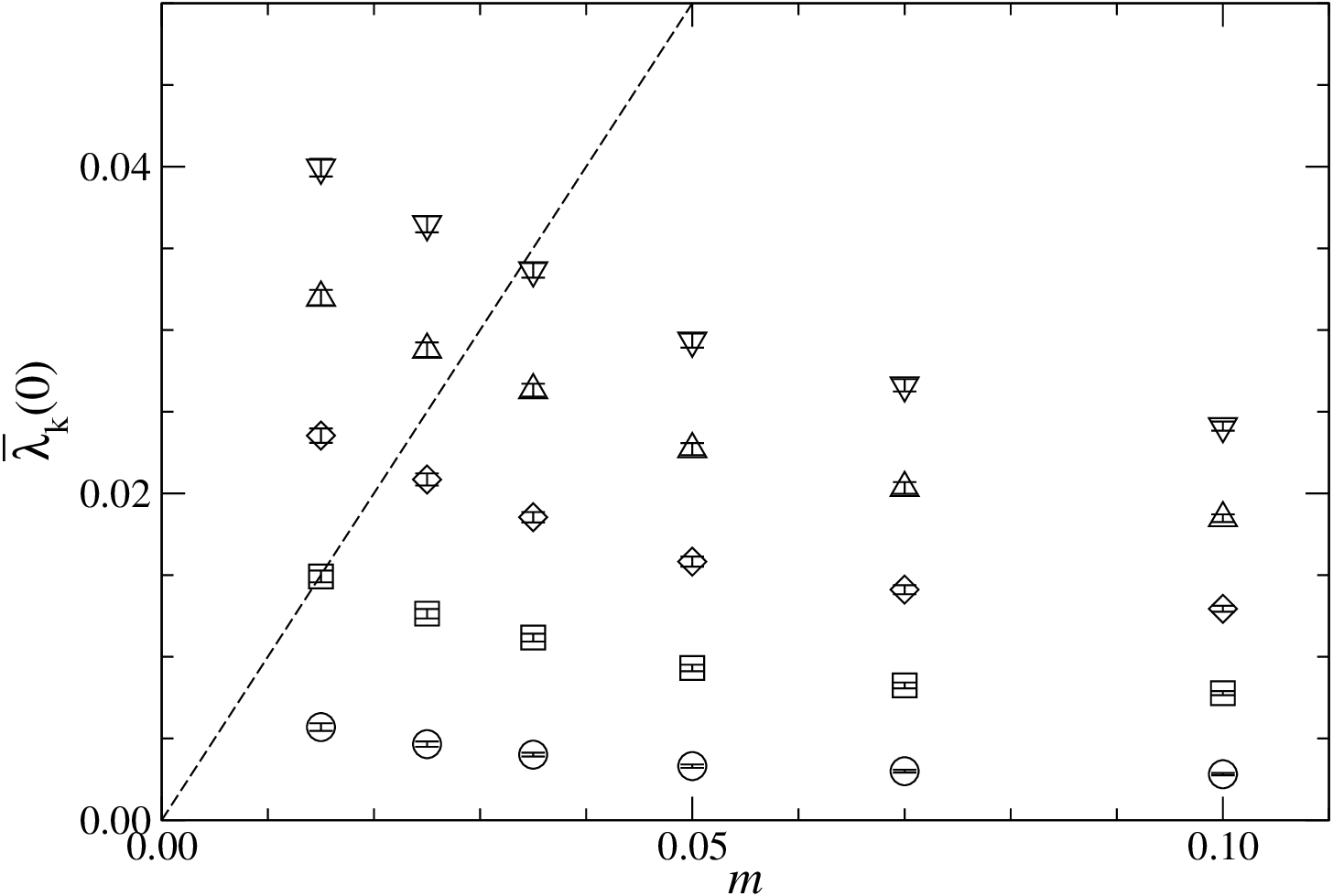}
   \vspace{-3mm}
   \caption{
      Projected eigenvalue $\lambda_k(0)$ as a function of $m$.
      We plot data for five lowest-lying modes.
      The dashed line shows $\lambda\!=\!m$.
   }
   \label{fig:run:hmc:lambda_proj}
\end{center}
\end{figure}


\begin{figure}[tbp]
\begin{center}
   \vspace{5mm}
   \includegraphics[width=0.48\linewidth,clip]{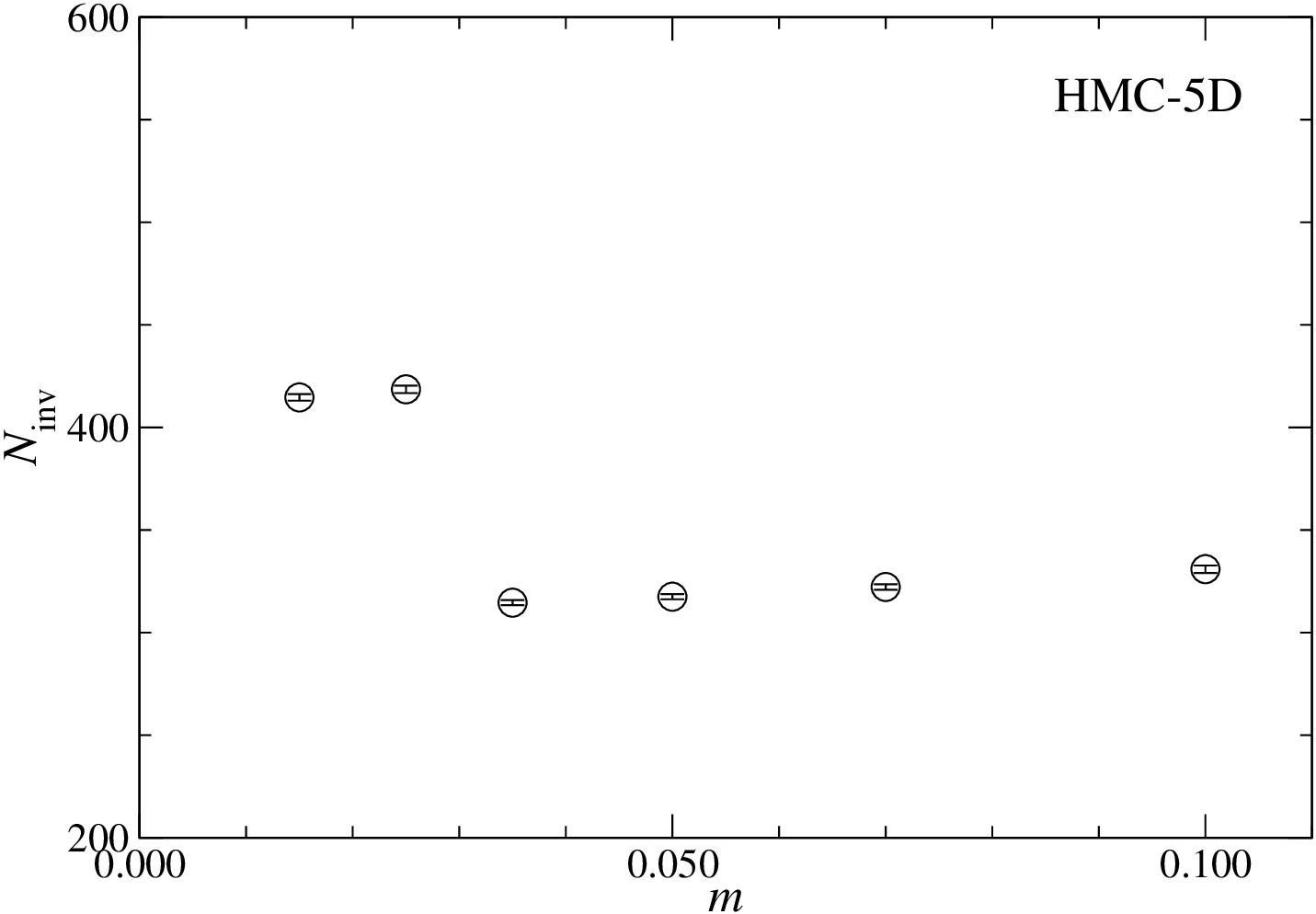}
   \hspace{2mm}
   \includegraphics[width=0.48\linewidth,clip]{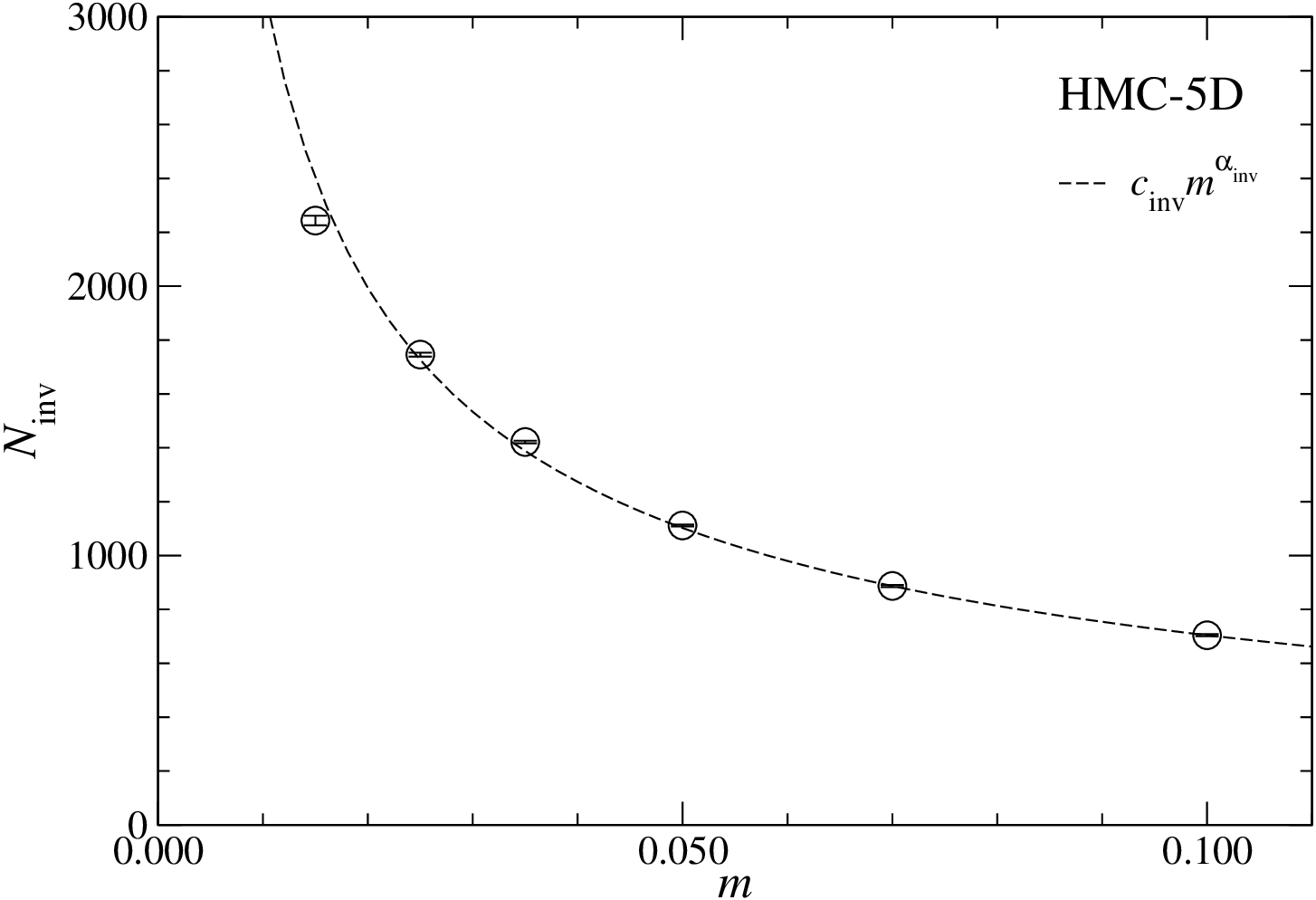}
   \vspace{-3mm}
   \caption{
      Quark mass dependence of CG iteration count $N_{\rm inv}$
      for 5D solver.
      Left and right panels show 
      data for $D^\prime(m^\prime)$ and $D^\prime(m)$, respectively.
   }
   \label{fig:run:hmc:Ninv_vs_m:md:5d}
\end{center}
\end{figure}

Our data of $N_{\rm inv}$ at $m \! \lesssim \! 0.025$ however 
clearly deviate from this fit.
%
This can be considered as a manifestation of finite size effects
as we approach the $\epsilon$-regime by decreasing $m$ 
with the fixed spatial extent $L$ 
\cite{e-regime:sigma:Nf2:RG+Ovr:JLQCD:1,e-regime:sigma:Nf2:RG+Ovr:JLQCD:2}.
Figure~\ref{fig:run:hmc:lambda_proj} actually shows that
the magnitude of $\tilde{\lambda}_k(0)$ in units of $m$ rapidly increases
toward smaller $m$.
Namely, at heavy quark masses $m \! \gtrsim \! 0.050$,
$D(m)^\dagger D(m)$ for CGNE has dense low-lying eigenvalues 
$|\lambda_k(m)|^2$ near $m^2$
and, hence, $m$ is a good parameter to characterize its condition number.
On the other hand, 
$|\lambda_k(m)|^2$ becomes sparse and deviates from $m^2$ as $m$ decreases.
It is likely that 
this rapid change in the low mode distribution 
distorts the $m$ dependence of $N_{\rm inv}$ 
from the simple scaling Eq.~(\ref{eqn:run:hmc:Ninv_vs_m}).
%

The influence of $\tilde{\lambda}_k(0)$ to $N_{\rm inv}$ 
is less clear in HMC-5D, since the matrix to be inverted is 
5D preconditioned matrix rather than $D(m)$.
We only note that,
as seen in Fig.~\ref{fig:run:hmc:Ninv_vs_m:md:5d},
$N_{\rm inv}$ for the preconditioner $D^\prime(m^\prime)$ 
is mainly determined by $m^\prime$ 
and $N_{\rm inv}$ for $D^\prime(m)$ 
shows a somewhat weaker power scaling with 
parameter listed in Table~\ref{tbl:run:hmc:Ninv:fit}.


\subsection{Properties of HMC}

\begin{figure}[tbp]
\begin{center}
   \vspace{5mm}
   \includegraphics[width=0.48\linewidth,clip]{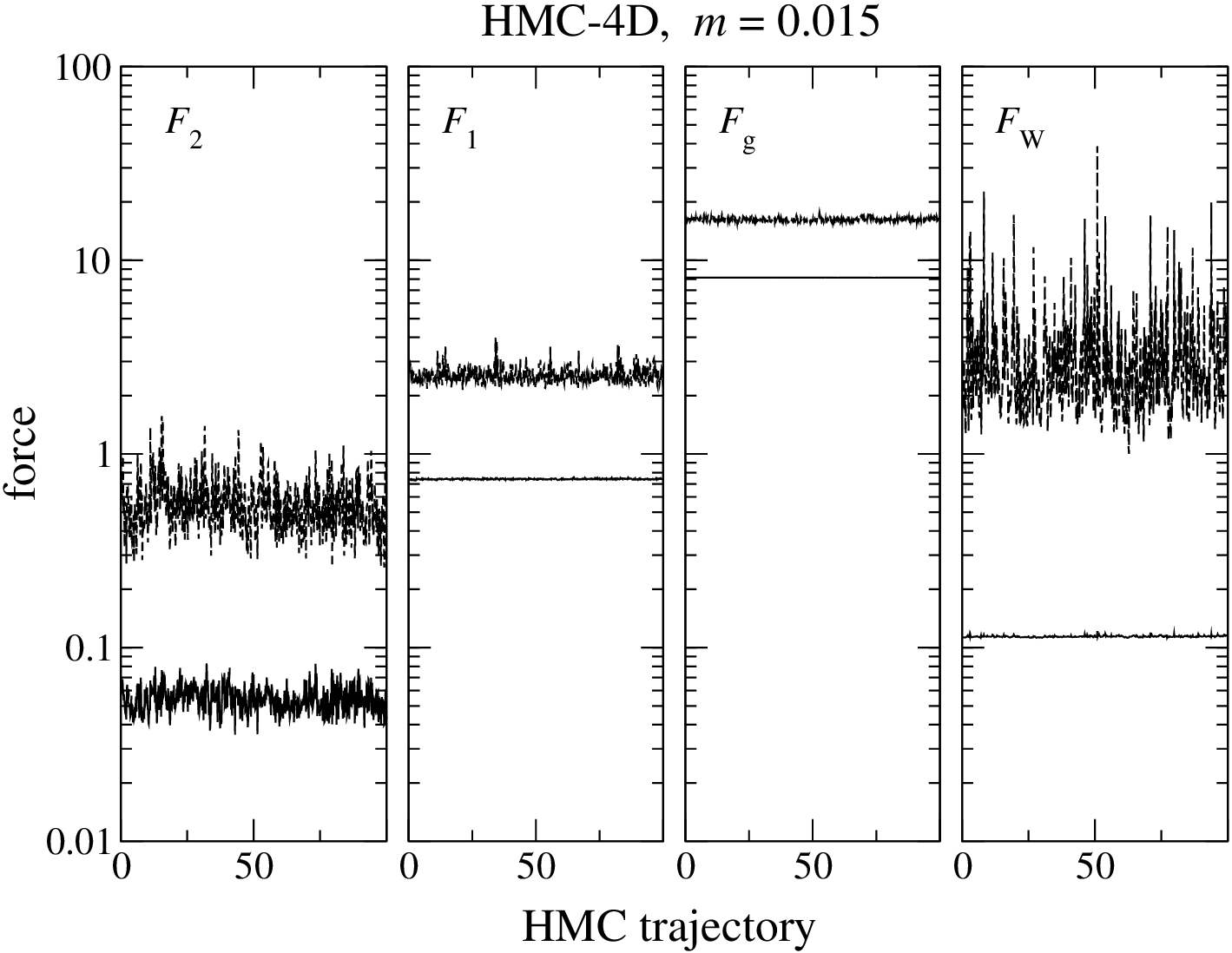}
   \hspace{2mm}
   \includegraphics[width=0.48\linewidth,clip]{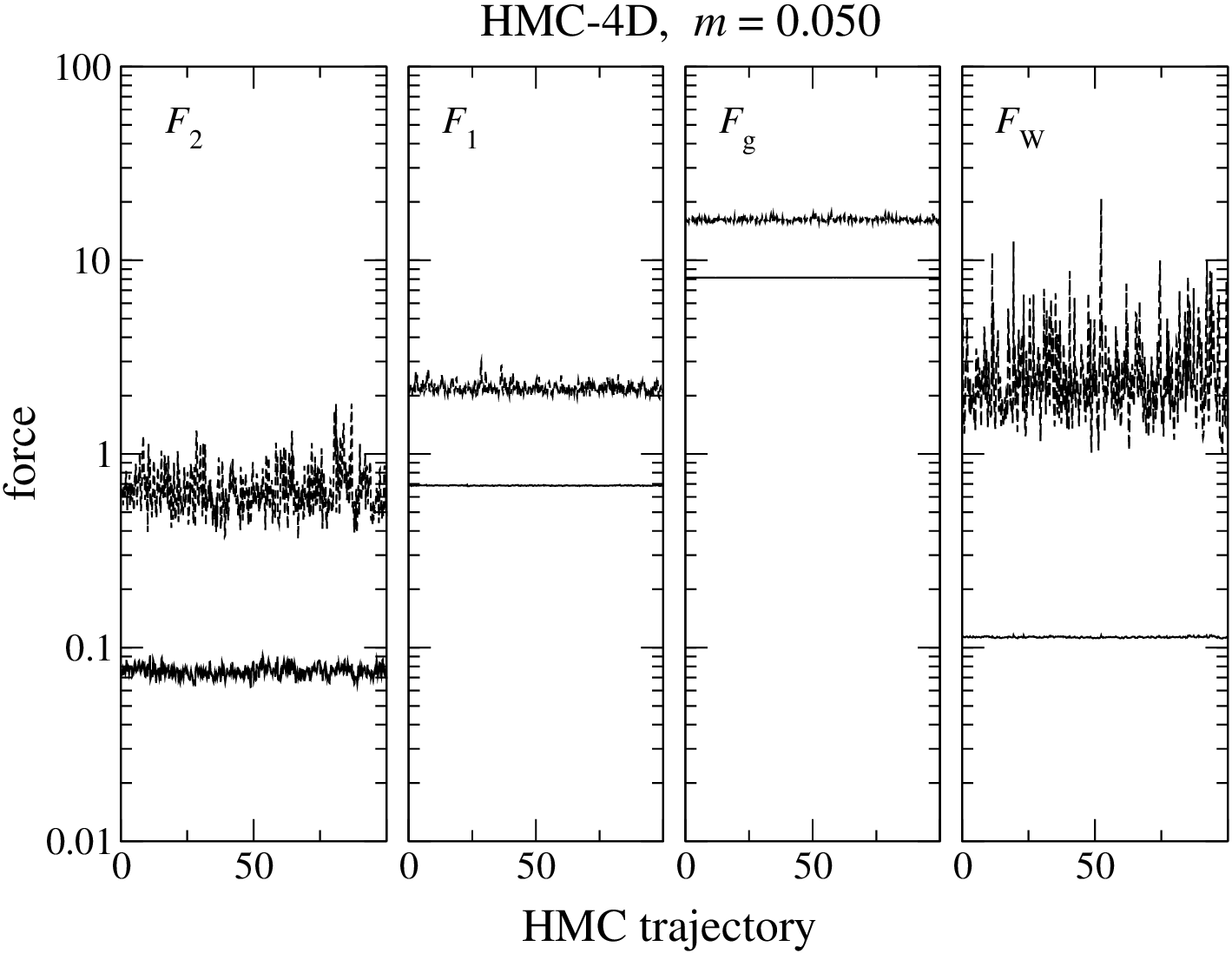}
   \vspace{-3mm}
   \caption{
      Time history of MD forces $F_2$, $F_1$, $F_g$ and $F_\rmW$
      at $m\!=\!0.015$ (four left-most panels)
      and 0.050 (four right-most panels) with HMC-4D.
      Two lines in each panel show the average and maximum value
      over the degrees of freedom
      (color, space-time direction and coordinates).
   }
   \label{fig:run:hmc:force:4d}
\end{center}
\end{figure}
\begin{figure}[tbp]
\begin{center}
   \vspace{5mm}
   \includegraphics[width=0.48\linewidth,clip]{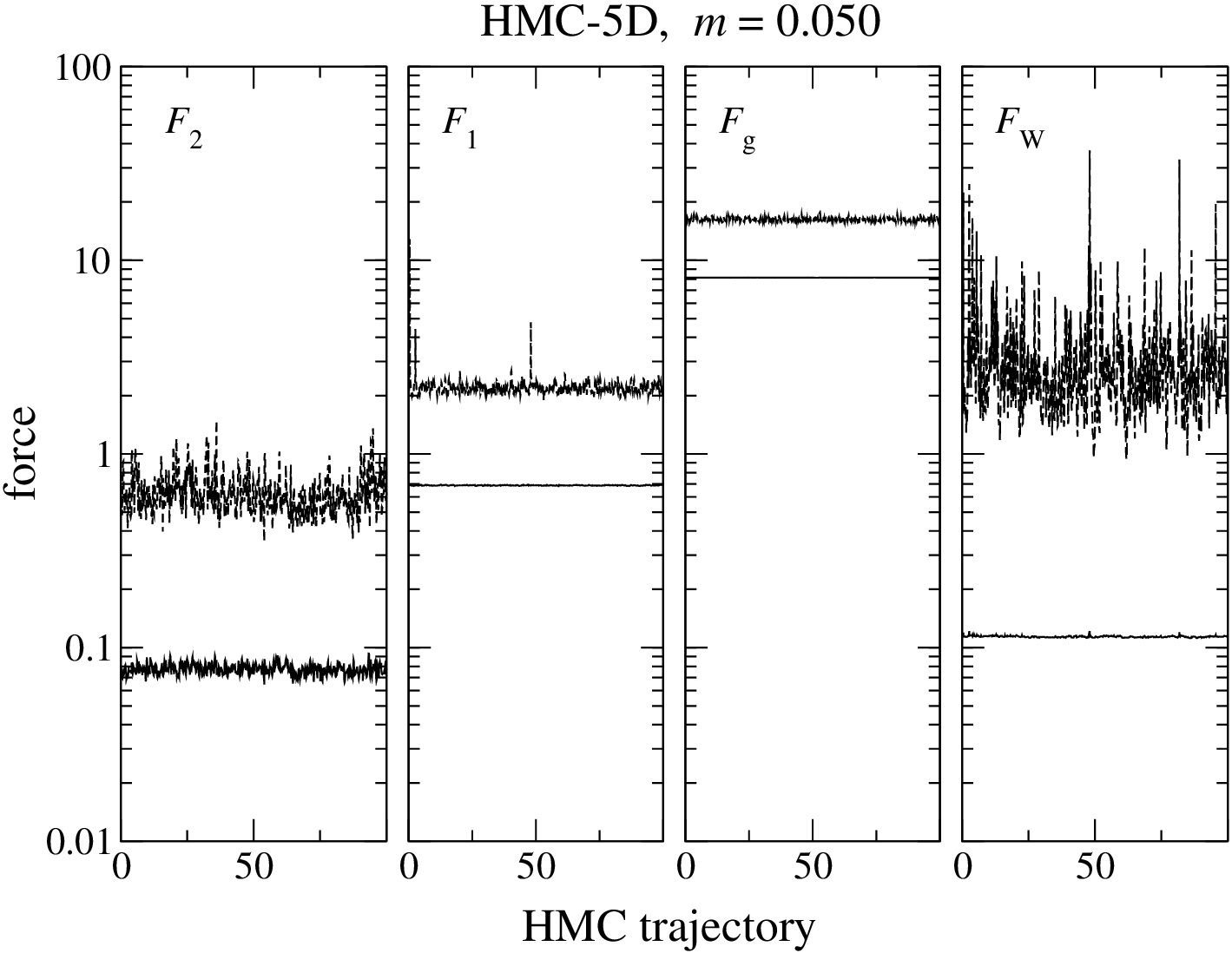}
   \vspace{-3mm}
   \caption{
      Time history of MD forces at $m\!=\!0.050$ with HMC-5D.
   }
   \label{fig:run:hmc:force:5d}
\end{center}
\end{figure}

The parameters for the Hasenbusch preconditioning, 
Eqs.~(\ref{eqn:hmc:weight:4d}) and (\ref{eqn:hmc:weight:5d}),
and the multiple time scale MD integration,
Eqs.~(\ref{eqn:MD:T_2})\,--\,(\ref{eqn:MD:U+W}), are listed 
in Tables~\ref{tbl:run:sim_param:4D}\,--\,\ref{tbl:run:sim_param:5D}.
The gauge force $F_g$ is known to be generally larger than 
the fermionic force(s).
Only $m^\prime$ needs a non-trivial tuning 
to make a hierarchic structure among the MD forces $F_2$, $F_1$ and $F_g$.
As discussed in Refs.\cite{mass_precond2,mass_precond+mtsMD:tmQCD},
$m^\prime$ should be decreased for smaller $m$
to avoid a too large condition number for the preconditioned 
Dirac operator $D(m)/D(m^\prime)$.
This is why $m^\prime$ is set to a smaller value at $m\!\leq\!0.025$
than others.
While we have not done further fine tuning of $m^\prime$, 
Figs.~\ref{fig:run:hmc:force:4d} and \ref{fig:run:hmc:force:5d} show that 
the forces $F_1$, $F_2$ and $F_g$ are well separated from each other 
with our choice of $m^\prime$.
This enables us to use the ratios of the step sizes, 
namely $r_\phi$ and $r_g$, around 4\,--\,8,
which considerably reduce the computational cost of the MD integration 
with the acceptance rate
kept in a reasonably high range $\gtrsim 0.7$.

The same figures show
that $F_\rmW$ from the extra-Wilson fermions exhibits 
large statistical fluctuations.
It becomes as large as $F_g$
probably due to the appearance of small eigenvalues of $\Hw$.
This is another reason why we update $F_\rmW$
in the inner-most loop in Eq.~(\ref{eqn:MD:U+W}),
in addition to its consistency with the updated gauge field
described in Sec.~\ref{sec:algorithm}.

\begin{table}[t]
   \vspace{3mm}
   \begin{center}
   \begin{tabular}{ll|ll|llll|lll}
   \hline
         &     & \multicolumn{2}{l}{HMC-4D}    & \multicolumn{2}{l}{HMC-5D}
   \\ \hline
   $m$   & $Q$ & $\Delta H$ & $e^{-\Delta H}$  & $\Delta H$ & $e^{-\Delta H}$ 
   \\ \hline
   0.015 &  0  & 0.0603(72) & 0.9957(64)       & 4618(4617) & 0.998(14)
   \\
   0.025 &  0  & 14(14)     & 0.9954(32)       & 0.1142(52) & 0.9997(59)
   \\
   0.035 &  0  & 0.2224(86) & 1.016(11)        & 0.068(15)  & 0.9964(42)
   \\
   0.050 &  0  & 0.1846(84) & 1.0008(83)       & 0.0519(39) & 0.9982(45)
   \\
   0.070 &  0  & 0.198(99)  & 0.9989(52)       & 0.0260(29) & 0.9950(24)
   \\
   0.100 &  0  & 0.0692(33) & 0.9963(34)       & 0.0207(33) & 0.9989(30)
   \\ \hline
   0.050 & -2  & 0.37(16)   & 0.978(15)        & 0.0432(80) & 1.0082(61)
   \\ \hline
   0.050 & -4  & --         & --               & 0.0601(55) & 0.9934(52)
   \\ \hline 
   \end{tabular}
   \caption{
      Average of $\Delta H$ and $e^{-\Delta H}$.
   }
   \vspace{0mm}
   \label{tbl:run:delta_H+P_HMC}
   \end{center}
\end{table}

\begin{figure}[h!]
\begin{center}
   \vspace{5mm}
   \includegraphics[width=0.48\linewidth,clip]{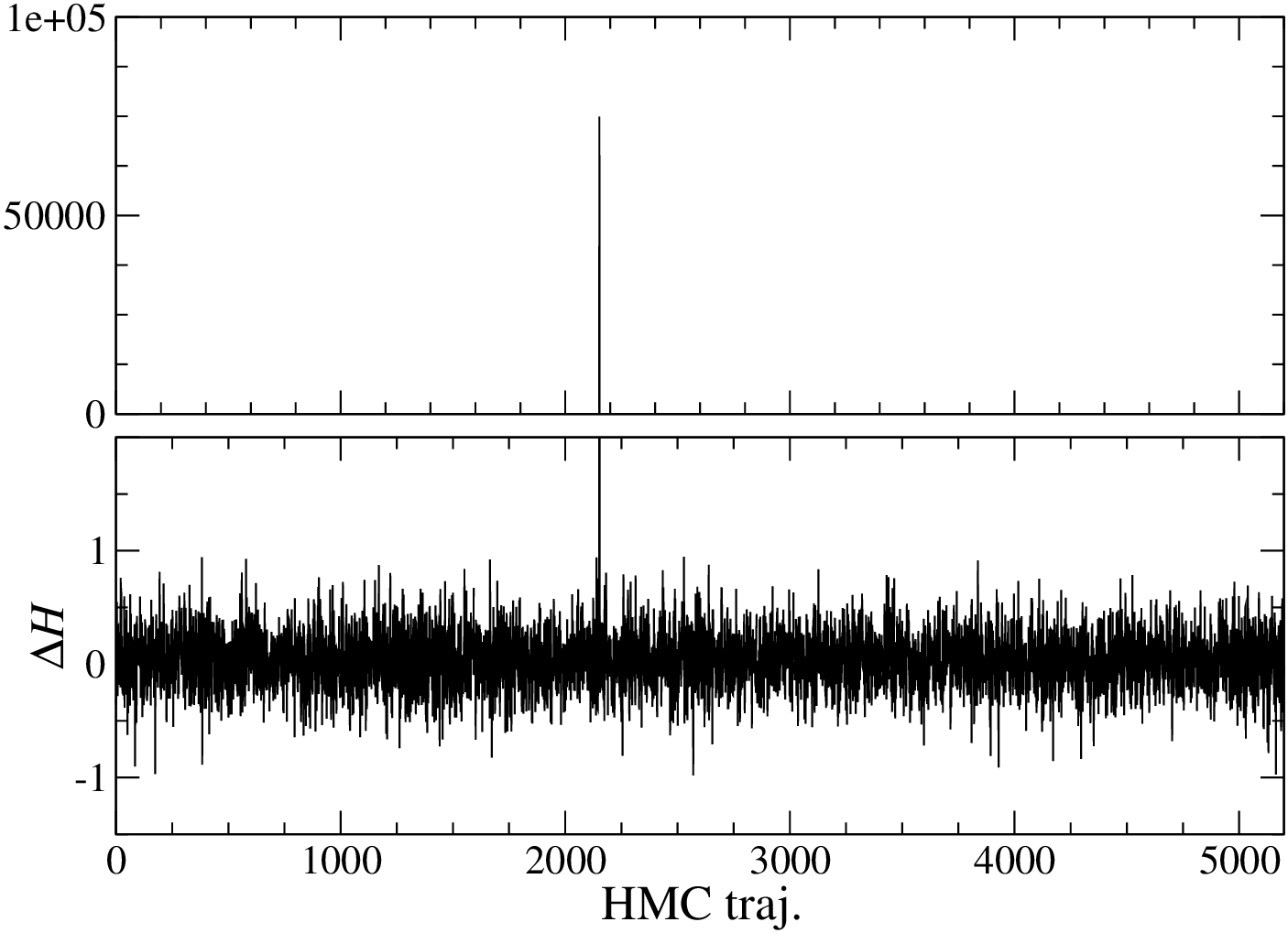}
   \vspace{-3mm}
   \caption{
      Time history of $\Delta H$ in simulation at $m\!=\!0.025$ with HMC-4D.
      We have a spike at 2151-th trajectory.
   }
   \label{fig:run:hmc:deltaH:spike}
\end{center}
\end{figure}

\begin{figure}[h!]
\begin{center}
   \vspace{5mm}
   \includegraphics[width=0.48\linewidth,clip]{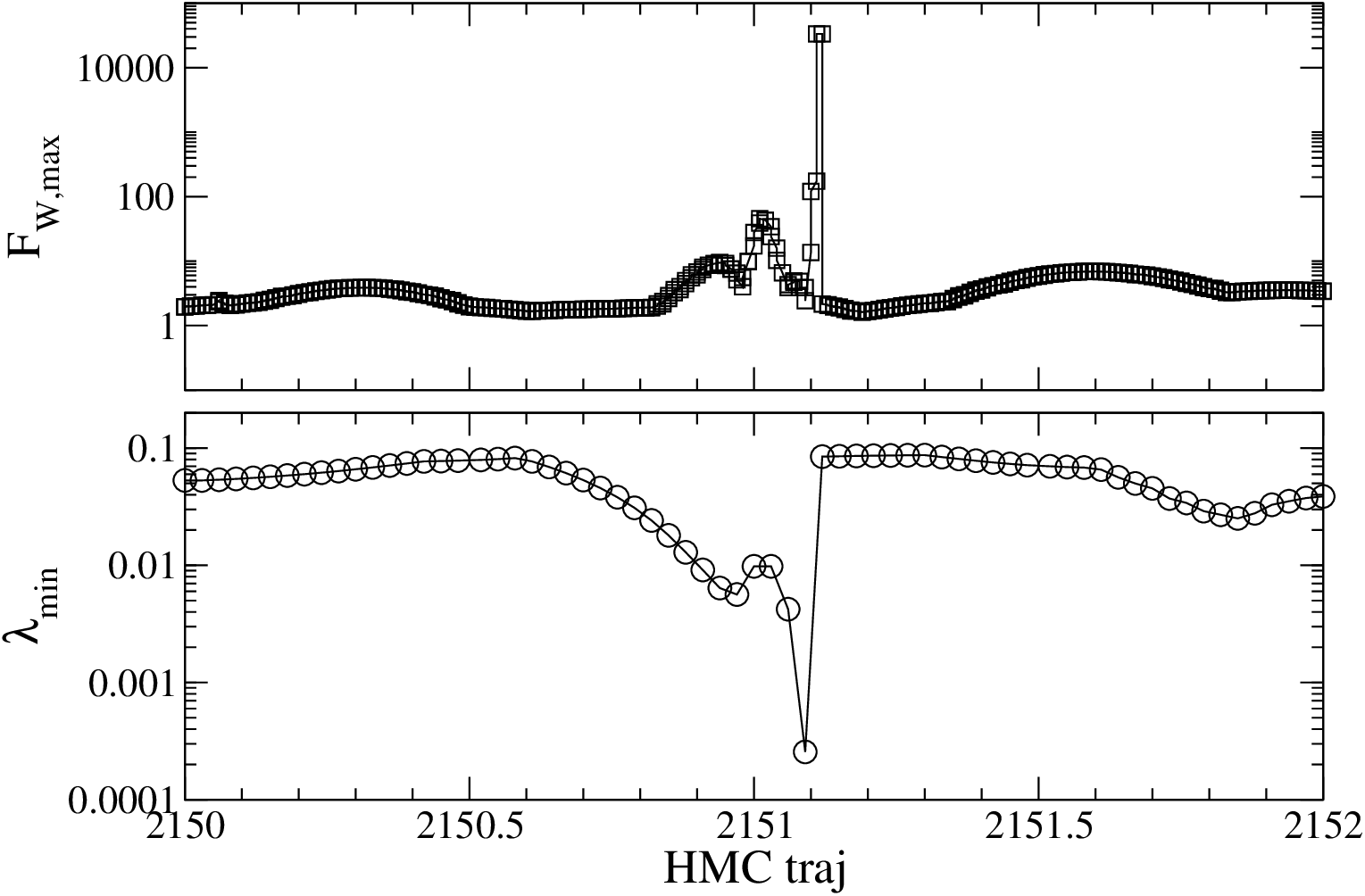}
   \vspace{-3mm}
   \caption{
      Maximum value of $F_\rmW$ and lowest eigenvalue of $\Hw$
      during MD evolution where a spike in $\Delta H$ is observed.
   }
   \label{fig:run:hmc:deltaH:spike:eval_F_max}
\end{center}
\end{figure}

\begin{figure}[h!]
\begin{center}
   \vspace{5mm}
   \includegraphics[width=0.48\linewidth,clip]{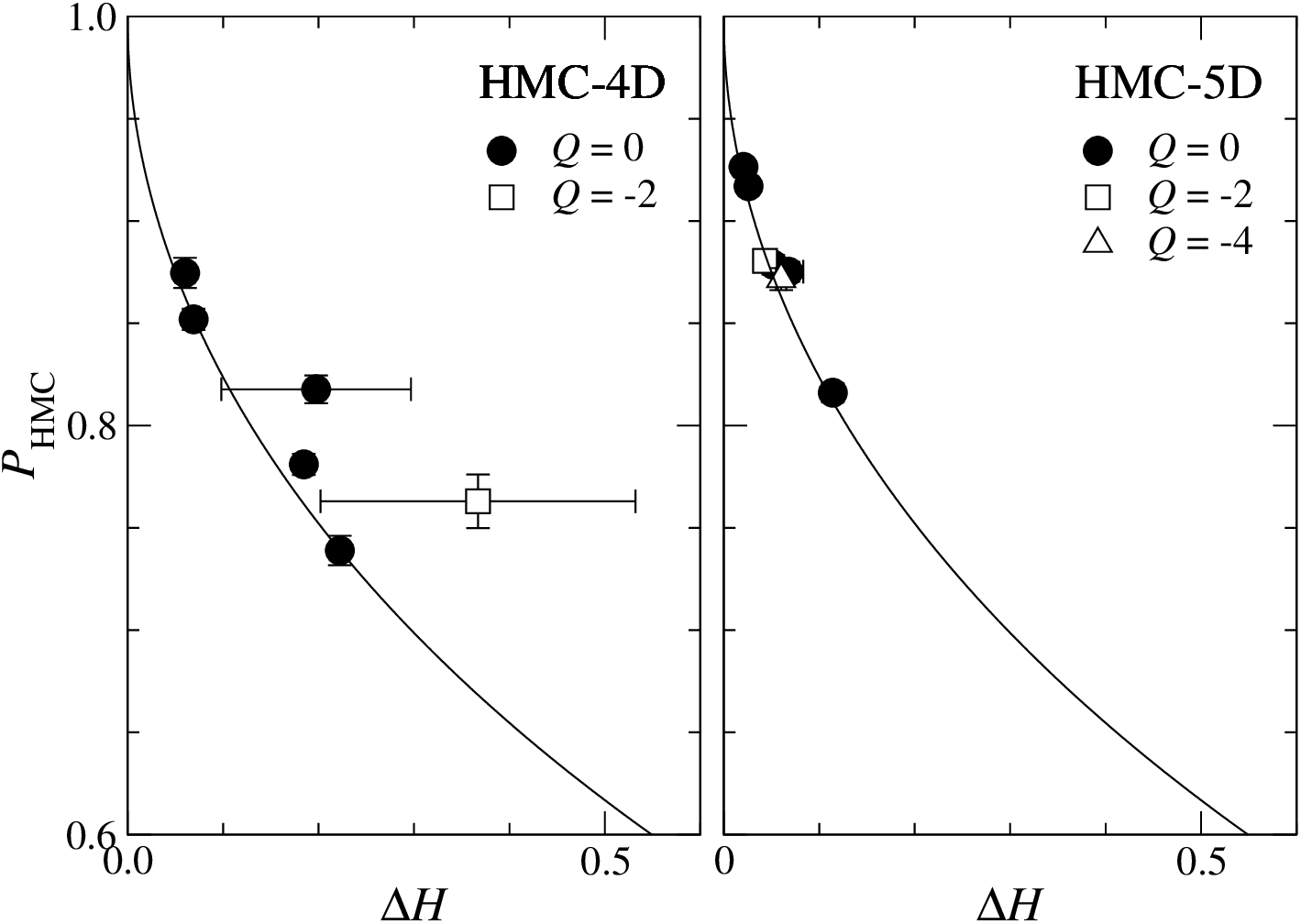}
   \vspace{-3mm}
   \caption{
      Acceptance rate $P_{\rm HMC}$ as a function of $\Delta H$
      in simulations with HMC-4D (left panel) and HMC-5D (right panel).
      The solid line is the expectation of Eq.~(\ref{eqn:run:hmc:P_HMC_erfc}).
      Data at $(m,Q)\!=\!(0.025,0)$ with HMC-4D and 
      at $(0.015,0)$ with HMC-5D are consistent with the expectation
      within their huge error and hence are omitted.
   }
   \label{fig:run:hmc:P_HMC_vs_dH}
\end{center}
\end{figure}

We denote the change in the Hamiltonian due to the discretized MD integration 
by $\Delta H$ and 
its average is summarized in Table~\ref{tbl:run:delta_H+P_HMC}.
The area-preserving property of MD leads to 
the following (in)equality
\bea
   e^{-\langle \Delta H \rangle}
   & \leq &
   \langle e^{-\Delta H} \rangle
   = 
   1,
   \label{eqn:run:hmc:area_preseve}
\eea
where 
we explicitly indicate the value averaged over HMC updating
by the bracket $\langle \cdots \rangle$.
The inequality predicts that averaged $\Delta H$ is positive 
and this is the case in our simulations.
Two of them are however dominated by huge spikes
shown in Fig.~\ref{fig:run:hmc:deltaH:spike}.
Similar spikes have been observed also in previous simulations 
with Wilson-type fermions \cite{dH_spike},
and they may be attributed to instability of HMC
with a large MD step size \cite{MD_instability}.
The spikes in our simulations have rather simple origin
as shown in Fig.~\ref{fig:run:hmc:deltaH:spike:eval_F_max}:
$\Hw$ can develop a very small lowest eigenvalue 
leading to a spike in $F_\rmW$ and hence in $\Delta H$.
This is why we take a larger value for $r_g$ in HMC-5D
to be more robust against the spikes 
with a smaller MD step size for the calculation of $F_\rmW$.
Thanks to the determinant $\det[ \Delta_\rmW ]$, however,
the number of such huge spikes is not large even in HMC-4D,
at most a few per 10,000 trajectories.
As a result,
the equality Eq.~(\ref{eqn:run:hmc:area_preseve}) is fulfilled
within 2\,\% accuracy
without introducing the replay trick \cite{DDHMC:1,replay_trick:2}.
We also note that $\Delta H$ dependence of $P_{\rm HMC}$ 
is consistent with the expected form 
of the complementary error function
\bea
   P_{\rm HMC}
   & = &
   \mbox{erfc}\left[\sqrt{\Delta H}/2\right].
   \label{eqn:run:hmc:P_HMC_erfc}
\eea
as plotted in Fig.~\ref{fig:run:hmc:P_HMC_vs_dH}.

\subsection{Simulation cost}

On a half rack of Blue Gene/L, 
the assembler code for the multiplication of $D_W$
achieves roughly 28\,\% efficiency of the peak performance
when all the data are in the L3 cache.
The sustained speed averaged over all HMC steps 
is about 15\,\% 
indicating significant overheads 
due to a limited bandwidth to the off-chip memory,
and to linear computations with quark vectors 
in the low mode preconditioning and so on.

\begin{figure}[tbp]
\begin{center}
   \vspace{5mm}
   \includegraphics[width=0.48\linewidth,clip]{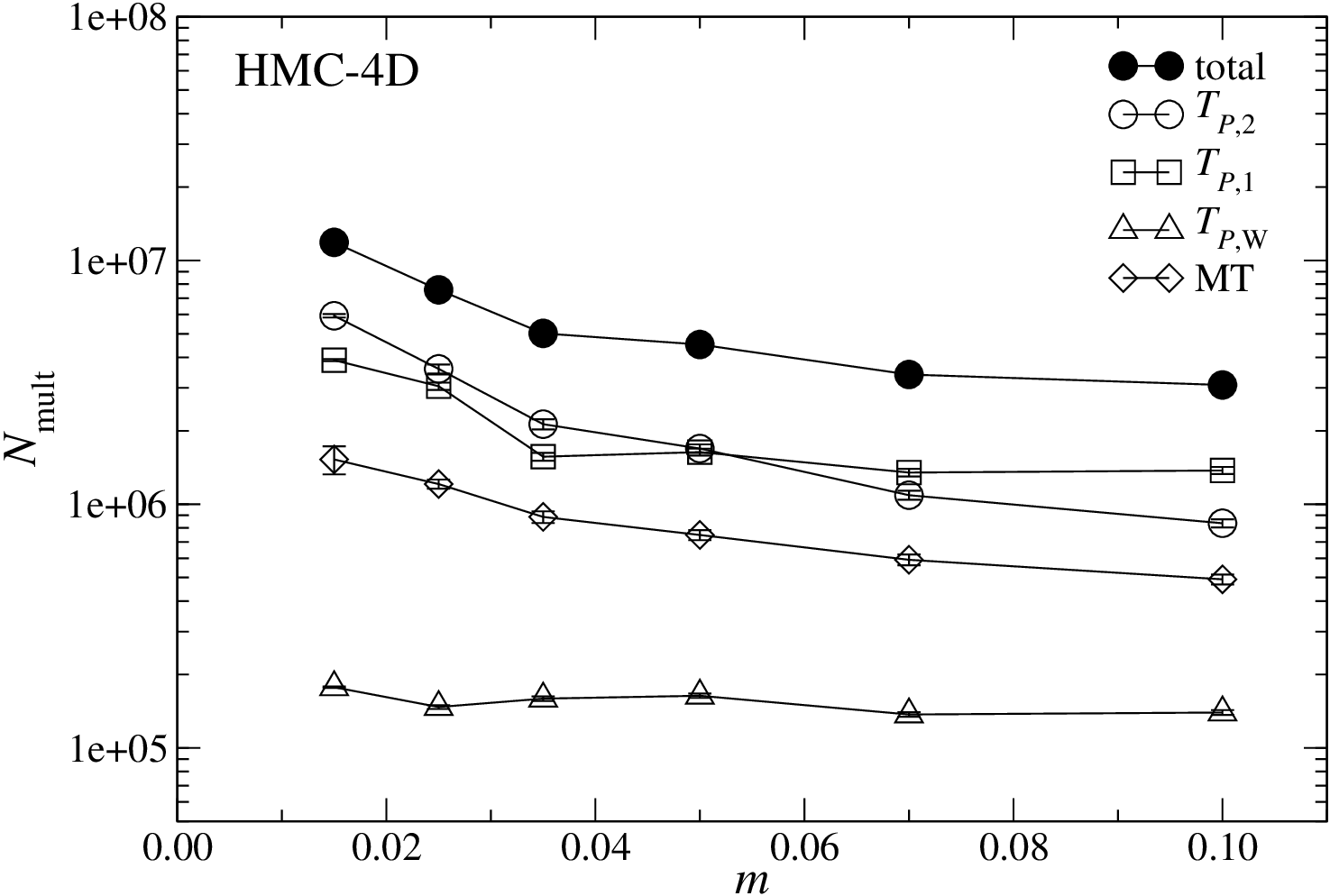}
   \hspace{2mm}
   \includegraphics[width=0.48\linewidth,clip]{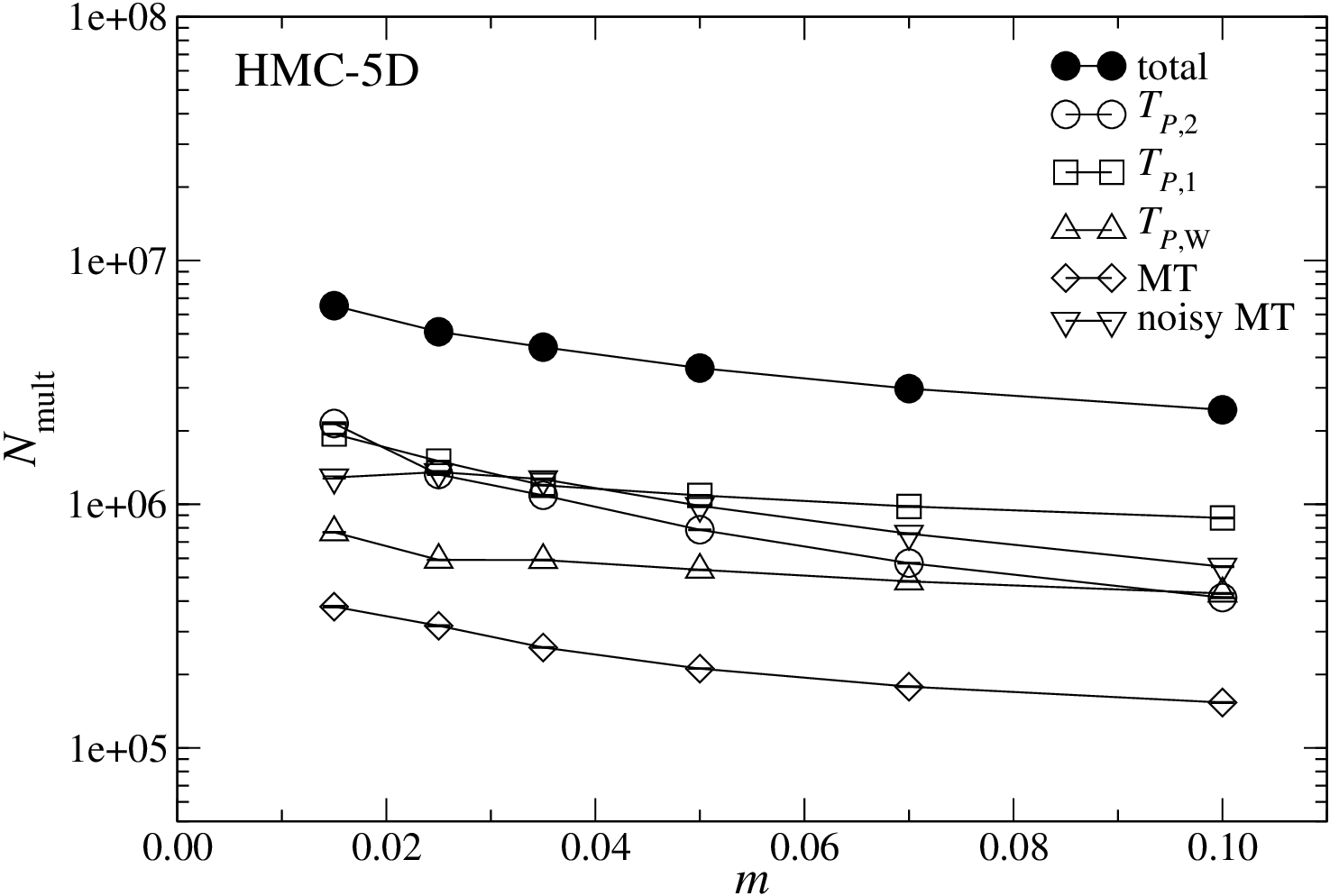}
   \vspace{-3mm}
   \caption{
      Number of $\Dw$ multiplications $N_{\rm mult}$ in HMC-4D (left panel)
      and HMC-5D (right panel).
      Open symbols are $N_{\rm mult}$ in calculations of MD forces 
      and in the Metropolis tests (MTs),
      whereas the filled symbol is their total.
   }
   \label{fig:run:hmc:Nmult:cntrib}
\end{center}
\end{figure}

In Table~\ref{tbl:run:hmc:Nmult},
we summarize the number of $D_\rmW$ multiplications $N_{\rm mult}$ 
per trajectory,
which serves as a machine independent measure of the simulation cost.
This is compared with $N_{\rm mult}$ at each HMC step 
in Fig.~\ref{fig:run:hmc:Nmult:cntrib}.
As expected, 
calculations of the overlap forces $F_1$ and $F_2$ 
spend a large part of the total CPU time 
especially at small quark masses $m \lesssim 0.050$.
Note also that 
the costs to calculate the two forces are of the same order:
in other words, they are reasonably balanced
with our choice of $m^\prime$ and $r_\phi$.
While $F_\rmW$ is calculated in the inner-most loop of our MD integration,
its computational cost turns out to be negligible in HMC-4D,
and is not large even in HMC-5D, where the overlap solver is accelerated by
the 5D algorithm.

\begin{table}[tbp]
   \vspace{3mm}
   \begin{center}
   \begin{tabular}{l|llllll|l|l}
   \hline
   $Q$         & \multicolumn{6}{l|}{0}   
               & -2         & -4
   \\ \hline
   $m$         & 0.015      & 0.025      & 0.035     
               & 0.050      & 0.070      & 0.100  
               & 0.050      & 0.050      
   \\ \hline
   HMC-4D      & 11.9(1)    & 7.6(4)     & 5.0(2)  
               & 4.5(2)     & 3.4(1)     & 3.1(1)
               & 5.08(2)    & -- 
   \\
   HMC-5D      & 6.53(4)    & 5.10(2)    & 4.41(2)
               & 3.62(1)    & 2.98(1)    & 2.44(1)
               & 3.61(2)    & 3.65(1) 
   \\ \hline
   \end{tabular}
   \caption{
      Number of $\Dw$ multiplications per trajectory in units of $10^6$.
   }
   \vspace{0mm}
   \label{tbl:run:hmc:Nmult}
   \end{center}
\end{table}

In simulations with HMC-5D, 
the noisy Metropolis test also needs a substantial fraction of the total time.
This is because 
it has to invoke the 4D solver with the strict stopping condition 
to calculate the probability Eq.~(\ref{algo:hmc:nmt}) accurately,
whereas other HMC steps are implemented with the much faster 5D solver.
We note that this step is removed 
in our latest $2\!+\!1$-flavor simulations
by incorporating the low-mode preconditioning into the 5D solver
\cite{Lat07:JLQCD:Matsufuru,Lat07:JLQCD:Hashimoto}.

\begin{figure}[tbp]
\begin{center}
   \vspace{5mm}
   \includegraphics[width=0.48\linewidth,clip]{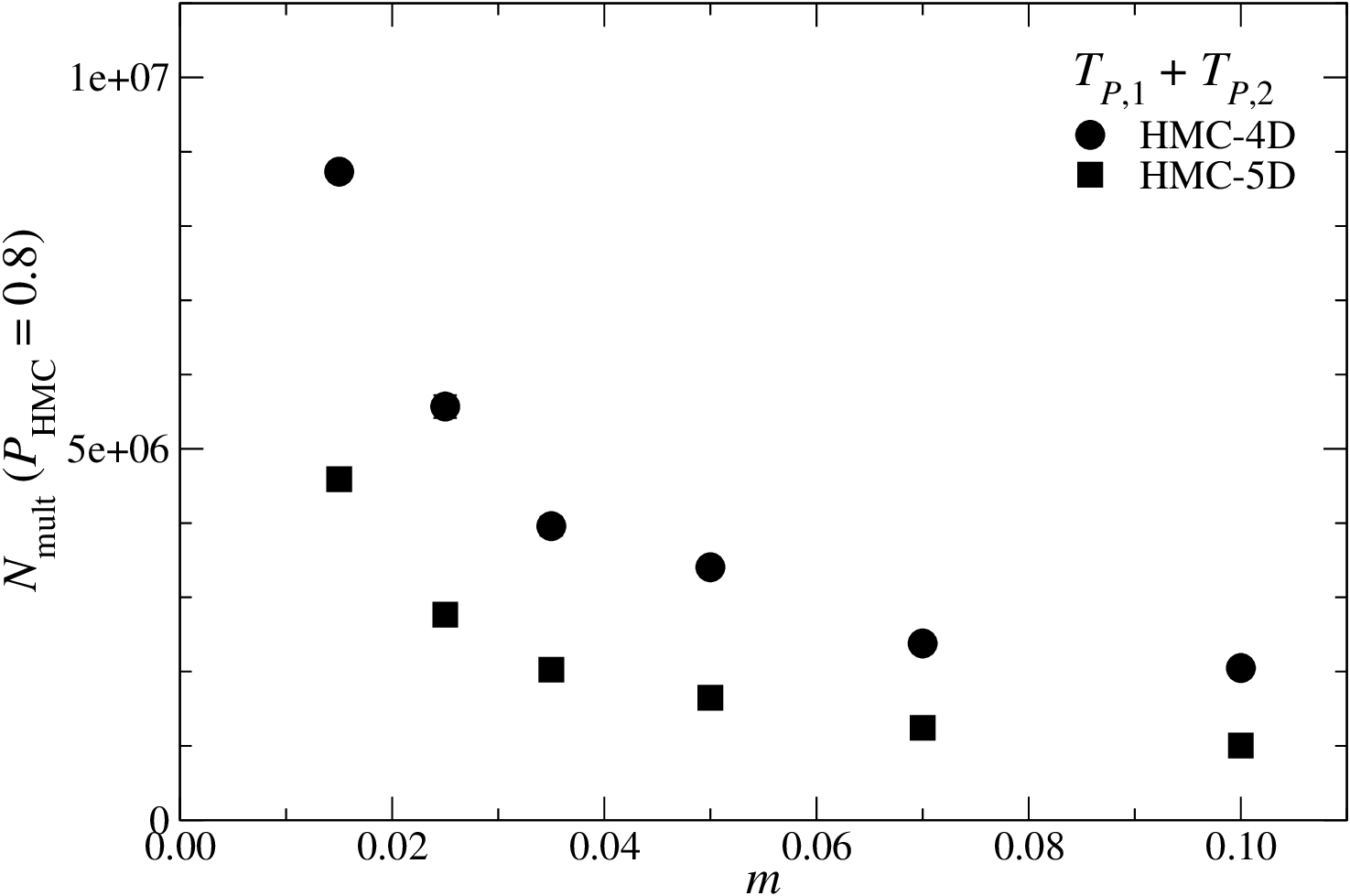}
   \hspace{2mm}
   \includegraphics[width=0.48\linewidth,clip]{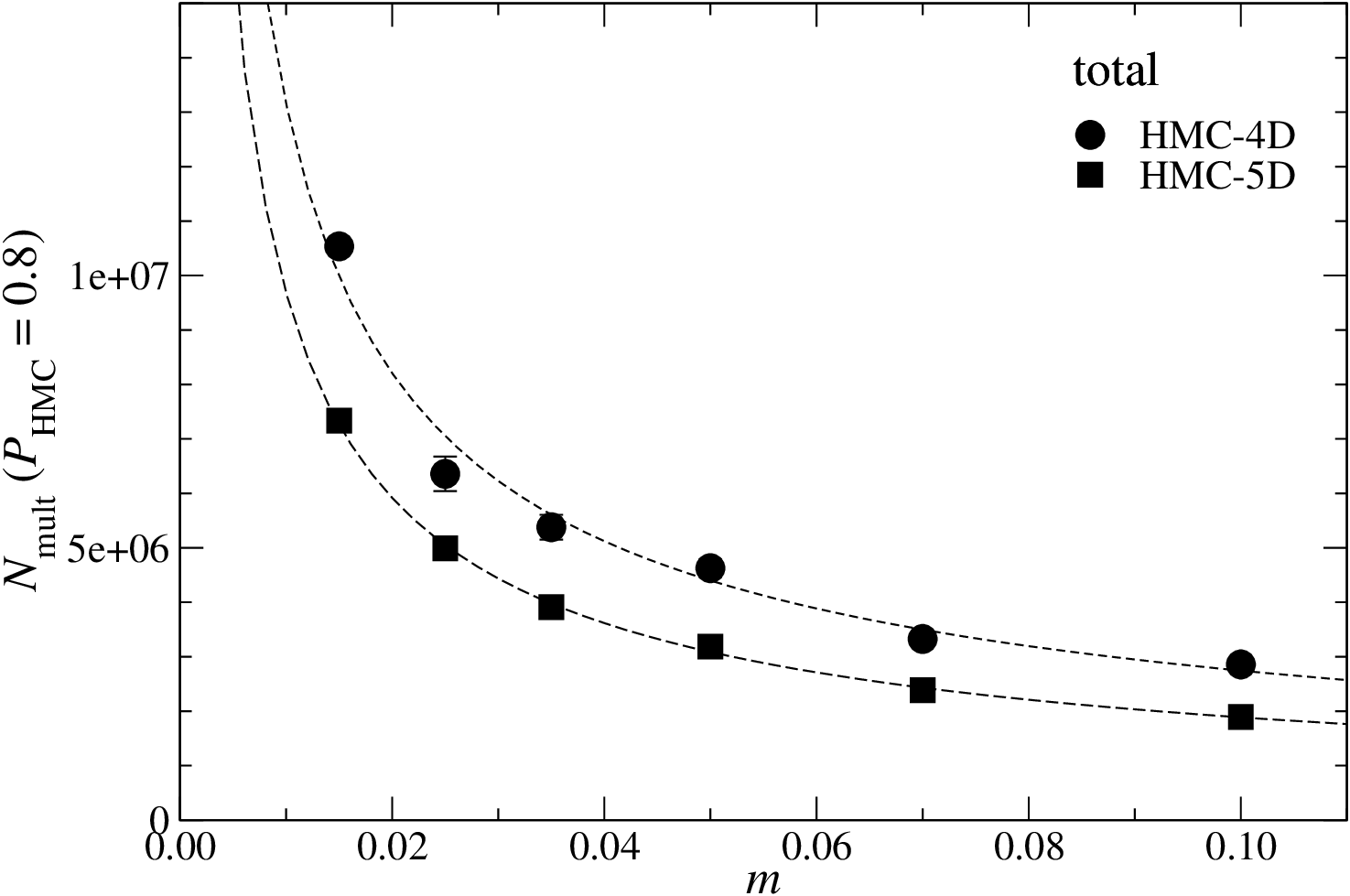}
   \vspace{-3mm}
   \caption{
      Comparison of $N_{\rm mult}$ between HMC-4D and HMC-5D.
      Left panel shows $N_{\rm mult}$ only for $T_{P,1}$ and $T_{P,2}$,
      whereas right panel is $N_{\rm mult}$ for all HMC steps.
   }
   \label{fig:run:hmc:Nmult:comp}
\end{center}
\end{figure}

Figure~\ref{fig:run:hmc:Nmult:comp} shows a comparison 
between HMC-4D and HMC-5D
in total $N_{\rm mult}$ and that for $T_{P,1}$ and $T_{P,2}$.
To take the difference in $P_{\rm HMC}$ into account,
$N_{\rm mult}$ in this figure is corrected by a factor
$\Delta \tau/\Delta \tau^\prime$,
where  
$\Delta \tau^\prime$ is the step size corresponding to $P_{\rm HMC}\!=\!0.8$
estimated by assuming Eq.~(\ref{eqn:run:hmc:P_HMC_erfc}) 
and $\Delta H \! \propto \! \Delta \tau^4$.
Due to the rough approximation and the lack of the low-mode preconditioning 
for $\sgn[\Hw]$ in $D^\prime$,
$\Delta \tau$ have to be decreased by roughly 50\% 
when the algorithm is switched from HMC-4D to HMC-5D 
with $P_{\rm HMC}$ kept fixed.
Even with this overhead,
we observe about factor of 2 reduction in 
$N_{\rm mult}$ for $T_{P,1}$ and $T_{P,2}$.
The noisy Metropolis test reduces the net gain 
to roughly 50\% at all the simulated quark masses.
We note in passing that 
the CPU time summarized 
in Tables~\ref{tbl:run:sim_param:4D} and \ref{tbl:run:sim_param:5D}
shows slightly better acceleration than in $N_{\rm mult}$
at $m \! \gtrsim \! 0.035$.
This is because 
the low-mode preconditioning leading to the overhead 
mentioned at the beginning of this subsection is switched off in the 5D solver.

\begin{table}[tbp]
   \vspace{3mm}
   \begin{center}
   \begin{tabular}{lll}
   \hline
   algorithm   & $c_{\rm mult}/10^6$ & $\alpha_{\rm mult}$ 
   \\ \hline
   HMC-4D      & 0.57(3)             & 0.68(2)
   \\
   HMC-5D      & 0.368(3)            & 0.710(3)
   \\ \hline
   \end{tabular}
   \caption{
      Fit parameters to Eq.~(\ref{eqn:run:hmc:Nmult_vs_m})
      for two algorithms HMC-4D and 5D.
   }
   \vspace{0mm}
   \label{tbl:run:hmc:Nmult:fit}
   \end{center}
\end{table}

For future reference, we fit $m$ dependence of the corrected $N_{\rm mult}$ 
into a simple power law
\bea
   N_{\rm mult} = c_{\rm mult}\,m^{-\alpha_{\rm mult}}.
   \label{eqn:run:hmc:Nmult_vs_m}
\eea
Fit parameters are summarized in Table~\ref{tbl:run:hmc:Nmult:fit}.
While data at small $m$ are subject to the finite size effects
as discussed in Sec.~\ref{sec:run:solver},
the fit parameters do not change significantly 
if we discard the data at $m\!\leq\!0.025$ from the fit.
Note that this is the cost per trajectory 
and the $m$ dependence of the autocorrelation,
which is not clear with our statistics, 
is not taken into account.
Thanks to the improved algorithms, 
$N_{\rm mult}$ has a much milder $m$ dependence than $m^{-2}$, 
which was employed to estimate the simulation cost 
with the standard HMC in Ref.\cite{Berlin_wall}.

In Table~\ref{tbl:run:hmc:Nmult},
we also list $N_{\rm mult}$ in the non-trivial topological sectors.
At least at the simulated quark mass $m\!=\!0.050$,
we have not observed a substantial $Q$ dependence of $N_{\rm mult}$.

\section{static quark potential}
\label{sec:pot}


We calculate the static quark potential to fix the lattice spacing 
through the Sommer scale \cite{r0}.
The temporal Wilson loops $W(r,t)$ are measured 
up to $t\!=\!N_t/2$ and $r\!=\!\sqrt{3}N_s/2$
with the spatial Wilson line parallel to one of the following six
directions
\bea
   (1,0,0), \ (1,1,0), \ (2,1,0), \ 
   (1,1,1), \ (2,1,1), \ (2,2,1).
   \label{eqn:pot:r-dir}
\eea
Gauge configurations separated by 10 HMC trajectories are smeared 
twenty times using a method proposed in Ref.\cite{smearing},
and we measure $W(r,t)$ every four smearing steps.
The computational cost of this measurement is not large:
it takes about 2 minutes per configuration 
on a single node of SR11000 with the sustained speed of 30\%.

\begin{figure}[tbp]
   \vspace{3mm}
   \begin{center}
      \includegraphics[width=0.48\linewidth,clip]{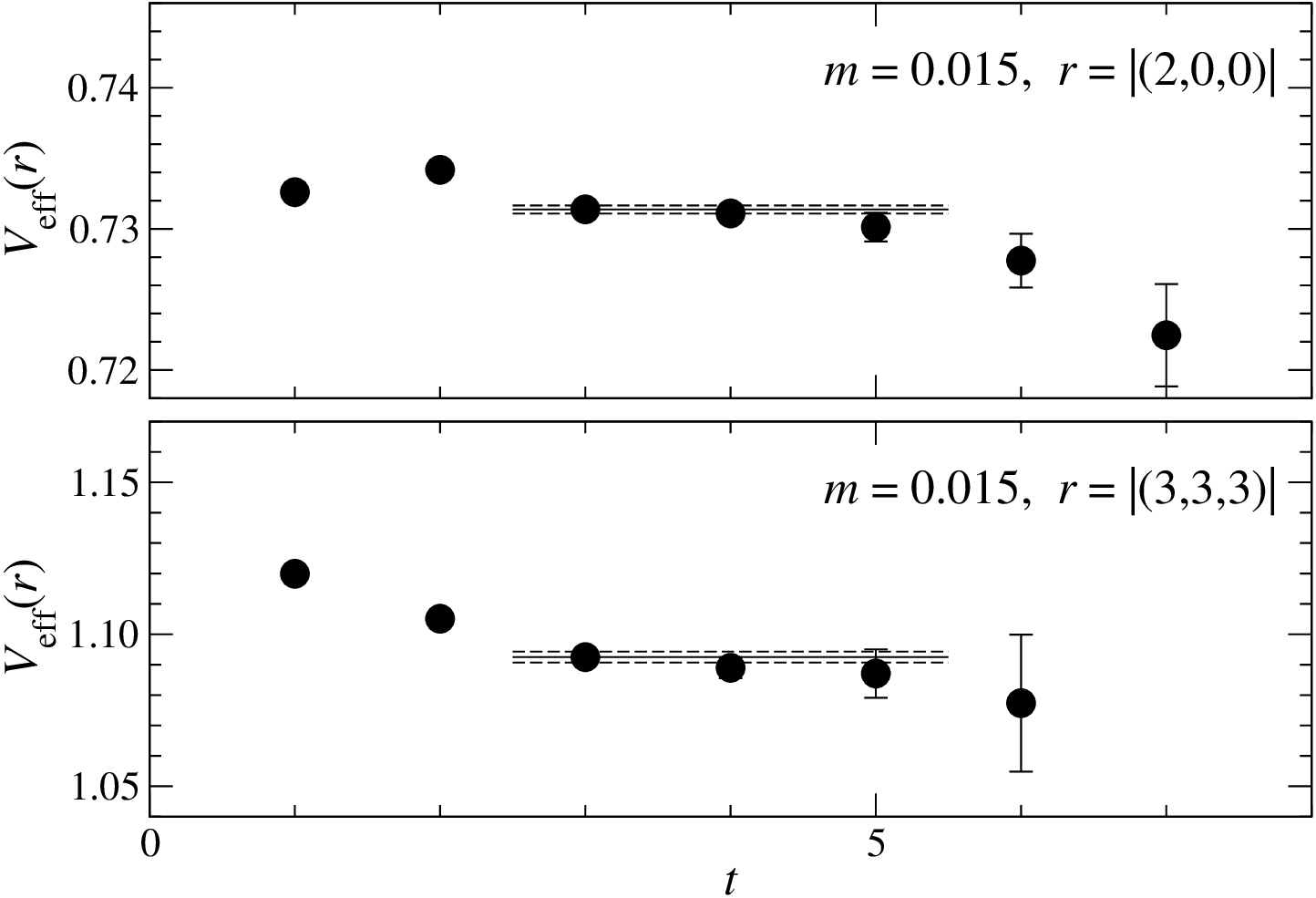}
      \hspace{2mm}
      \includegraphics[width=0.48\linewidth,clip]{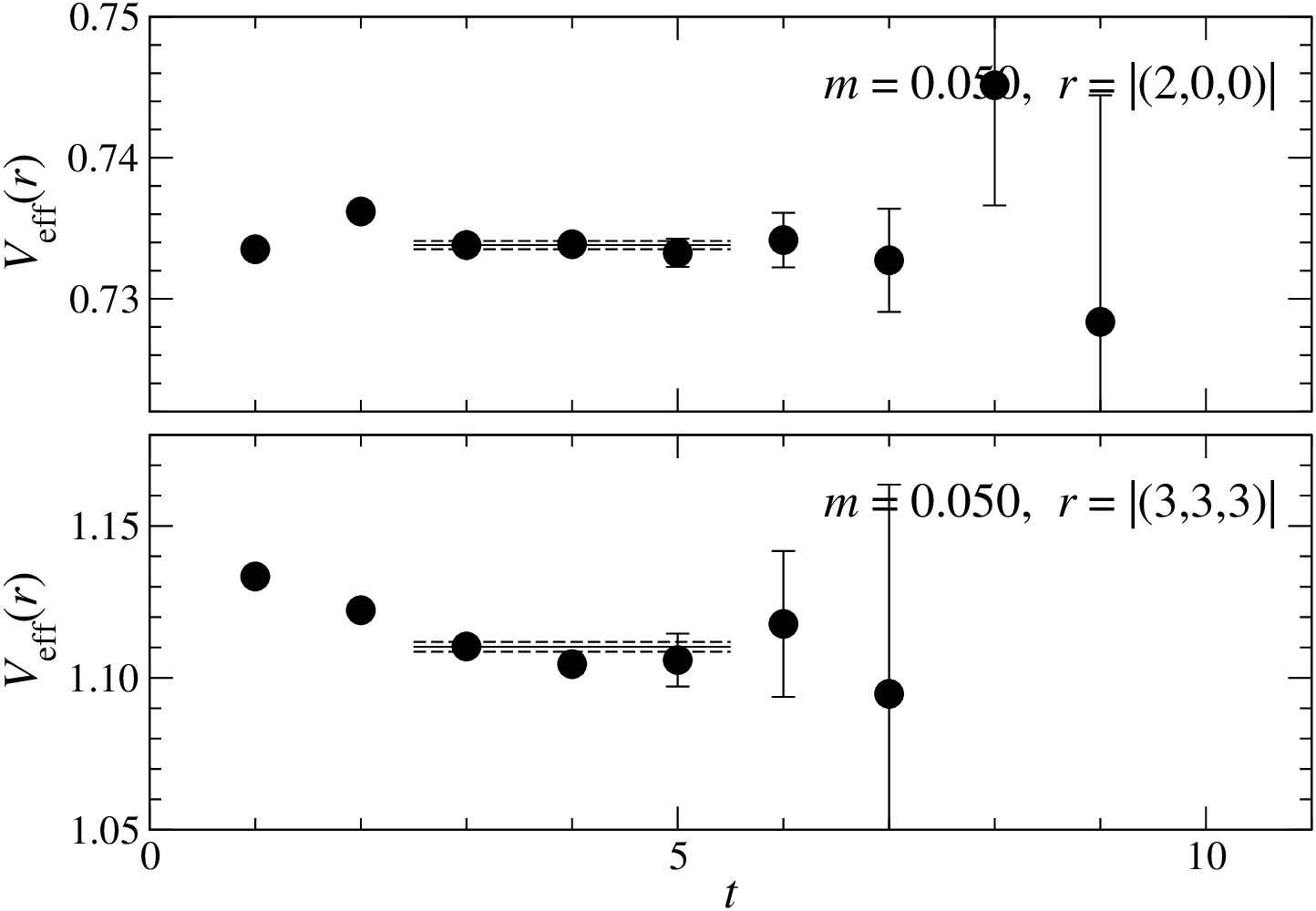}
   \end{center}
   \vspace{-3mm}
   \caption{
      Effective potential $V_{\rm eff}(r)$ at $r\!=\!2$ and $3\sqrt{3}$.
      Left and right panels show data at $m\!=\!0.015$ and 0.050.
   }
   \label{fig:pot:em}
\end{figure}

\begin{figure}[tbp]
   \vspace{3mm}
   \begin{center}
      \includegraphics[width=0.48\linewidth,clip]{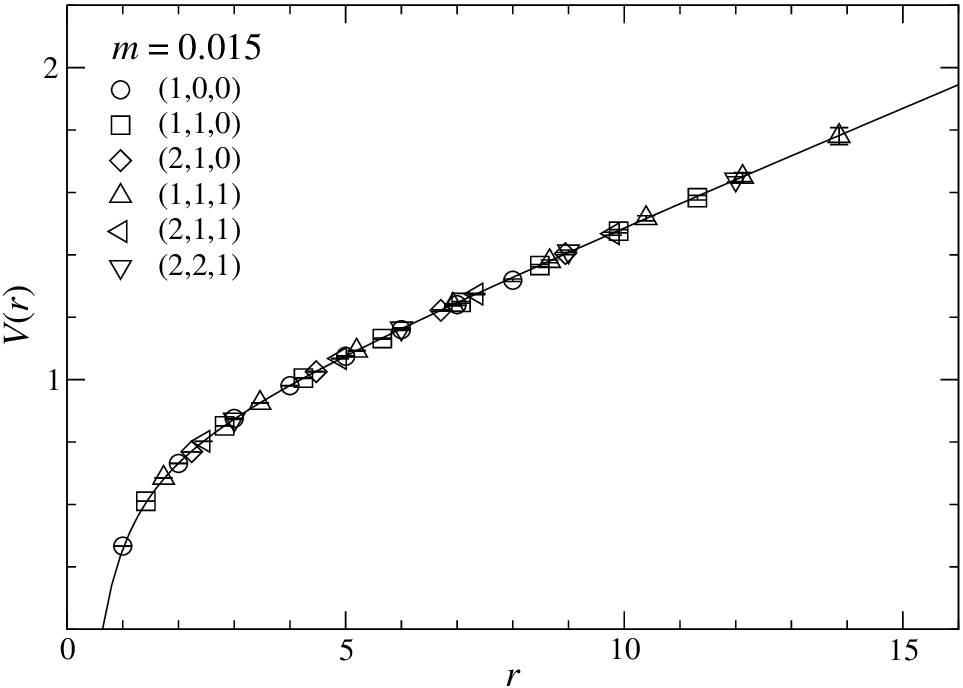}
      \hspace{2mm}
      \includegraphics[width=0.48\linewidth,clip]{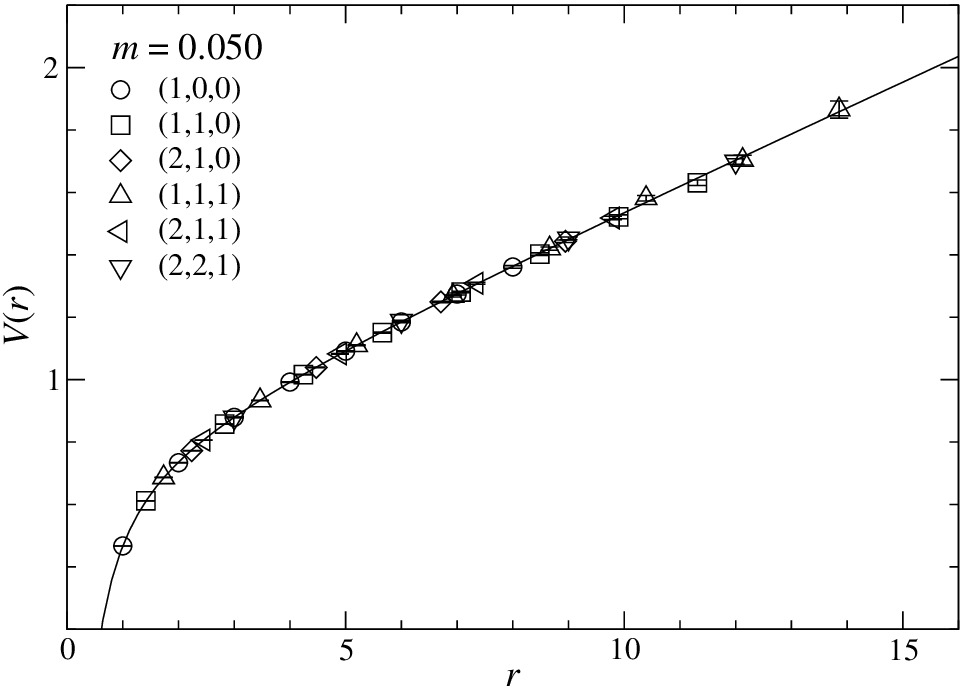}
   \end{center}
   \vspace{-3mm}
   \caption{
      Static quark potential $V(r)$ 
      at $m\!=\!0.015$ (left panel) 0.050 (right panel).
      The solid line shows the fit of Eq.~(\ref{eqn:pot:fit2}),
      whereas 
      figure legend represents
      the direction of the spatial Wilson line 
      listed in Eq.~(\ref{eqn:pot:r-dir}).
   }
   \label{fig:pot:V_vs_r}
\end{figure}



We determine the static potential $V(r)$ from the correlated fit
\bea
   W(r,t) 
   & = &
   C(r)\,\exp[-V(r)\,t]
   \label{eqn:pot:fit1}
\eea
at the number of the smearing steps 
which gives the maximum value of the overlap to the ground state $C(r)$.
The fit range $[t_{\rm min},t_{\rm max}]$ is set to $[3,5]$
by inspecting $t$-dependence of the effective potential
\bea
   V_{\rm eff}(r)
   & = &
   \ln[ W(r,t)/ W(r,t+1) ].
   \label{eqn:pot:Veff}
\eea
Examples of $V_{\rm eff}(r)$ are shown in Fig.~\ref{fig:pot:em},
and $V(r)$ is plotted as a function of $r$ in Fig.~\ref{fig:pot:V_vs_r}.

%


\begin{figure}[tbp]
   \vspace{3mm}
   \begin{center}
      \includegraphics[width=0.48\linewidth,clip]{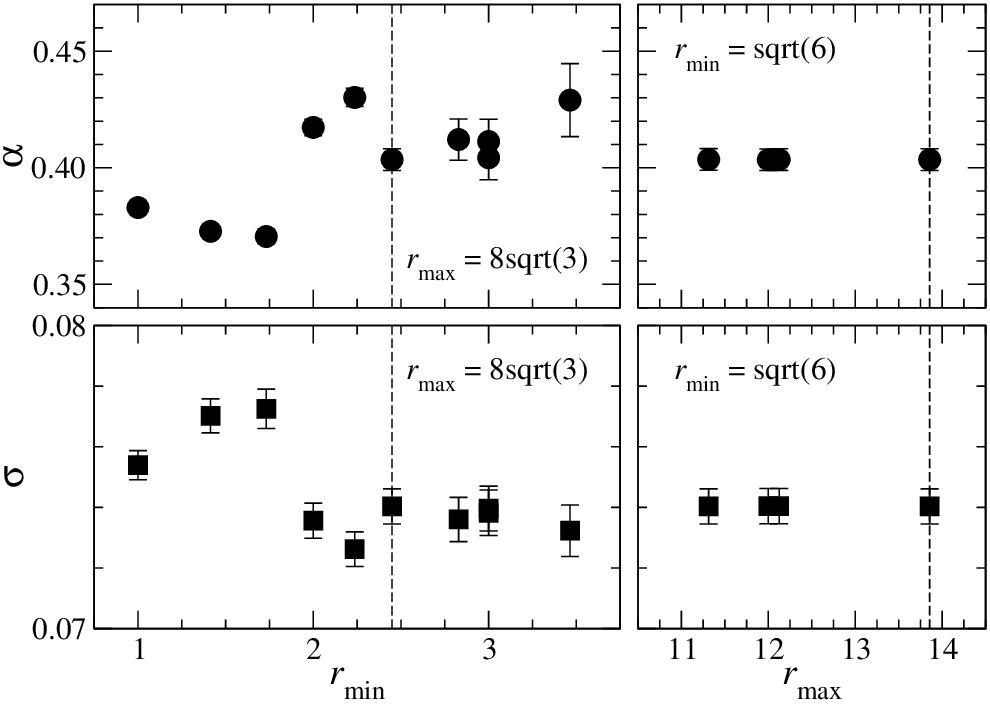}
   \end{center}
   \vspace{-3mm}
   \caption{
      Fit range $[r_{\rm min},r_{\rm max}]$ dependence of fit parameters
      $\alpha$ and $\sigma$ in Eq.~(\ref{eqn:pot:fit2})
      at $m\!=\!0.015$.
      Our choice $r_{\rm min}\!=\!\sqrt{6}$ and $r_{\rm max}\!=\!8\sqrt{3}$ 
      is shown by dashed lines.
   }
   \label{fig:pot:rmin-dep}
\end{figure}

\begin{table}[tbp]
\begin{center}
   \caption{
      Fit parameters in Eq.~(\ref{eqn:pot:fit2}) and 
      $r_0$ from Eq.~(\ref{eqn:pot:r0}).
      The first and second errors are statistical and systematic,
      respectively.
   }
   \vspace{3mm}
   \begin{tabular}{lll|lll|l}
      \hline
      $\beta$  & $m$      & $Q$      &
      $V_0$    & $\alpha$ & $\sigma$ & $r_0$
      \\ \hline
      2.30 & 0.015 &  0 & 0.786(3)(15)   & 0.403(5)(28)   & 
                          0.0740(6)(26)  & 4.103(14)(51)  
      \\
      2.30 & 0.025 &  0 & 0.776(4)(21)   & 0.389(5)(36)   &
                          0.0763(7)(37)  & 4.064(13)(59)
      \\
      2.30 & 0.035 &  0 & 0.769(4)(22)   & 0.381(5)(38)   &
                          0.0780(6)(34)  & 4.032(10)(49)
      \\
      2.30 & 0.050 &  0 & 0.760(3)(16)   & 0.375(5)(30)   &
                          0.0812(7)(26)  & 3.963(11)(42)
      \\
      2.30 & 0.070 &  0 & 0.756(4)(25)   & 0.373(6)(42)   &
                          0.0832(7)(40)  & 3.917(11)(40)
      \\
      2.30 & 0.100 &  0 & 0.749(4)(18)   & 0.368(5)(30)   &
                          0.0864(7)(30)  & 3.852(10)(35)
      \\
      \hline
      2.30 & 0.050 & -2 & 0.759(7)(24)   & 0.370(11)(44)  &
                          0.0803(12)(34) & 3.993(16)(29)
      \\	       	     
      \hline
      2.30 & 0.050 & -4 & 0.758(6)(20)   & 0.368(10)(36)  &
                          0.0811(12)(29) & 3.976(19)(36)   
      \\
      \hline
   \end{tabular}
   \label{tbl:pot:fit2}
\end{center}
\end{table}

We do not observe any clear sign of the string breaking
even at our smallest quark mass $\sim m_s/6$
possibly due to small overlap 
of the spatial Wilson line to the two static-light meson state.
We therefore fit $V(r)$ to the conventional form with 
the perturbative Coulomb and the linear confinement terms
\bea
   V(r)
   & = &
   V_0 - \alpha/r + \sigma\,r.
   \label{eqn:pot:fit2}
\eea
The fit range is set to 
$[r_{\rm min},r_{\rm max}]\!=\![\sqrt{6},8\sqrt{3}]$ at all quark masses
from the stability of $\alpha$ and $\sigma$ against the choice of the fit range 
shown in Fig.~\ref{fig:pot:rmin-dep}.
Fit results are summarized in Table~\ref{tbl:pot:fit2}.
Systematic errors due to the choice of the fit ranges
are estimated from the (maximum) change in the fit parameters
by shifting $[t_{\rm min},t_{\rm max}]$ to $[4,6]$ or
varying $r_{\rm min}$ and $r_{\rm max}$ in ranges
$r_{\rm min}\!\in\![2,3]$ and $r_{\rm max}\!\in\![8\sqrt{2},8\sqrt{3}]$.
These are added in quadrature in Table~\ref{tbl:pot:fit2}.
The fit curves are shown in Fig.~\ref{fig:pot:V_vs_r}.


\vspace{3mm}
\begin{figure}[tbp]
   \begin{center}
      \includegraphics[width=0.48\linewidth,clip]{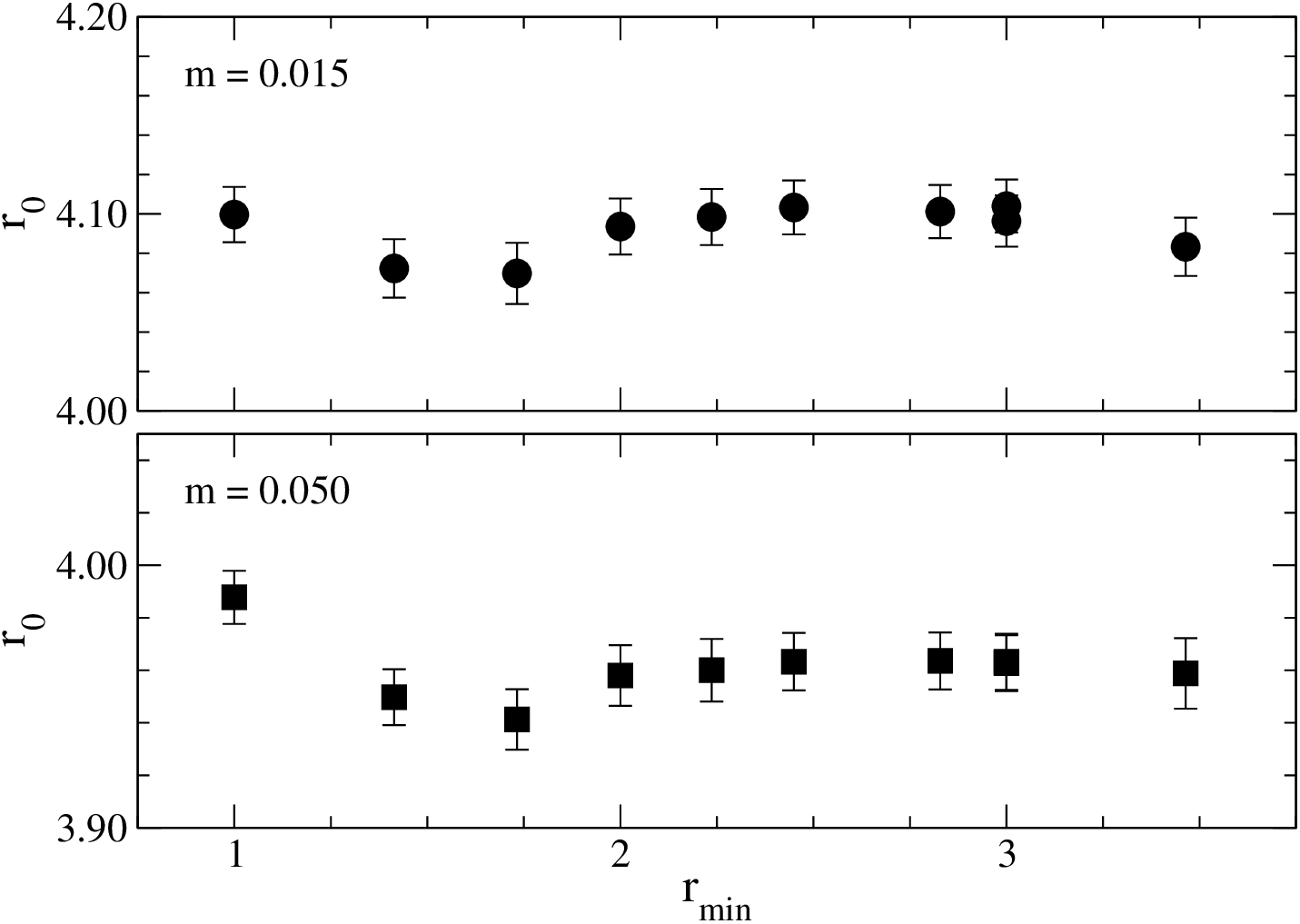}
   \end{center}
   \vspace{-3mm}
   \caption{
      Sommer scale $r_0$ as a function of $r_{\rm min}$
      at $m\!=\!0.015$ (top panel) and 0.050 (bottom panel).
      We set $r_{\rm max}\!=\!8\sqrt{3}$.
   }
   \label{fig:pot:r0_vs_rmin}
\end{figure}

The Sommer scale $r_0$ is defined through 
the derivative of $V(r)$ in the intermediate region of $r$ \cite{r0}
\bea
   r_0^2 \left. dV(r)/dr \right|_{r=r_0} 
   & = &
   1.65.
   \label{eqn:pot:r0:def}
\eea
We fix $r_0$ in our simulations 
through the parametrization Eq.~(\ref{eqn:pot:fit2})
\bea
   r_0 
   & = &
   \sqrt{\frac{1.65-\alpha}{\sigma}}
   \label{eqn:pot:r0}
\eea
instead of the numerical derivative.
In Fig.~\ref{fig:pot:rmin-dep}, 
we observe that 
$r_{\rm min}$ dependence of $\alpha$ is large and correlated to 
that of $\sigma$ at $r_{\rm min}\! \lesssim \! 2$.
It turns out that these uncertainties 
partially cancel each other in the ratio Eq.~(\ref{eqn:pot:r0})
leading to a mild $r_{\rm min}$ dependence of $r_0$ 
shown in Fig.~\ref{fig:pot:r0_vs_rmin}.
Therefore, as intended in Ref.\cite{r0},
$r_0$ provides a more reliable estimate of the lattice scale 
than the previously-used input $\sqrt{\sigma}$
even through the parametrization Eq.~(\ref{eqn:pot:fit2}) 
over the wide region of $r$.
Our numerical results are summarized in Table~\ref{tbl:pot:fit2}.
We note that 
$r_0$ from three topological sectors are consistent with each other
within their statistical accuracy. Its $Q$ dependence is therefore
ignored in the following analysis.

\begin{table}[tbp]
\begin{center}
   \caption{
      Fit parameters in Eqs.~(\ref{eqn:pot:a_vs_m}) and 
      (\ref{eqn:pot:ainv_vs_m}). 
      The first and second errors are statistical and systematic.
   }
   \vspace{3mm}
   \begin{tabular}{l|llll}
      \hline
      fit form  &  $\chi^2/{\rm dof}$ &
      $c_0^{(\prime)}$  & $c_1^{(\prime)}$  &$c_2^{(\prime)}$  
      \\ \hline
      Eq.~(\ref{eqn:pot:a_vs_m}) &
      1.60      & 0.1184(3)(17)  & 0.0914(49)(57)   & -- 
      \\
      Eq.~(\ref{eqn:pot:a_vs_m}) &
      0.47      & 0.1172(6)(17)  & 0.145(24)(8)     & -0.45(20)(2)
      \\
      Eq.~(\ref{eqn:pot:ainv_vs_m}) &
      2.00      & 8.43(2)(12)    & -5.93(32)(50)    & --
      \\
      Eq.~(\ref{eqn:pot:ainv_vs_m}) &
      0.46      & 8.52(4)(13)    & -10.0(1.6)(0.8)  & 34(13)(3)
      \\
      \hline
   \end{tabular}
   \label{tbl:pot:a_vs_m}
\end{center}
\end{table}

\vspace{3mm}
\begin{figure}[tbp]
   \begin{center}
      \includegraphics[width=0.48\linewidth,clip]{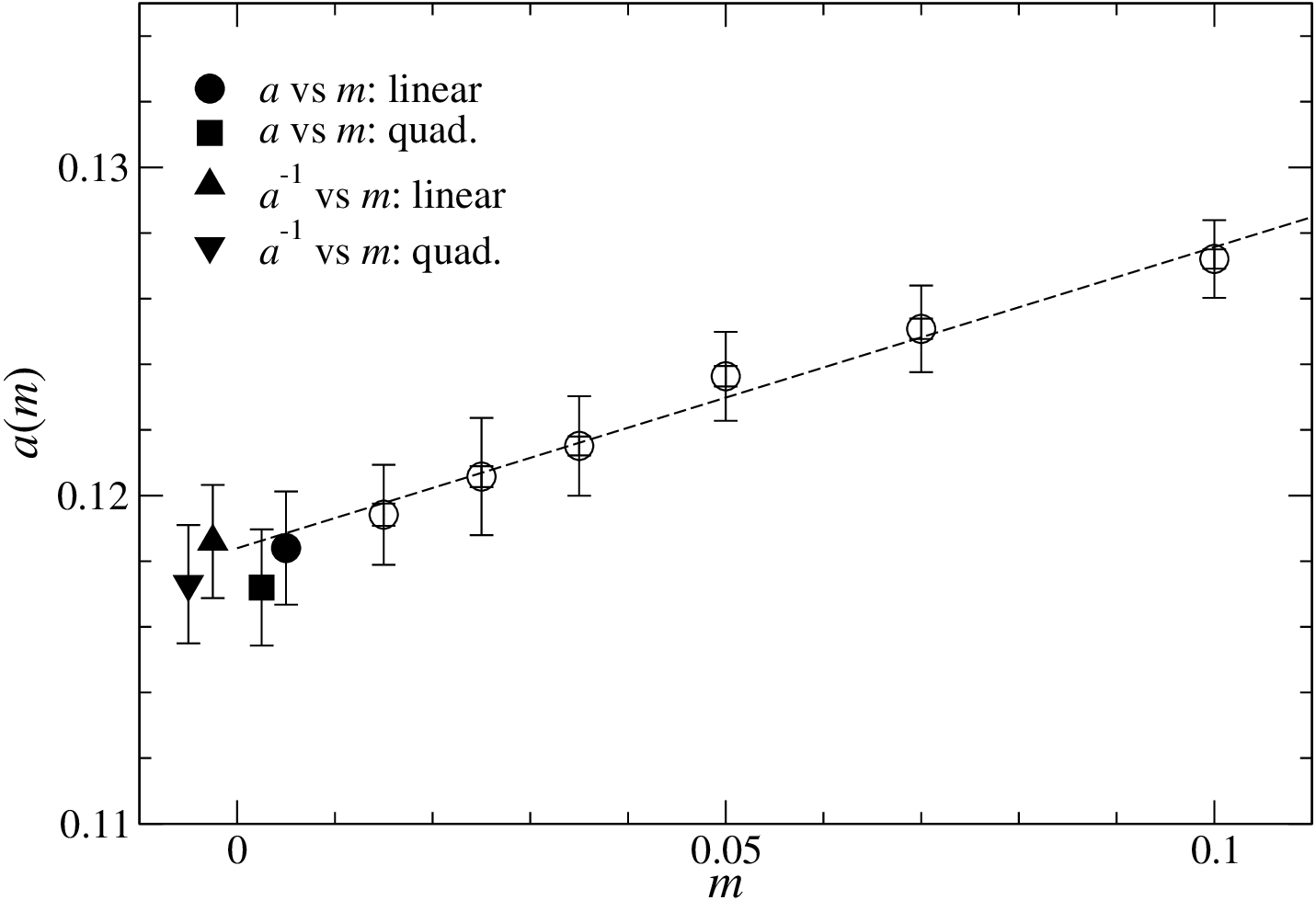}
   \end{center}
   \vspace{-3mm}
   \caption{
      Chiral extrapolation of $a(m)$ 
      with linear form of Eq.~(\ref{eqn:pot:a_vs_m}) (circles).
      Extrapolated values with other fitting forms 
      in Eqs.~(\ref{eqn:pot:a_vs_m}) (square) and 
      (\ref{eqn:pot:ainv_vs_m}) (triangles) are also plotted.
      Two error bars for these symbols show statistical and total errors.
   }
   \label{fig:pot:a_vs_m}
\end{figure}

We employ an input $r_0\!=\!0.49$~fm to fix the scale.
A quantity $a(m)\!=\!0.49/r_0(m)$ is then extrapolated to the chiral limit 
testing the following fitting functions up to quadratic order
\bea
   a(m)
   & = &
   c_0 + c_1\,m \ (+ \, c_2\,m^2),
   \label{eqn:pot:a_vs_m}
   \\
   a(m)^{-1} 
   & = &
   c_0^\prime + c_1^\prime\,m \ (+ \, c_2^\prime\,m^2).
   \label{eqn:pot:ainv_vs_m}
\eea
Fit parameters are summarized in Table~\ref{tbl:pot:a_vs_m}.
Since we have accurate data in the wide range of $m$,
the lattice spacing in the chiral limit $a\!=\!a(0)$ 
is very stable against the choice of the fitting function
as plotted in Fig.~\ref{fig:pot:a_vs_m}.
We obtain 
\bea
   a
   & = &
   0.1184(3)(17)(12)~{\mbox{fm}},
   \label{eqn:pot:a}
\eea
where the central value is from the linear form of Eq.~(\ref{eqn:pot:a_vs_m})
and the first error is statistical.
The second error is due to the choice of the fit ranges 
for Eqs.~(\ref{eqn:pot:fit1}) and (\ref{eqn:pot:fit2}).
The third represents the uncertainty due to the choice of the chiral 
extrapolation form and estimated by the maximum deviation in $a$ 
from other three forms 
Eqs.~(\ref{eqn:pot:a_vs_m}) (quadratic) and (\ref{eqn:pot:ainv_vs_m}).


\begin{figure}[tbp]
\vspace{3mm}
   \begin{center}
      \includegraphics[width=0.48\linewidth,clip]{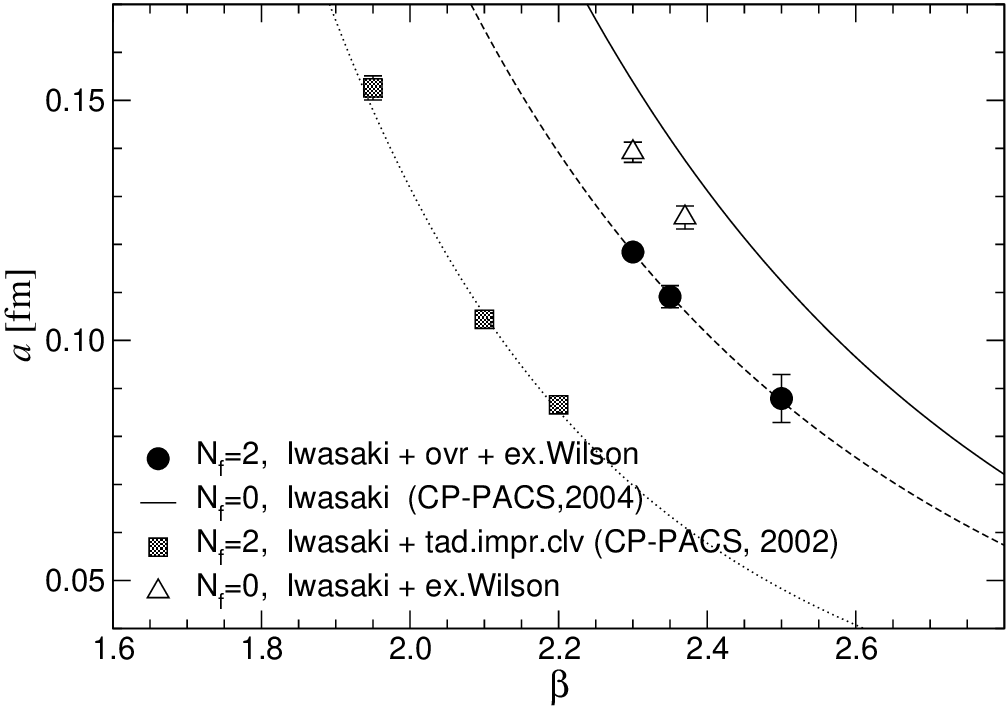}
   \end{center}
   \vspace{-3mm}
   \caption{
      Lattice spacing determined from $r_0$ in simulations 
      with Iwasaki gauge action.
      Circles are our estimate Eq.~(\ref{eqn:pot:a}) and 
      those in Ref.\cite{e-regime:sigma:Nf2:RG+Ovr:JLQCD:1}
      and our preparatory study.
      These are compared with 
      results in pure gauge theory \cite{SSF:Nf0:RG} (solid line), 
      only with extra fermions \cite{Lat06:JLQCD:Yamada} (triangles),
      and with an improved Wilson fermions 
      \cite{Spectrum:Nf2:RG+Clv:CP-PACS} (squares).
   }
   \label{fig:pot:a_vs_beta}
\end{figure}

\subsection{$\beta$ shift}

Inclusion of dynamical quarks into simulations 
generally makes us decrease $\beta$ 
to keep the lattice spacing fixed.
The magnitude of the $\beta$ shift depends on the fermion formulation.
A sizable negative shift, 
or too large bare coupling in other words,
could cause problematic lattice artifacts:
for instance,
one may suffer from 
a remnant of the fundamental-adjoint phase transition \cite{FAPT}.
In practice, 
some evidence of non-trivial phase structure has been found 
in previous unquenched simulations 
even at relatively fine lattice spacing $a \! \approx \! 0.1$~fm
\cite{PhaseDiag:Nf2:Plq+Wlsn,PhaseDiag:Nf2:Plq+tmW,PhaseDiag:Nf3:Plq+Clv}.

In Fig.~\ref{fig:pot:a_vs_beta},
we compare the lattice spacing determined from $r_0$ 
in our and previous simulations with the Iwasaki gauge action.
The $\beta$ shift due to the extra-Wilson fermions is not expected 
to be large, since effects of their high modes are cancelled  
in the ratio Eq.~(\ref{eqn:exW+extmW}).
This is supported by the one-loop calculation of the 
vacuum polarization function in Ref.\cite{exW+extmW:JLQCD},
and Fig.~\ref{fig:pot:a_vs_beta} provides a non-perturbative confirmation.

The figure also shows that 
the dynamical overlap fermions lead to small $\beta$ shift,
which is in a good accordance with 
an one-loop calculation in Ref.\cite{beta_shift}.
The net shift is substantially smaller than that from the tadpole-improved 
clover fermions.
Therefore the $\beta$ shift is less problematic 
in dynamical overlap simulations even with the unphysical fermions,
and this is also likely the case in three flavor QCD.

\section{Locality}
\label{sec:locality}


The locality of the overlap operator $D$ is closely related to 
the properties of low-lying modes $(\lambda_{\rmW,k},u_{\rmW,k})$ of $\Hw$.
It is proved in Ref.\cite{locality:HPL}
that $D$ is exponentially local 
$|D(x,y)| \! \propto \! e^{-|x-y|/l}$
if $|\lambda_{\rmW,k}|$ has a positive lower bound.
This does exist in our simulations by the use of the auxiliary determinant
$\det[ \Delta_\rmW ]$.

The central concern is therefore the size of the localization range $l$,
which should be smaller than the QCD scale $\Lambda_{\rm QCD}^{-1}$.
In Refs.\cite{locality:GS,locality:GSS},
it is argued that 
the range of $D$ is characterized by two sets of eigenmodes of $\Hw$:
i) localized low-lying modes, whose maximum eigenvalue is denoted
by $\bar{\lambda}_\rmW$ in the following,
and ii) extended modes with higher eigenvalues.
It leads to a conjecture
\bea
   |D(x,y)|
   & \sim &
   \bar{\lambda}_\rmW \rho(\bar{\lambda}_\rmW) 
   \exp\left[ - \frac{|x-y|}{2l_{\rmW,l}(\bar{\lambda}_\rmW)} \right]
   + 
   C\, \exp\left[ -\lambda_{\rmW,c}|x-y| \right],
   \label{eqn:local:exp_local}
\eea
where $l_{\rmW,l}$ 
is the localization length of the localized modes,
and $\rho$ represents the spectral density.
The parameter $\lambda_{\rmW,c}$ is the so-called mobility edge,
which separates the localized and extended modes.
The prefactor of the first term follows from a steep rise in $\rho$
observed in Refs.\cite{locality:GS,locality:GSS}.
The extended modes govern the localization properties of $D$
through $\lambda_{\rmW,c}$ 
provided that 
$C\! \gg \! \bar{\lambda}_\rmW \, \rho(\bar{\lambda}_\rmW)$
and $\lambda_{\rmW,c} \! \lesssim \! (2l_{\rmW,l})^{-1}$.

\begin{figure}[tbp]
\begin{center}
   \vspace{5mm}
   \includegraphics[width=0.48\linewidth,clip]{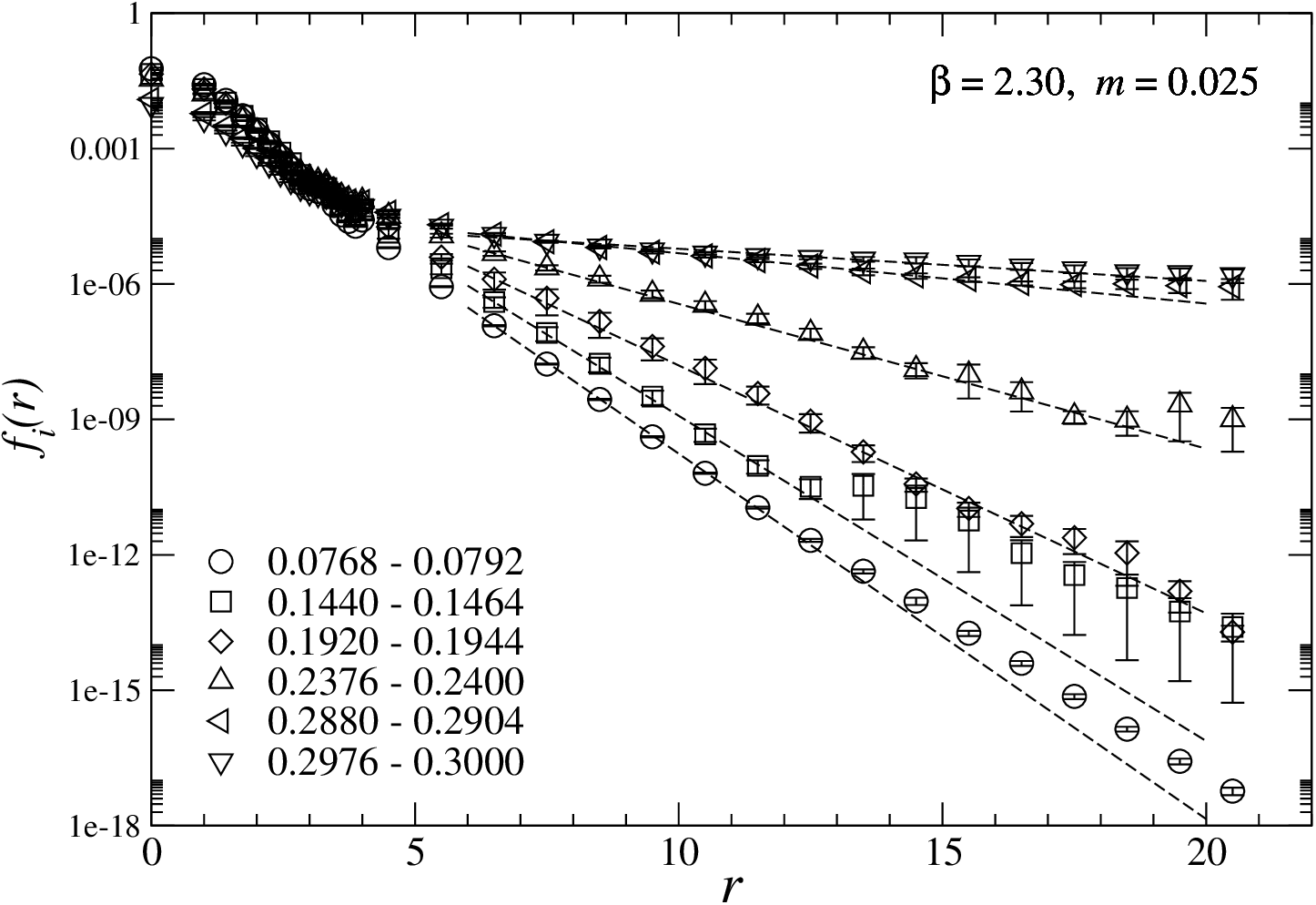}
   \vspace{-3mm}
   \caption{
      Function $f_{\rmW,i}(r)$ at $m\!=\!0.025$.
      The figure legend shows windows in $|\lambda_\rmW|$.
      The dashed lines show the fit Eq.~(\ref{eqn:local:local_length}).
   }
   \label{fig:local:decay_func:Hw}
\end{center}
\end{figure}

\begin{figure}[tbp]
\begin{center}
   \vspace{5mm}
   \includegraphics[width=0.48\linewidth,clip]{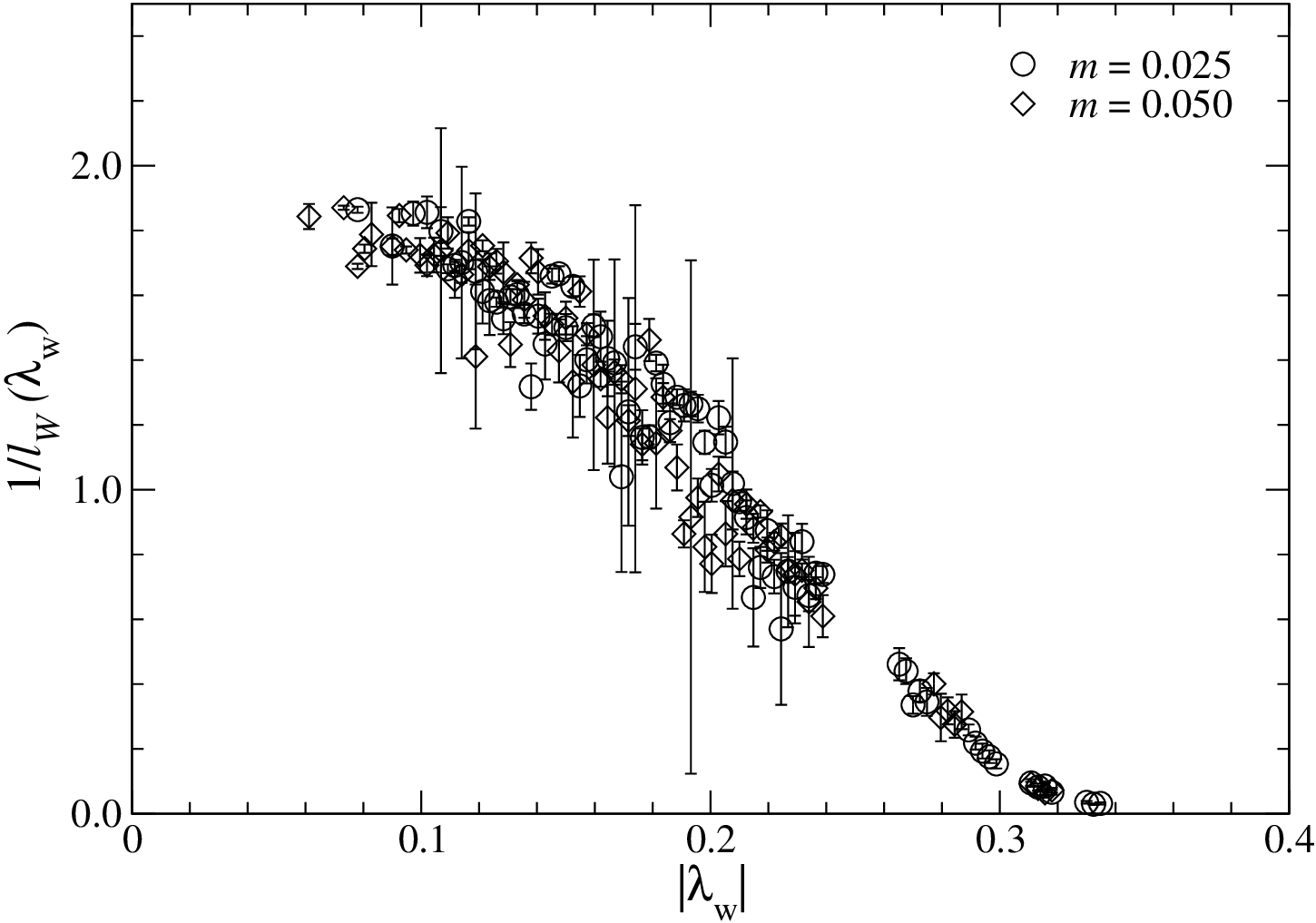}
   \vspace{-3mm}
   \caption{
      Inverse of localization length for low-lying modes of $\Hw$ 
      as a function of their eigenvalue $|\lambda_\rmW|$.
   }
   \label{fig:local:local_length:Hw}
\end{center}
\end{figure}

We estimate $\lambda_{\rmW,c}$ in our simulations in the following steps.
First, we locate the lattice site $y_k$, where the $k$-th lowest mode has 
its maximum magnitude $\eta_k(x)\!=\!u_{\rm W,k}(x)^\dagger u_{\rm W,k}(x)$. 
Then a function characterizing its decay is obtained by the average
\bea
   f_{\rmW,k}(r)
   & = &
   \frac{1}{N_{\rm pt}(r)}
   \sum_{x,|x-y_k|=r} \eta_k(x),
   \label{eqn:local:decay_func:Hw}
\eea
where $N_{\rm pt}(r)$ represents the number of lattice points which have
the same distance $r$ from $y_k$.
Since the spectrum of $\Hw$ depends on the gauge configuration,
we consider a range $0 \! \leq \! |\lambda_\rmW| \! \leq \! 0.3$,
which is divided into windows with its size of 
$\Delta \lambda_\rmW \!=\! 0.3/125$,
and $f_{\rm W,k}(r)$ is averaged over the eigenmodes in each window.
We calculate $f_{\rmW,i}(r)$, where $i$ is now a window index,
at $m\!=\!0.025$ and 0.050 
using 10\,--\,40 configurations separated by 10 trajectories.
Because of the small statistics,
we set the bin size to 1 configuration,
which possibly underestimates the statistical error quoted in this section.
An example of $f_{\rm W,i}(r)$ is plotted 
in Fig.~\ref{fig:local:decay_func:Hw}.
Generally speaking,
low modes decay exponentially at large $r$
and the decay rate decreases as $|\lambda_\rmW|$ increases.

The localization length at $i$-th window $l_\rmW(|\lambda_\rmW|_i)$ 
is determined by fitting $f_{\rmW,i}(r)$ at large $r$ to 
\bea
   f_{\rmW,i}(r) 
   & = & 
   c_i
   \exp\left[ - \frac{r}{l_\rmW(|\lambda_\rmW|_i)} \right].
   \label{eqn:local:local_length}
\eea
Its $|\lambda_\rmW|$ dependence is plotted 
in Fig.~\ref{fig:local:local_length:Hw}.
The mobility edge $\lambda_{\rmW,c}$ is then estimated as 
$|\lambda_{\rmW}|$ at which $l_\rmW(|\lambda_\rmW|)^{-1}$ vanishes.
It turns out that 
$\lambda_{\rmW,c}$ has small $m$ dependence but is roughly $0.33$.
We obtain $\lambda_{\rmW,c}^{-1} \sim 550$~MeV in physical units
from our estimate of $a$ in Eq.~(\ref{eqn:pot:a}).


\begin{figure}[tbp]
\begin{center}
   \vspace{5mm}
   \includegraphics[width=0.48\linewidth,clip]{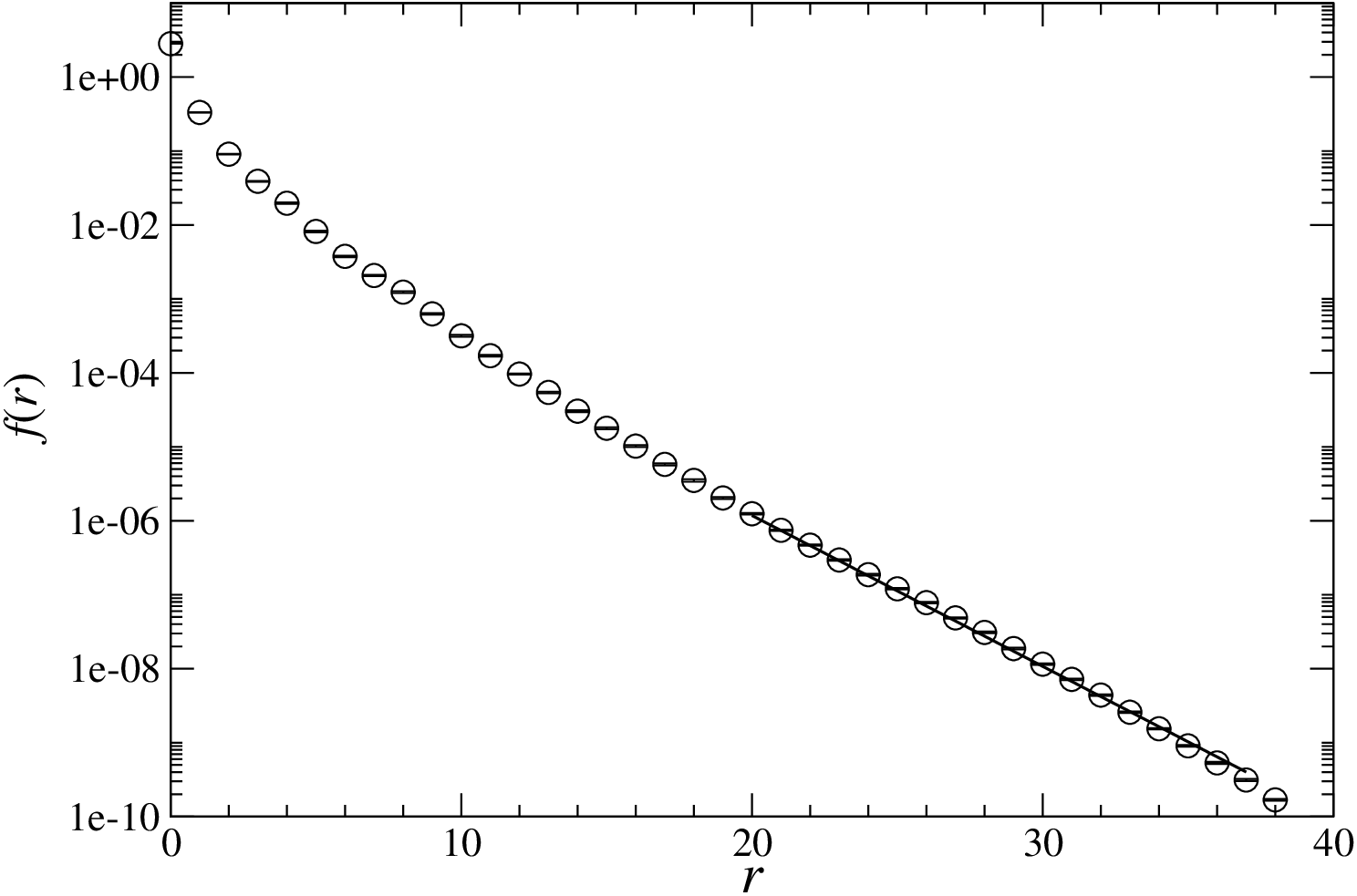}
   \vspace{-3mm}
   \caption{
      Function $f(r)$ at $m\!=\!0.025$ as a function of distance $r$.
      The solid line shows the exponential fit 
      $f(r) \! \propto \! e^{-r/l}$.
   }
   \label{fig:local:local_range:ov}
\end{center}
\end{figure}

At $m\!=\!0.025$,
we also study the localization properties directly from 
the overlap operator multiplied to a point like quark vector
\bea
   f(r)
   & = &
   \max_{x,\ |x-y|_1=r}
   \left\{
      \sum_{x^\prime} D(x,x^\prime)\delta(x^\prime-y)
   \right\}.
   \label{eqn:local:decay_func:ov}
\eea
Here we use the taxi driver distance $|x-y|_1\!=\!\sum_\mu |x_\mu-y_\mu|$
to avoid underestimating the localization range $l$.
We obtain 
\bea
   l^{-1} 
   & = & 
   796(2)~\mbox{MeV}
\eea
by an exponential fit $f(r) \propto e^{-r/l}$
shown in Fig.~\ref{fig:local:local_range:ov}.
Both of $\lambda_{\rmW,c}$ and $l$ therefore 
suggest that 
the overlap operator is exponentially local with 
a localization range smaller than $\Lambda_{\rm QCD}^{-1}$ 
in our simulations.

\section{Conclusions}
\label{sec:conclusion}

In this article, 
we simulate two-flavor QCD with dynamical overlap quarks 
on the reasonably large ($1.9$~fm) and fine ($a\!=\!0.12$~fm) lattice.
The high statistics of 10,000 trajectories are accumulated 
at sea quark masses down to $m_s/6$.
The key step leading to such large-scale simulations is 
the suppression of the (near-)zero modes of $\Hw$ 
by the auxiliary determinant.
This enables us to use relatively cheap approximation of $\sgn[\Hw]$
and also to avoid the substantial overhead to deal with the 
discontinuity of the overlap action.
The use of the 5D CG algorithm, the Hasenbusch mass preconditioning
and the multiple time scale MD integration also reduces the simulation
cost to a large extent.

Dynamical overlap simulations are still computationally demanding 
compared to the domain-wall fermions \cite{Lat07:JLQCD:Matsufuru}.
The complexity of the overlap formulation however suggests that
there is much room of improvement in the implementation of HMC.
The low-mode preconditioning for the 5D solver 
is developed after this study and implemented 
in our latest runs \cite{Lat07:JLQCD:Matsufuru,Lat07:JLQCD:Hashimoto}.
Further improvement in the solver algorithm, 
especially in the 5D solver (or alternatives) 
to invert $(D^\dagger D)$,
is a central concern for pushing simulations to larger volumes.
The test of MD integration schemes with less discretization error
and/or a further tuning of the HMC parameters and the unit trajectory length
are also interesting subjects to be studied.

We are now studying various non-perturbative aspects of 
two-flavor QCD using the generated gauge ensembles.
The chiral condensate is one of the most fundamental parameters in ChPT
and has been determined in Refs.\cite{e-regime:sigma:Nf2:RG+Ovr:JLQCD:1,e-regime:sigma:Nf2:RG+Ovr:JLQCD:2}.
Studies of the low-lying hadron spectrum \cite{Lat07:JLQCD:Noaki},
the kaon $B$ parameter \cite{Lat07:JLQCD:Yamada}
and the pion form factor \cite{Lat07:JLQCD:Kaneko}
are in progress with paying particular attention to 
the consistency of their chiral behavior with ChPT.
Finite size corrections based on ChPT are 
also important issue in these studies.
Our calculation of the topological susceptibility 
\cite{chi_t:Nf2:RG+Ovr:JLQCD+TWQCD}
is an important step to study the nature of the QCD vacuum
in fixed topological sectors.
The pion mass splitting through the vector and axial-vector current 
correlators \cite{Lat07:JLQCD:Shintani}
is an example for which the exact chiral symmetry is crucial
and might be difficult to study even with the domain-wall fermions.
Finally,
our simulations have been already extended to three-flavor QCD
\cite{Lat07:JLQCD:Matsufuru,Lat07:JLQCD:Hashimoto}
for fully realistic studies of QCD.

\section*{Acknowledgment}

Numerical simulations are performed on Hitachi SR11000 and 
IBM System Blue Gene Solution 
at High Energy Accelerator Research Organization (KEK) 
under a support of its Large Scale Simulation Program
(No.~06-13 and 07-16).
We thank J.~Doi, H.~Samukawa and S.~Shimizu 
of IBM Japan Tokyo Research Laboratory 
for assembler coding on the Blue Gene computer.
This work is supported in part 
by the Grant-in-Aid of the Ministry of Education 
(No.~17340066, 17540259, 17740171, 18034011, 18340075, 18740167, 
18840045, 19540286 and 19740160).
The work of HF is also supported by Nishina Memorial Foundation.


\end{document}